\documentclass[apj]{emulateapj}

\slugcomment{The Astrophysical Journal, 835:55 (17pp), 2017 January 20}

\shorttitle{The 1.1~mm Survey of the SMC}
\shortauthors{Takekoshi et al.}

\begin{document}
\title{The 1.1~mm Continuum Survey of the Small Magellanic Cloud:\\Physical Properties and Evolution of the Dust-selected Clouds\footnote{{\it Herschel} is an ESA space observatory with science instruments provided by European-led Principal Investigator consortia and with important participation from NASA.}}
\author{
Tatsuya~Takekoshi\altaffilmark{1,2,3},
Tetsuhiro~Minamidani\altaffilmark{1,2,4,5},
Shinya~Komugi\altaffilmark{3,6,7},
Kotaro~Kohno\altaffilmark{6,8},
Tomoka~Tosaki\altaffilmark{9},
Kazuo~Sorai\altaffilmark{1,4},
Erik~Muller\altaffilmark{3},
Norikazu~Mizuno\altaffilmark{3,5,10},
Akiko~Kawamura\altaffilmark{3},
Toshikazu~Onishi\altaffilmark{11},
Yasuo~Fukui\altaffilmark{12},
Hajime~Ezawa\altaffilmark{3,5},
Tai~Oshima\altaffilmark{2,5,13},
Kimberly~S.~Scott\altaffilmark{14,15},
Jason~E.~Austermann\altaffilmark{14,16},
Hiroshi~Matsuo\altaffilmark{5,13},
Itziar~Aretxaga\altaffilmark{17},
David~H.~Hughes\altaffilmark{17},
Ryohei~Kawabe\altaffilmark{2,3,5,10,18},
Grant~W.~Wilson\altaffilmark{14},
and Min~S.~Yun\altaffilmark{14}
}
\altaffiltext{1}{Department of Cosmosciences, Graduate School of Science, Hokkaido University, Sapporo 060-0810, Japan}
\altaffiltext{2}{Nobeyama Radio Observatory, National Astronomical Observatory of Japan (NAOJ), National Institutes of Natural Sciences (NINS), 462-2, Nobeyama, Minamimaki, Minamisaku, Nagano 384-1305, Japan}
\altaffiltext{3}{Chile Observatory, National Astronomical Observatory of Japan (NAOJ), National Institutes of Natural Sciences (NINS), 2-21-1, Osawa, Mitaka, Tokyo 181-8588, Japan}
\altaffiltext{4}{Department of Physics, Faculty of Science, Hokkaido University, Sapporo 060-0810, Japan}
\altaffiltext{5}{Department of Astronomical Science, School of Physical Science, SOKENDAI (The Graduate University for Advanced Studies), 2-21-1, Osawa, Mitaka, Tokyo 181-8588, Japan}
\altaffiltext{6}{Institute of Astronomy, The University of Tokyo, 2-21-1 Osawa, Mitaka, Tokyo 181-0015, Japan}
\altaffiltext{7}{Kogakuin University, 2665-1 Nakano, Hachioji, Tokyo 192-0015, Japan}
\altaffiltext{8}{Research Center for the Early Universe, University of Tokyo, 7-3-1 Hongo, Bunkyo, Tokyo 113-0033, Japan}
\altaffiltext{9}{Joetsu University of Education, Joetsu, Niigata 943-8512, Japan}
\altaffiltext{10}{Joint ALMA Observatory, Vitacura, Santiago, Chile.}
\altaffiltext{11}{Department of Physical Science, Osaka Prefecture University, Gakuen 1-1, Sakai, Osaka 599-8531, Japan}
\altaffiltext{12}{Department of Astrophysics, Nagoya University, Chikusa-ku, Nagoya 464-8602, Japan}
\altaffiltext{13}{Advanced Technology Center, National Astronomical Observatory (NAOJ), National Institutes of Natural Sciences (NINS), 2-21-1, Osawa, Mitaka, Tokyo 181-8588, Japan}
\altaffiltext{14}{Department of Astronomy, University of Massachusetts, Amherst, MA 01003, USA}
\altaffiltext{15}{North American ALMA Science Center, National Radio Astronomy Observatory, 520 Edgemont Road, Charlottesville, Virginia 22903, USA}
\altaffiltext{16}{National Institute of Standards and Technology, Boulder, CO 80305, USA}
\altaffiltext{17}{Instituto Nacional de Astrof\'{i}sica, \'{O}ptica y Electr\'{o}nica (INAOE), 72000 Puebla, Mexico}
\altaffiltext{18}{Division of Radio Astronomy, National Astronomical Observatory of Japan, 2-21-1 Osawa, Mitaka, Tokyo 181-8588 Japan}

\begin{abstract}
The first 1.1~mm continuum survey toward the Small Magellanic Cloud (SMC) was performed using the AzTEC instrument installed on the ASTE 10-m telescope.
This survey covered 4.5~deg$^2$ of the SMC with $1\sigma$ noise levels of 5--12~mJy~beam$^{-1}$, and 44~extended objects were identified.
The 1.1~mm extended emission has good spatial correlation with {\it Herschel} 160~$\micron$, indicating that the origin of the 1.1~mm extended emission is thermal emission from a cold dust component.
The 1.1~mm objects show dust temperatures of 17--45~K and gas masses of $4\times10^3$--$3\times10^5~M_\sun$, assuming single-temperature thermal emission from the cold dust with an emissivity index, $\beta$, of 1.2 and a gas-to-dust ratio of 1000.
These physical properties are very similar to those of giant molecular clouds (GMCs) in our galaxy and the Large Magellanic Cloud.
The 1.1~mm objects also displayed good spatial correlation with the {\it Spitzer} 24~$\micron$ and CO emission, suggesting that the 1.1~mm objects trace the dense gas regions as sites of massive star formation.
The dust temperature of the 1.1~mm objects also demonstrated good correlation with the 24~$\micron$ flux connected to massive star formation.
This supports the hypothesis that the heating source of the cold dust is mainly local star-formation activity in the 1.1~mm objects.
The classification of the 1.1~mm objects based on the existence of star-formation activity reveals the differences in the dust temperature, gas mass, and radius, which reflects the evolution sequence of GMCs.
\end{abstract}





\section{Introduction}
\label{sec1}
Giant molecular clouds (GMCs) are aggregations of dense ($\sim10^2~\mathrm{cm^{-3}}$) and cold ($<100~\mathrm{K}$) molecular gas in galaxies, and the principle formation site of high-mass stars and clusters.
The studies for unveiling the properties of the GMCs have mainly been developed through observations of the CO rotational lines toward our galaxy and nearby galaxies \citep[e.g.,][]{1985ApJ...289..373S, 1987ApJ...319..730S, 2008ApJS..178...56F, 2008ApJS..175..485M, 2011AJ....141...73M, 2009ApJS..184....1K, 2010ARA&A..48..547F, 2012ApJ...761...37M}.
Although CO line observations play an important role in revealing the general characteristics of GMCs, CO lines do not necessarily trace the whole molecular gas mass content, because CO molecules in the envelope of GMCs are photo-dissociated by the interstellar radiation field. Meanwhile, $\mathrm{H_2}$ molecules which are the main ingredient of the GMCs are better protected by a stronger self-shielding effect against ultraviolet (UV) radiation when compared to CO molecules \citep[e.g.,][]{1985ApJ...291..722T,1999RvMP...71..173H}. It should also be noted that it takes significant time to form CO molecules, especially in a low-metallicity environment \citep[e.g.,][]{2012MNRAS.426..377G}.

Recent observational studies based on $\gamma$-rays \citep{2005Sci...307.1292G}, dust extinction \citep{2012A&A...543A.103P}, and [C{\small II}] 158~$\micron$ emissions \citep{2013A&A...554A.103P} reveal that approximately 30\% of the molecular gas mass in our galaxy is not traced by the CO lines.
The results are consistent with a theoretical estimate of this ``dark molecular gas'' component \citep{2010ApJ...716.1191W}.
A numerical study by \citet{2012MNRAS.426..377G} suggests that, in a low-metallicity environment, the fraction of the dark molecular gas component increases rapidly with decreasing metallicity.
Therefore, it is essential to investigate GMCs in low-metallicity environments by means other than CO line observations.
Dust continuum observations using a submillimeter and millimeter imaging array can provide an alternative method to investigate the amount of the dark molecular gas component.

The Small Magellanic Cloud (SMC) is a dwarf galaxy that provides a unique opportunity to study the physics of the interstellar medium (ISM) because of its proximity and low metallicity.
The distance to the SMC is $\sim60$~kpc \citep[e.g.,][]{2000A&A...359..601C,2005MNRAS.357..304H}, making it one of the nearest galaxies along with the Large Magellanic Cloud (LMC).
This proximity makes it possible to observe more detailed structures of the ISM in comparison to other nearby galaxies.
The SMC is also characterized by its metallicity of $\sim1/5~Z_\sun$ \citep[e.g.,][] {2000A&A...364..455L,1999ApJ...518..246K}, a high gas-to-dust ratio of $\sim$1000 \citep[e.g.,][]{2007ApJ...658.1027L,2011A&A...536A..17P,2014ApJ...797...85G}, and strong UV field produced by active star-formation activity \citep[e.g.,][]{1980A&A....90...73V}.
These peculiarities make the SMC an ideal laboratory to investigate the physics of the ISM under the extreme conditions that may be similar to those in galaxies forming in the early Universe \citep[e.g.,][]{2013ApJ...778..102O}.

The GMC CO survey of the SMC was conducted by the NANTEN 4-m telescope \citep{2001PASJ...53L..45M} with a linear resolution of 50~pc.
Follow-up observations by Mopra 22-m telescope provide high-resolution (12~pc) CO images toward the NANTEN sources \citep{2010ApJ...712.1248M,2013IAUS..292..110M}.
Another method to reveal the spatial distribution of the ISM involves observing thermal emissions from the interstellar dust.
Continuum data by TopHat \citep{2003ApJ...596..273A}, COBE/DIRBE \citep{2002ApJ...576..762L,2003ApJ...596..273A, 2010A&A...519A..67I}, and {\it Planck} \citep{2011A&A...536A..17P} are available at submillimeter and millimeter wavelengths with a resolution of $\ga5\arcmin$.
The 1.2 mm \citep[SIMBA/SEST, ][]{2004A&A...425L...1R, ferreira2004, 2007A&A...471..103B} and 870~$\micron$ \citep[LABOCA/APEX, ][]{2010A&A...524A..52B,2015MNRAS.448.1847H} bands have been observed only toward famous star-forming regions in the SMC.
The {\it Spitzer} Space Telescope provides the infrared maps of the SMC, revealing the spatial distribution of the $\mathrm{H_2}$ gas component in comparison with H{\small I} data \citep{2007ApJ...658.1027L,2011ApJ...741...12B}.
{\it AKARI} also provides far-infrared images (65, 90, 140, and $160~\micron$) of the SMC as a part of the all-sky survey \citep{2015PASJ...67...50D,2015PASJ...67...51T}.
A high resolution and wide field survey at submillimeter wavelengths, sufficient to investigate detailed ISM properties, was also conducted by {\it Herschel} \citep[100--500~$\micron$]{2010A&A...518L..71M,2013AJ....146...62M}, and the cold ISM properties were investigated using image-based spectral energy distribution (SED) fitting in the Magellanic Clouds \citep{2014ApJ...797...85G,2015MNRAS.450.2708L,2016ApJ...825...12J}.
Millimeter-wavelength images at 1.4, 2.1, and 3.0 mm, combined from South Pole Telescope and Planck data, were also presented by \citet{2016ApJS..227...23C}.

In this study, we conducted a 1.1~mm continuum survey using the AzTEC instrument on the ASTE telescope to investigate the physical properties of the GMCs in the SMC with a resolution comparable to {\it Herschel}.
The 1.1~mm continuum thermal emission from the coldest component of interstellar dust ($\la$20~K) robustly correlates with ISM column density via the Rayleigh-Jeans approximation.
Therefore, the high-resolution and high-sensitivity continuum map at 1.1~mm can provide the distribution of the GMCs in the SMC.

This paper is organized as follows: Section \ref{sec2} describes the AzTEC/ASTE observations and the data analysis methods.
Section \ref{sec3} presents the image and detected objects obtained by the 1.1~mm continuum survey of the SMC.
In Section \ref{sec4}, we estimate the missing flux caused by the data analysis used to recover extended objects.
The result of the spectral energy distribution analysis used to estimate the physical parameters of the detected objects, is shown in Section \ref{sec5}.
In Section \ref{sec6}, we discuss the nature of the 1.1~mm objects, focusing on the origin of the dust temperature and the comparison with other observation data and the GMC properties.
The evolutionary sequence of the 1.1~mm objects compared with the star-formation activity is also discussed.
Finally, Section \ref{sec7} summarizes the result and the future prospects of our study.
Throughout the paper, we adopt a distance to the SMC of 60~kpc.

\section{Observation and data reduction} 
\label{sec2}
\subsection{AzTEC/ASTE observation}
\begin{deluxetable*}{lcccc}
\tablecaption{Observed fields of AzTEC/ASTE SMC wide observations.\label{wide:obsregion}}
\startdata
\tableline
Field&SW&NE&Wing&N88\\  \tableline
R.A. (J2000)&$00^\mathrm{h}50^\mathrm{m}00^\mathrm{s}$&$01^\mathrm{h}01^\mathrm{m}00^\mathrm{s}$&$01^\mathrm{h}13^\mathrm{m}00^\mathrm{s}$&$01^\mathrm{h}24^\mathrm{m}00^\mathrm{s}$\\
Dec. (J2000)&$-73\arcdeg 10\arcmin 00\arcsec$&$-72\arcdeg 20\arcmin 00\arcsec$&$-73\arcdeg 00\arcmin 00\arcsec$&$-73\arcdeg 15\arcmin 00\arcsec$\\
Position angle&\multicolumn{4}{c}{$20\arcdeg$}\\
Field size&\multicolumn{4}{c}{$1\fdg 1 \times 1\fdg 1$} \\
Scan method&\multicolumn{4}{c}{Orthogonal raster scan} \\ 
1$\sigma$ noise (mJy beam$^{-1}$)&5.1&6.0&6.4&11.6\\
\enddata
\end{deluxetable*}

\begin{figure*}
\plotone{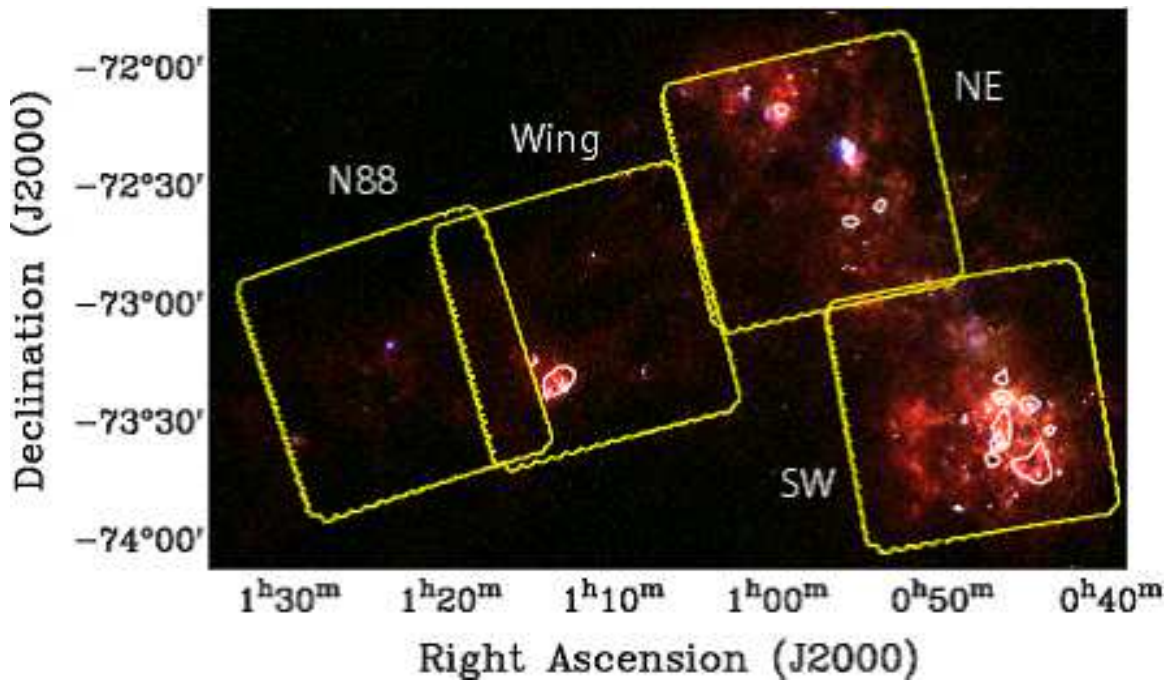}
\caption{The observing regions of the AzTEC/ASTE 1.1~mm continuum survey of the SMC. Yellow boxes show the observing four patches of SW, NE, Wing and N88 regions. The background image is a three-color composite of {\it Herschel} $500 \micron$ (red), $160 \micron$ (green), and {\it Spitzer} $24 \micron$ (blue) data and the white contours are CO($J=1$--$0$) intensity ($0.5~\mathrm{K~km~s^{-1}}$) by NANTEN. \label{fig:obsregion}}
\end{figure*}

Continuum observations at 1.1~mm toward the SMC were conducted with the AzTEC instrument \citep{2008MNRAS.386..807W} installed on the ASTE telescope \citep{2004SPIE.5489..763E, ezawa2008new} in the Atacama desert, Chile.
AzTEC is a 144-pixel bolometer camera with a bandwidth of 49 GHz and a center frequency of 270 GHz.
The angular resolution is $28\arcsec$ at FWHM. 
The observations covered a total of a 4.5 $\deg^2$ field of the SMC by connecting four patches of scans that are denominated as southwest (SW), northeast (NE), Wing, and N88, as shown in Figure \ref{fig:obsregion}.
Table \ref{wide:obsregion} shows the summary of each observing field.
Each scan was performed by orthogonal raster scans of a $1\fdg 1 \times 1\fdg 1$ field (position angle of $20\degr$).
Observations were performed from October 7 to December 26, 2008.
The total observing time is 42 hours with an on-source time of $\sim$30 hours.
The N88 field was observed under bad weather conditions ($\tau_{220 \mathrm{GHz}} \ga 0.1$), whereas the remaining fields were observed with good atmospheric transparency ($\tau_{220 \mathrm{GHz}} \la 0.1$).
Quasars J2355-534 and J2326-502 were observed every two hours to measure the pointing offsets during the observations.
The resulting pointing accuracy after applying the pointing correction is was better than $3\arcsec$ at the 1$\sigma$ confidence level \citep{2008MNRAS.390.1061W}.
Uranus was observed once every night for the purpose of absolute flux calibration and point spread function measurement of each bolometer pixel.
Flux calibration has an uncertainty of eight percent \citep{2008MNRAS.390.1061W,2010AJ....139.1190L}.

\subsection{Data reduction}
The correlated noise removal was performed using a principal component analysis (PCA) cleaning \citep{2005ApJ...623..742L}.
The time fluctuation of the bolometer signal is dominated by atmospheric emission, which is spatially extended on the sky and has a strong correlation among the bolometer pixels.
On the other hand, point source emission and detector noise do not have a correlation among the bolometer pixels. 
This means that the principal components that have large eigenvalues can be removed as atmospheric emission.
Therefore, PCA cleaning is an effective way to remove the atmospheric emission and to retrieve faint point-source signals.
The detailed treatment of PCA cleaning in the AzTEC IDL routine, which we use, is described in \citet{2008MNRAS.385.2225S} and \citet{2012MNRAS.423..529D}.

The PCA cleaning method is only sensitive to point sources; as extended astronomical objects have a correlation among bolometer pixels in the same manner as the atmosphere emission.
FRUIT \citep{2010AJ....139.1190L} is a method to recover extended components by performing the PCA cleaning iteratively.
In the FRUIT procedure, the signals retrieved by the PCA cleaning are used as an astronomical source model.
Time-stream data that have the source models subtracted are analyzed by the PCA cleaning, and astronomical signals that are retrieved are added to the source models.
This process continues iteratively until no significant signals are found.
The sky maps are constructed by adding the final source models to the source-free maps.

As a result of the FRUIT imaging, noise levels of 5--12 mJy beam$^{-1}$ are achieved (see also Table \ref{wide:obsregion} for each field).
The point response function (PRF), after applying FRUIT, gives a Gaussian FWHM of $40\arcsec$, which corresponds to af physical scale of 12~pc at the distance of the SMC.

\section{Result}
\label{sec3}
\subsection{The map}
\begin{figure*}
\plotone{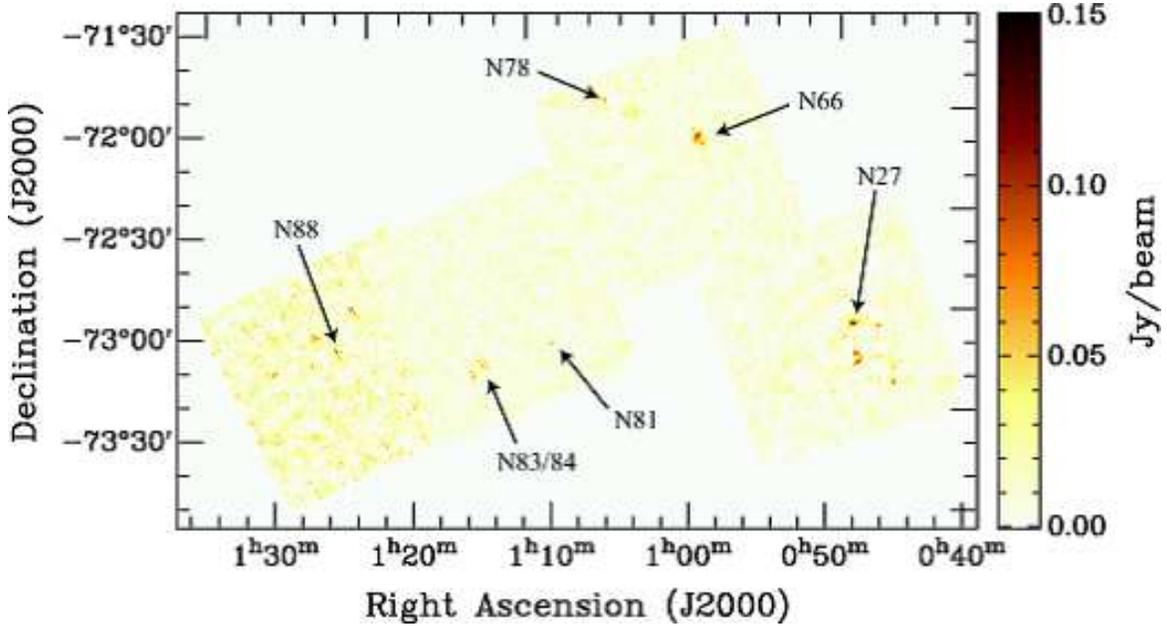}
\caption{The 1.1~mm continuum map of the SMC analyzed by the FRUIT procedure.\label{fig:aztec_fruit}}
\end{figure*}

Figure \ref{fig:aztec_fruit} shows the 1.1~mm continuum flux map of the whole SMC reduced using the FRUIT analysis.
Most of the representative star-forming regions in the SMC, such as N27, N66, N81, N83, N84, and N88, were detected with sufficient S/N ratios of $>10$ at the peak positions. The PCA map is also shown in Appendix \ref{sec:PCA}.

\subsection{Source catalog}
Extended sources were identified by contouring significant emission ($\sim5\sigma$) that is brighter than 30~mJy~beam$^{-1}$ in the SW, NE, and Wing fields, and 60~mJy~beam$^{-1}$ in the N88 field.
In addition, sources smaller than the effective resolution of $40\arcsec$ were eliminated from the catalog.
Two slightly significant sources ($5.8\sigma$ in the Wing field and $6.5\sigma$ in the N88 field) that have no object corresponding to the 24--160~$\micron$ range and the $3.5\sigma$ PCA objects were excluded from the catalog, because they would be false detections.

The extended source catalog of each field is shown in Table \ref{table:catalog_fruit}.
We have identified a total of 44~sources, including 24, 14, 5, and 1~sources in the SW, NE, Wing, and N88 fields, respectively.
The errors of the peak fluxes were given by the map noise level at the source positions and were not considered as systematic uncertainty in the calibration.
The errors of the total flux were propagated normally from the map noise level and the calibration error.
The source radii were estimated under the assumption that the structure is spherically symmetrical, and the deconvolved radii were calculated by assuming that the PRF is the Gaussian profile with $40\arcsec$ FWHM.

The detected 1.1~mm objects show a peak flux range of 35--149~mJy~beam$^{-1}$, a total flux range of 24--1683~mJy, and a source radius range of 6--40~pc ($0.3\arcmin$--$2.3\arcmin$).
SW-1, SW-2, and NE-1 have relatively large radii compared to the other detected objects.
With the exception of these objects, the ranges of the peak flux, total flux, and radius are 35--148~mJy~beam$^{-1}$, 24--400~mJy, and 6--23~pc ($0.3\arcmin$--$1.3\arcmin$), respectively.

The identification of the objects retrieved by PCA is summarized in Appendix \ref{sec:PCA}.
Note that our PCA map is sensitive enough to serendipitously uncover distant gravitationally amplified galaxies \cite[e.g.,][]{2013ApJ...774L..30T}.

\section{Evaluation of missing flux}
\label{sec4}
Although the FRUIT analysis is an effective way to recover the flux of extended structures, not all are recovered.
This requires estimating the missing fluxes of the 1.1~mm objects to obtain reliable physical properties.
Therefore, we evaluate the missing flux by embedding test sources with a Gaussian flux distribution and {\it Herschel} map into the time-stream data.

\subsection{Gaussian source analysis}
To evaluate the reproducibility performance of FRUIT for extended structures, we embedded 20 two-dimensional, axisymmetric Gaussian flux distribution sources with the combinations of four FWHMs of $1\arcmin$, $2\arcmin$, $3\arcmin$, and $4\arcmin$, and five peak fluxes of 30, 60, 90, 120 and 150~mJy, in a source-free region of the SMC Wing field.
The peak flux and FWHM ranges were  selected in order to simulate objects in the FRUIT maps.
Peak fluxes of 30, 60, 90, 120, and 150~mJy $\mathrm{beam^{-1}}$ correspond to an S/N ratio of $\sim5\sigma$, $10\sigma$, $15\sigma$, $20\sigma$, and $25\sigma$, respectively, for the SW, NE, and Wing fields.

\begin{figure}
\plotone{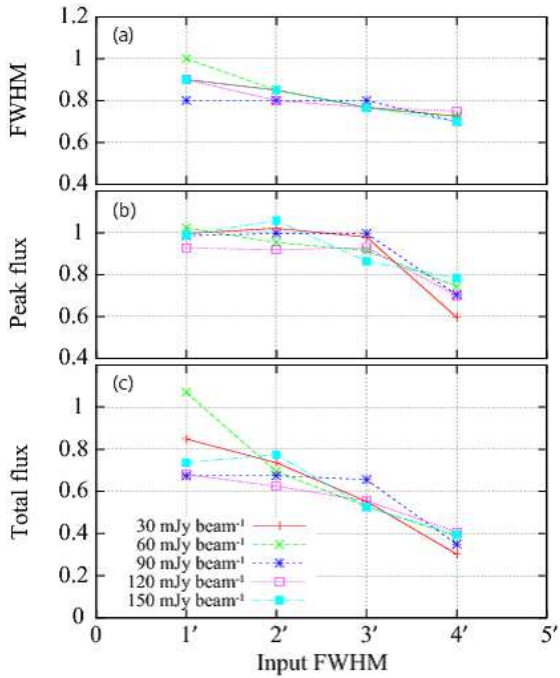}
\caption{The reproducibility of the Gaussian sources after the FRUIT procedure. (a) the ratio of output and input FWHM. (b) the ratio of output and input peak flux. (c) the ratio of output and input total flux.\label{result:gaussian}}
\end{figure}

Figure \ref{result:gaussian} shows the result of the Gaussian fits before and after FRUIT was applied.
We ascertained three facts through this analysis.
Firstly, about 80\% of the FWHMs of Gaussian distribution were recovered but tended to have slightly lower recovery rates with larger input FWHMs.
Secondly, the peak flux was recovered to approximately 90--100\% below the FWHM of $3\arcmin$.
Thirdly, the recovery rate of the total flux decreased as the FWHM increased, and 60--70\% was recovered for $2\arcmin$--$3\arcmin$ FWHM objects.
Considering the source size of the 1.1~mm objects, all the objects tended toward an $\sim80\%$ smaller FWHM value and recovered their peak flux correctly.
On the other hand, the recovered total flux depends on the source size.
All 1.1~mm objects showed FWHMs $\la3\arcmin$ from the peaks, suggesting that a total flux of  $\ga50$\% was recovered.
In particular, the peak structure of SW-1, 2, and NE-1 showed an FWHM of $2\arcmin$--$3\arcmin$, which is expected to recover the total flux of 50--70\%.
The other smaller objects also would recover a total flux of $>60$\%.
These results are also consistent with the previous FRUIT simulation studies by \citet{2011PASJ...63..105S,2015ApJS..217....7S} and \citet{2011PASJ...63.1139K}.

\subsection{{\it Herschel} data analysis}
As an alternative approach for investigating the missing flux of the extended emission, we estimated the missing flux of the 1.1~mm data after applying the FRUIT procedure using the {\it Herschel} maps \citep{2013AJ....146...62M}.
Both 1.1~mm and {\it Herschel} 100--500~$\micron$ radiation are expected to arise from the cold dust component in the SMC, thus the {\it Herschel} maps should have a distribution similar to the 1.1~mm distribution.
We constructed filtered {\it Herschel} maps, which have the diffuse emission filtered out in the same way as the 1.1~mm map using the following steps.
Firstly, appropriately scaled {\it Herschel} data were embedded into the 1.1~mm time-stream data that removed the astronomical signal component estimated by the FRUIT analysis of the 1.1~mm data.
Here we used flux scaling factors of 0.036, 0.023, 0.047, 0.085, and 0.118 for 100, 160, 250, 350, and 500~$\micron$, respectively, to make the peak flux comparable to the 1.1~mm images.
Secondly, the time-stream data, in which the {\it Herschel} maps were embedded, are analyzed using the FRUIT procedure.
Finally, the analyzed {\it Herschel} maps were rescaled by dividing by the scaling factor.
These filtered {\it Herschel} maps were expected to have the extended flux filtered out in a similar way to that seen in the 1.1~mm images.
Therefore, we can make a comparison with the 1.1~mm and the filtered {\it Herschel} data directly, and estimate the dust temperature of the detected objects without the bias of missing flux by using this dataset.

\begin{figure*}
\plotone{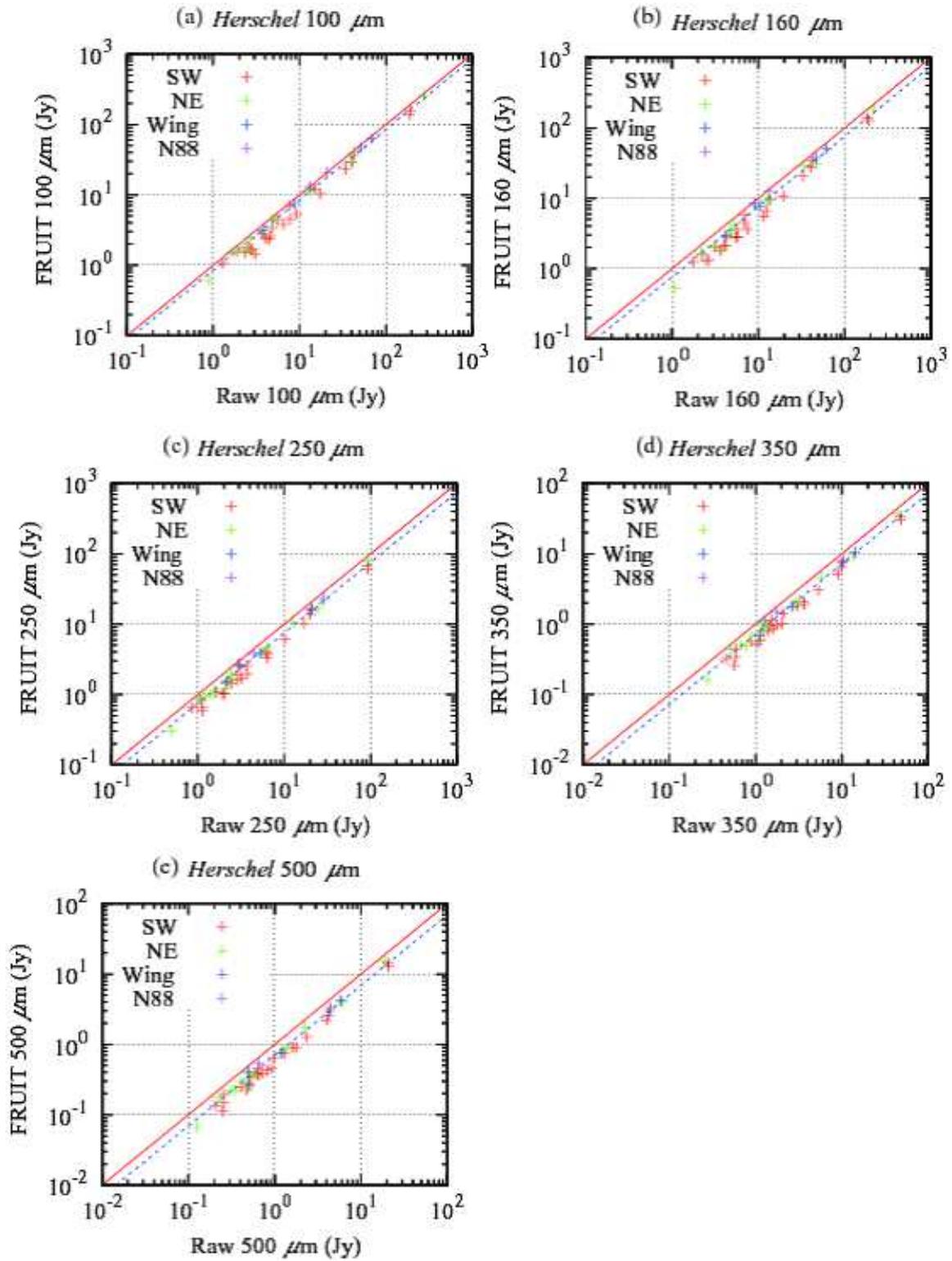}
\caption{The {\it Herschel} 100--500~$\micron$ fluxes of the 1.1~mm objects before and after FRUIT analysis. The red, green, blue, and purple crosses are the 1.1~mm objects in SW, NE, Wing, and N88 fields, respectively. The dashed blue line is the best-fit function of $S_\mathrm{FRUIT} (\mathrm{Jy}) = a S_\mathrm{Raw}$, and the solid red line is $S_\mathrm{FRUIT} = S_\mathrm{Raw}$.\label{fig:Herschelfruit}}
\end{figure*}

Figure \ref{fig:Herschelfruit} shows the {\it Herschel} fluxes as estimated by the contours of  the 1.1~mm objects before and after FRUIT analysis.
The distribution of the raw and filtered fluxes are fit by: $$S^\mathrm{FRUIT}_\lambda  (\mathrm{Jy}) =  a S^\mathrm{Raw}_\lambda,$$ where $a$ is 0.85, 0.76, 0.75, 0.72, and 0.69 for 100, 160, 250, 350, and 500~$\micron$, respectively, with standard deviations of 0.09, 0.07 0.08, 0.08, and 0.07, which may arise from additional photometric errors caused by FRUIT.
This result indicates that the total fluxes are systematically decreased in comparison to the intrinsic fluxes by applying the FRUIT procedure.
Since the 1.1~mm fluxes are affected by the same filtering as the filtered {\it Herschel} data, it is possible to derive dust temperatures that are not biased due to the missing flux of the diffuse component when we use both the 1.1~mm and filtered {\it Herschel} data. 
We are not interested in the missing diffuse flux because this component is expected to be more diffuse than the size of the GMCs.
Thus, we estimate the dust temperature and the lower limits of dust masses of the 1.1~mm objects using the 1.1~mm and {\it Herschel} data fluxes estimated from the filtered images.

The total $1\sigma$ photometric errors of the 1.1~mm object, given in Table \ref{table:catalog_fruit}, were estimated by taking into account the photometric calibration errors, the random error caused by FRUIT filtering, and the random error caused by the noise level of the FRUIT map.
The photometric calibration errors are 8\% at 1.1~mm \citep{2008MNRAS.390.1061W}, and 8\% and 10\% in the SPIRE (250, 350, and 500~$\micron$) and PACS  (100 and 160~$\micron$) bands \citep{2013AJ....146...62M, 2014ApJ...797...85G}.
The random error caused by FRUIT filtering is, at most, approximately 10\%.
As a result, the $1\sigma$ photometric error was estimated to be 12.8\% in the 1.1~mm and SPIRE bands and 14.1\% in the PACS bands.
Subsequently, we added the random error caused by the map noise, which depends on the size of the objects and the noise level of the object area, to estimate the total $1\sigma$ photometric errors.

\section{Analysis for physical property estimates}
\label{sec5}
\begin{figure*}
\plotone{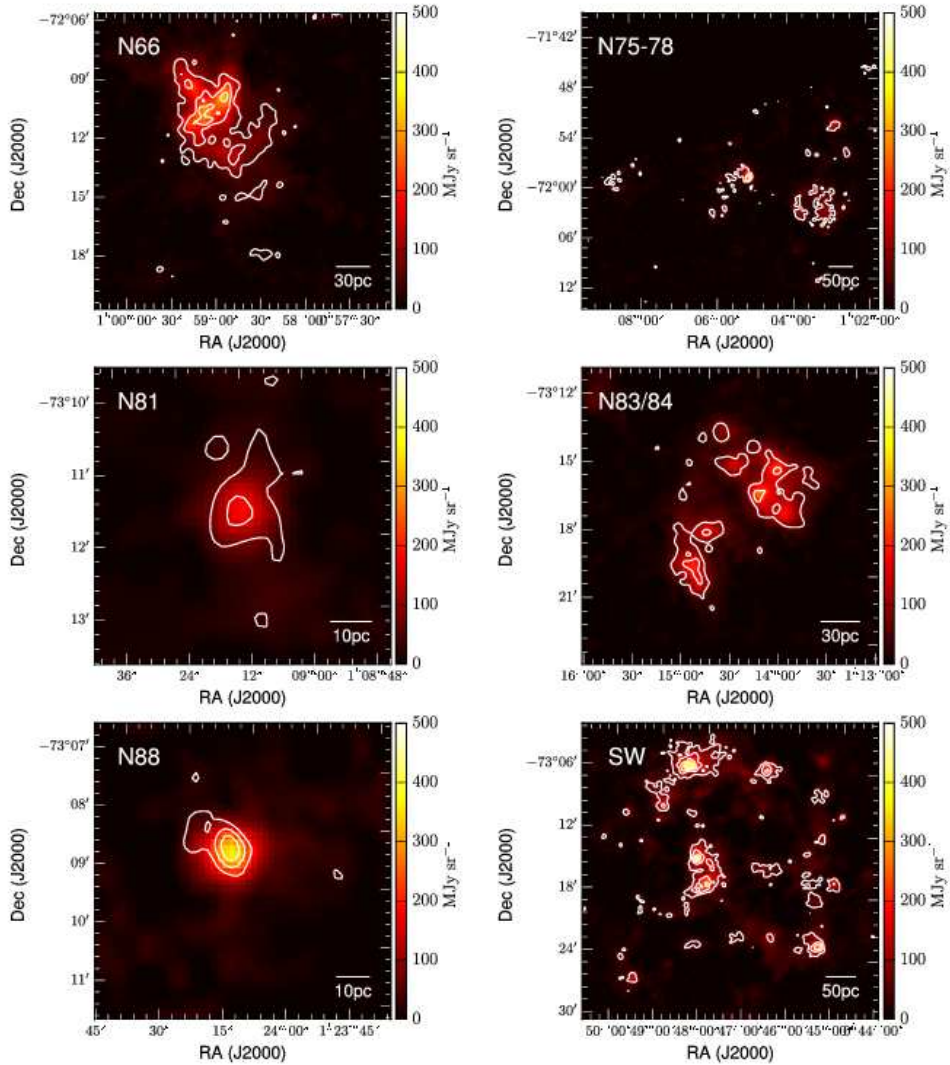}
\caption{The filtered {\it Herschel} PACS 160~$\micron$ images of the representative star-forming regions in the SMC. The contours of AzTEC 1.1~mm continuum are shown from 30~mJy~$\mathrm{beam^{-1}}$ with 30~mJy~$\mathrm{beam^{-1}}$ intervals for N66, N75--78, N81, N83/84, SW, and from 60 mJy~$\mathrm{beam^{-1}}$ with 30~mJy~$\mathrm{beam^{-1}}$ intervals for N88.\label{fig:aztec_on_herschel}}
\end{figure*}

Figure \ref{fig:aztec_on_herschel} shows a comparison of the filtered 160~$\micron$ and 1.1~mm flux distributions in representative star-forming regions.
The 1.1~mm emission has a good spatial correlation with the distribution of 160~$\micron$ emission, supporting that the origins of both the 160~$\micron$ and 1.1~mm fluxes are thermal emission from the cold dust component.
This justifies deriving the physical parameters of the 1.1~mm objects using SED analysis of dust continuum emission at 1.1~mm and 160~$\micron$.

\subsection{SED analysis method}
The flux densities of the identified 1.1~mm objects are shown in Table \ref{table:catalog_flux}.
The physical properties of the 1.1~mm objects were derived by the SED analysis under the assumption of single-temperature thermal dust emission using flux densities at the AzTEC and {\it Herschel} bands after removing the contribution of free-free emission.
At first, the contributions of free-free emission to the 1.1~mm (and {\it Herschel} bands) fluxes were estimated by fitting the free-free spectra to the 4.8 and 8.64~GHz continuum data \citep[ATCA+Parkes]{2010AJ....140.1511D} convolved to $40\arcsec$.
We estimated the photometric flux error of the 1.1~mm objects by using the image photometric errors of 1\% and the image noise levels.
The spectral index is flat for most objects, so we assume that synchrotron emission does not contribute to the radio data \citep{2010A&A...524A..52B}.
Assuming an electron temperature of $10^4$ K in the ionized regions, the flux density of the free-free emission is scaled by $1-0.16\ln{(\nu/10^{11} \mathrm{Hz})}$ \citep{1992ApJ...392L..35R}.
Secondly, thermal emission from a cold dust component was estimated using the fluxes at 1100, 500, 350, 250, 160, and 100~$\micron$ after subtracting the free-free contribution.
The total flux, $S_\lambda$, of thermal emission from the cold dust can be calculated using:
$$S_\lambda = \kappa_{\mathrm{dust}, \lambda} B_\lambda(T_{\mathrm{dust}}) M_{\mathrm{dust}} D^{-2},$$
where $\kappa_{\mathrm{dust}, \lambda}$ is the emissivity of dust, $B_\lambda$ is the Planck function, $M_{\mathrm{dust}}$ is total dust mass, $T_{\mathrm{dust}}$ is the dust temperature, and $D$ is the distance to the SMC.
In the SED analysis, $M_{\mathrm{dust}}$ and $T_{\mathrm{dust}}$ are calculated from the 1.1~mm and {\it Herschel} fluxes.
The total gas mass $M_{\mathrm{gas}}$ (including $\mathrm{H_2}$, H{\small I}, and He gas component) was estimated assuming a gas-to-dust mass ratio $GDR$:
$$M_{\mathrm{gas}} = GDR \times M_{\mathrm{dust}}.$$

We used plausible dust parameters of the SMC obtained by far-infrared to millimeter wavelength surveys, including {\it Spitzer}, {\it Planck}, and {\it Herschel} data \citep[e.g.,][]{2007ApJ...658.1027L,2011A&A...536A..17P,2014ApJ...797...85G}.
The emissivity of the dust, $\kappa_{\mathrm{dust}, \lambda}=12.5 \times (160 \micron/\lambda)^\beta \mathrm{cm^2 g^{-1}}$, is assumed\citep{2007ApJ...657..810D,2014ApJ...780..172D}.
Recent data at millimeter and submillimeter wavelengths have pointed out the shallow index of the emissivity, $\beta=1.2\pm0.3$ \citep{2003ApJ...596..273A,2007ApJ...658.1027L,2010A&A...523A..20B,2010A&A...519A..67I,2011A&A...536A..17P}.
Here we assumed the index $\beta=1.2$, for simplicity.
We also assumed that the $GDR=1000$, which is a typical value in the SMC \citep{2007ApJ...658.1027L,2011A&A...536A..17P,2014ApJ...797...85G}, to estimate the total gas mass, although these studies showed an uncertainty of a factor of two.

The number density and column density of molecular hydrogen ($\mathrm{H_2}$) were estimated by assuming that the 1.1~mm objects are dominated by the molecular component.
This assumption usually causes overestimates of the molecular component.
However, the use of the spatially filtered maps for this analysis, in which most of the diffuse components (corresponds to H{\small I} gas) are removed, supports the validity of the assumption above.
This assumption is also consistent with the robust spatial correlation between the 1.1~mm objects and CO emission (described in detail in Section \ref{sec621}).
We estimated the amount of $\mathrm{H_2}$ molecules using the total-to-$\mathrm{H_2}$ molecular gas mass ratio of 1.36, by taking into account the relative abundance of helium and heavier elements \citep{2010ApJ...712.1248M}.
The number density was estimated by assuming spherically symmetric clouds with radii of $R$.

\subsection{Physical properties of the 1.1~mm objects}
Computing the SED fits for each object naturally leads to computations of the dust mass, dust temperature, and other supplemental physical parameters.
The SED of SW-1 is shown in Figure \ref{gmc:sed:sw1} as an example, and that of the others in Appendix \ref{sec:sed}.
Free-free contributions were subtracted to some extent from the 1.1~mm flux densities.
The AzTEC and {\it Herschel} fluxes correspond to a single gray body component of the 30~K dust, which is the dominant dust mass component.
The 24 and 70~$\micron$ fluxes show the excess from the cold dust component spectrum that would originate from warm dust components, such as very small grains \citep[e.g.,][] {1990A&A...237..215D}.

\begin{figure}
\plotone{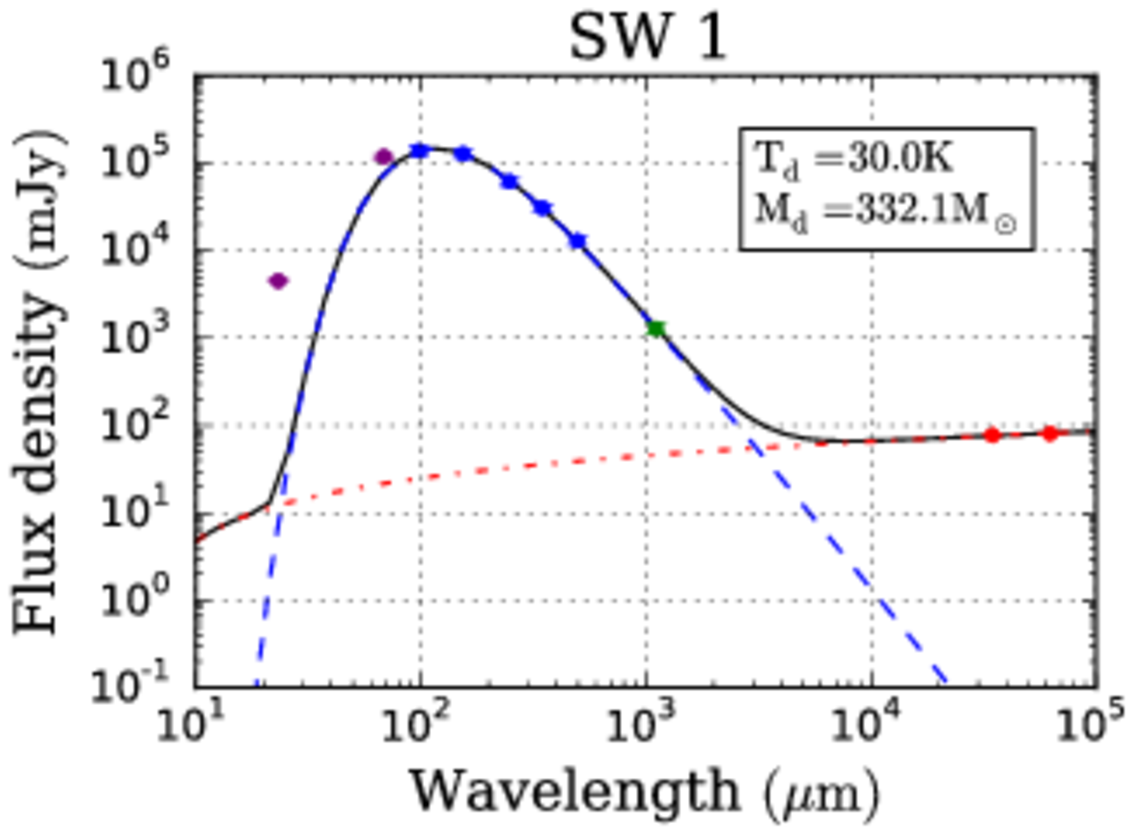}
\caption{The SED of SW-1. The fitting points for cold dust SED are shown in green (1.1~mm) and blue (100, 160, 250, 350, and 500~$\micron$). The fitting points for free-free model SED are shown in red (4.8 and 8.64~GHz). The purple points (70 and 24~$\micron$) are not used for the fitting. The dashed blue and dashed-dotted red lines show the SED model for cold dust and free-free emission, respectively. The solid line shows the total SED for cold dust and free-free emission.  \label{gmc:sed:sw1}}
\end{figure}

The derived physical parameters from the SED fits are shown in Table \ref{gmc:sedfit}.
Most of the 1.1~mm objects have free-free contaminations smaller than 10\%, but seven sources (including N88-1) have a free-free contribution of over 10\% for the 1.1~mm fluxes.
We noted that NE-1, which is associated with the active star-forming region, N66, has a free-free contamination of over 20\% for the 1.1~mm flux.
The dust temperatures displayed a range of 17--45~K, and an average and standard deviation of $28.7\pm4.4$~K\footnote{N88 is excluded from the statistics in order to avoid a bias due to the significant difference of map sensitivity.}, which is higher than the global dust temperature of 21.6$\pm$1.9~K\, estimated over the whole SMC, with an emissivity index of $\beta=1.2$ \citep{2011A&A...536A..17P}.
\citet{2015MNRAS.450.2708L} derived the spatially resolved ($\sim45\arcsec$) dust temperature distribution using {\it Herschel} data and assuming a fixed emissivity index $\beta=1.5$.
The derived temperature range was 21--35~K, which is similar to the temperature estimated using our dataset and assuming $\beta=1.5$ ($25.0\pm3.6$~K).
The gas masses are distributed in the range of $4.1\times 10^3$--$3.4\times 10^5~M_\sun$.
The massive 1.1~mm objects, SW-1, SW-2, and NE-1, that have gas masses of $>10^5~M_\sun$, correspond to the famous star-forming regions Lirs~49, SMCB-2~N, and N66, respectively.
\citet{2010A&A...524A..52B} estimated the gas masses of $3.1\times10^5~M_\sun$ for Lirs~49 and $4.6\times10^5~M_\sun$ for SMCB-2~N, which are comparable to the result of our SED fits, considering the differences in the dust temperature and $GDR$.

\section{Discussion}
\label{sec6}
\subsection{Heating sources of cold dust}
\begin{figure}
\plotone{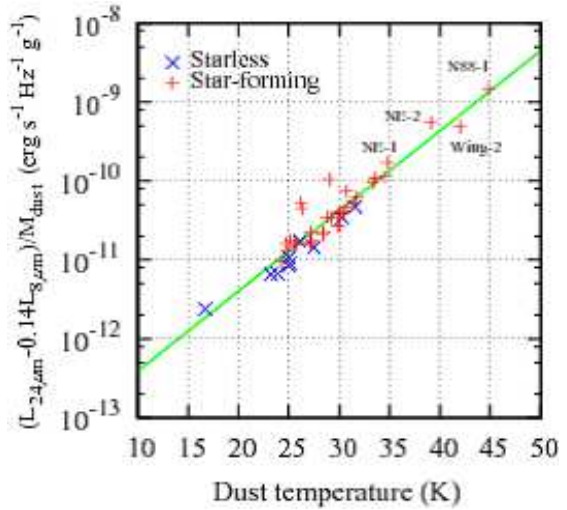}
\caption{The relationship between dust temperature and 24~$\micron$ luminosity per dust mass for 1.1~mm selected objects.
The green line is the best-fit function of these parameters: $(L_{24\micron}-0.14L_{8\micron})/M_{\mathrm{dust}}=3.768\times 10^{-14} (\mathrm{erg\ s^{-1} Hz^{-1} g^{-1}}) \exp(0.2336 (T_{\mathrm{dust}}/1\mathrm{K})$.
Blue cross and red plus show the starless and star-forming objects, respectively.\label{wide:TL}}
\end{figure}

To understand the heating mechanism of the cold dust, we investigated the relations between the dust temperature and the star-formation activity of the 1.1~mm objects.
The {\it Spitzer} 24~$\micron$ data is widely used as a tracer of embedded star-formation activity \citep[e.g.,][]{2007ApJ...666..870C}.
Thus, it is useful to determine the significance of the compact and massive star-forming activity that accompanies the strong UV radiation field to the dust temperatures of the 1.1~mm objects.
Some part of the 24~$\micron$ flux is contributed by stochastically heated PAH emission, which accompanies the diffuse cold dust component, and we removed the PAH contribution using an empirical relation, $L_{24\micron}-0.14L_{8\micron}$, obtained by \citet{2007ApJ...657..810D}.
The ratio of the 24~$\micron$ luminosity, connected to massive star-formation activity, to the dust mass corresponds to the intensity of the radiation field that contributes to the dust heating.
It is, therefore, expected that the ratio shows good correlation with the dust temperature.
Figure \ref{wide:TL} shows the relation between the dust temperature and the ratio of the corrected 24~$\micron$ luminosity to the dust mass.
These parameters are well fitted to an exponential function and have a correlation coefficient of $r^2=0.881$.
This strongly supports the idea that the origin of the variety of dust temperatures is the interstellar radiation field from young and massive stars formed inside each 1.1~mm object.

We also found that the dust temperature of the Wing region is significantly higher than the other regions. 
The average dust temperatures and the standard deviations are $27.8\pm2.8$, $28.5\pm5.4$, and $33.8\pm5.1$~K in the SW, NE, and Wing regions, respectively. 
The differences between the Wing and other regions are significant, with the p-values of $0.59$ (SW and NE), $0.01$ (SW and Wing), and $0.08$ (NE and Wing) according to the Mann-Whitney U test \citep[e.g.,][]{2003psa..book.....W}.
The corresponding objects of the Wing region (N81 and N83/84) are known as very active H{\small II} regions, accompanying O-type stars \citep{1999A&A...344..848H,1987A&A...178...25T}. 
In the Wing region, the number density of low-mass stars is smaller than that in the SW and NE regions, where the stellar bar is located.
Therefore, it is likely that the contribution to the dust temperature comes mainly from the strong radiation field caused by local young massive stars.
This situation is different from that in some nearby galaxies that show radial gradients of the dust temperature, mainly caused by low-mass stellar populations \citep[e.g.,][]{2010A&A...518L..56E,2011PASJ...63.1139K,2012MNRAS.425..763G,2012MNRAS.426..892G,2014A&A...561A..95T}.
From a different perspective, \citet{2014ApJ...797...85G} revealed variations of the dust emissivity spectral index, $\beta$, in different regions of the SMC.
This may also explain the discrepancy of the dust temperature between the Wing and NE/SW regions when assuming the higher $\beta$ value of 1.5 in the Wing region ($29.0\pm 4.0$~K).

\subsection{The 1.1~mm objects as star-formation sites}
\subsubsection{Comparisons with the other wavelength data\label{sec621}}
We focus on the nature of the detected 1.1~mm objects by investigating the correlations with gas and star-formation tracers.
We compared the 1.1~mm data with the survey data of atomic hydrogen (H{\small I}), CO, and continuum emission at 24 $\micron$.
The H{\small I} data is the combination of the Australia Telescope Compact Array and the 64-m Parkes telescope \citep{1999MNRAS.302..417S}.
The CO ($J=$1--0) survey data was taken by NANTEN \citep{2001PASJ...53L..45M}.
The 24 $\micron$ data was provided by the SAGE-SMC program of the {\it Spitzer} Space Telescope \citep{2011AJ....142..102G}.
The spatial resolutions of these data are $98\arcsec$, $2\farcm6$, and $6\arcsec$ for H{\small I}, CO, and 24~$\micron$, respectively.

\begin{figure*}
\plotone{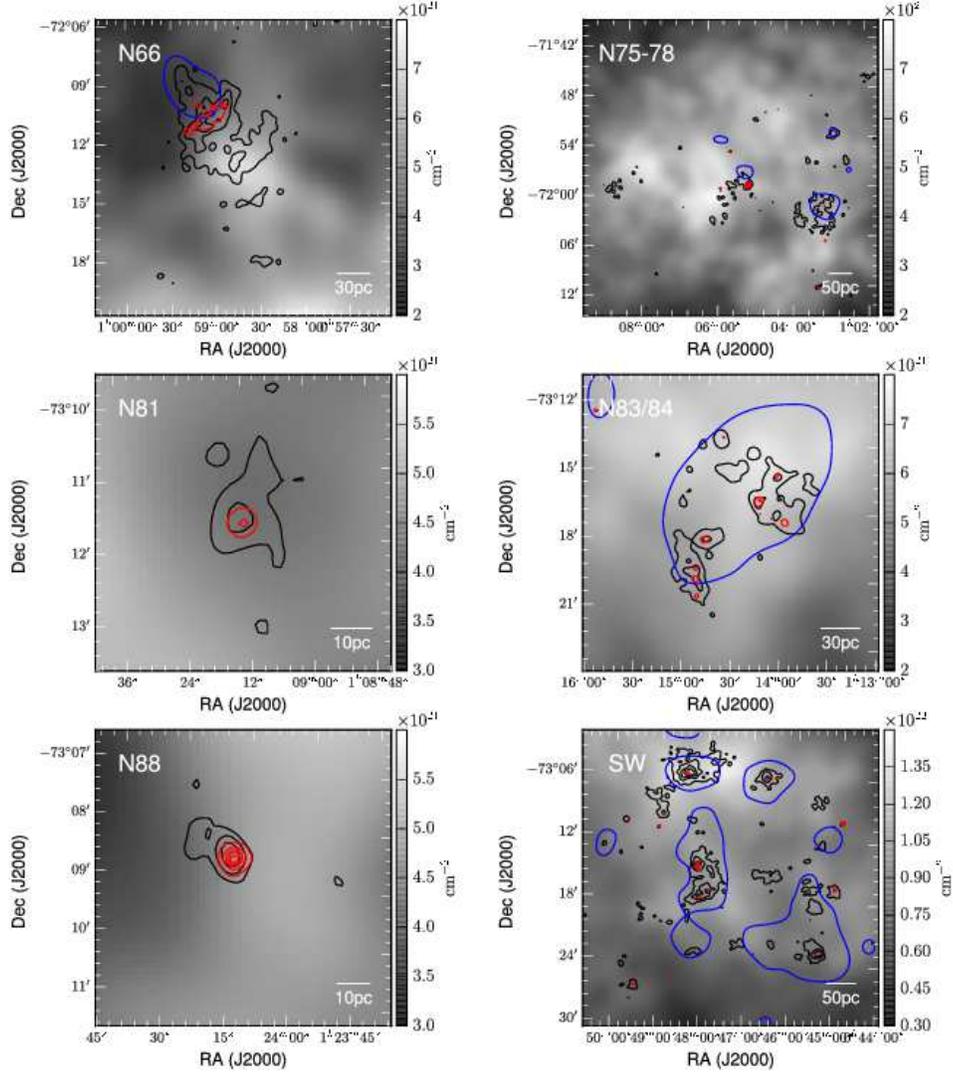}
\caption{The AzTEC 1.1~mm contours (black, 30+30~mJy~$\mathrm{beam^{-1}}$ for N66, N75--78, N81, N83/84, SW, and 60+30~mJy~$\mathrm{beam^{-1}}$ for N88) are overlaid on the H{\small I} column density image (gray scale) in the representative star-forming regions. {\it Spitzer} MIPS~24~$\micron$ (50+300~MJy~$\mathrm{sr^{-1}}$) and NANTEN CO (0.5~$\mathrm{K~km~s^{-1}}$, no CO data in N81 and N88) are also shown in red and blue contours, respectively. }
\label{fig:aztec_on_HI}
\end{figure*}

Figure \ref{fig:aztec_on_HI} shows the comparison between H{\small I}, CO, 1.1~mm, and 24~$\micron$ in the N66, N75--78, N81, N83/84, N88, and SW regions.
We describe the comparison for each region below:
\begin{itemize}
\item{N66: the most active star-forming region in the SMC. The NANTEN CO emission is seen in the northern part of the 1.1~mm object.
\citet{2000A&A...359.1139R} reported CO ($J=2$--$1$) detection in the northern and central parts, with the central part being weaker than the northern part.
In contrast, the southern part shows a high H{\small I} column density.
We note that the H{\small I} profile shows two velocity components at $120$ and $160~\mathrm{km~s^{-1}}$, and CO emission is associated with the $160~\mathrm{km~s^{-1}}$ component.
The 24~$\micron$ emission is located in the central region of the 1.1~mm object.}
\item{N75--78: there are some compact 1.1~mm objects. The peaks of the 1.1~mm objects do not always exactly co-located with peaks of the H{\small I} column density, but they exist in the high H{\small I} column density region. The 1.1~mm emission also seems to be a more compact structure than the H{\small I} emission.}
\item{N81, N83/84 and N88: these are known as the active star-forming sites associated with the H{\small II} regions. The peaks of the 1.1~mm objects in these regions are in very good agreement with the 24~$\micron$ emission.}
\item{SW: this region is located on the main part of the SMC bar, and shows bright CO emission, elevated star-formation activity, and a high H{\small I} column density.
The CO emission appears to be associated with the 1.1~mm objects.
Mopra observation shows that the NANTEN CO clouds are resolved into smaller CO objects \citep{2013IAUS..292..110M}.
The 24 $\micron$ emission is located on the peak of the 1.1~mm objects. The large 1.1~mm objects, such as SW-1 and 2, are in agreement with the high column H{\small I} region, but the 1.1~mm structures appear to be sharper than the H{\small I} column density.
The other 1.1~mm objects also have smaller structures than the H{\small I} distribution.
We note that H{\small I} spectra show complex velocity structures distributed around $120$ and $160~\mathrm{km~s^{-1}}$, and CO emission is mainly associated with the $120~\mathrm{km~s^{-1}}$ component.}
\end{itemize}

The three main findings from the comparisons between the 1.1~mm objects and the other tracers in the representative regions in the SMC are summarized as follows:
\begin{enumerate}
\item{The spatial correlation between the 1.1~mm emission and the H{\small I} column density is not necessarily good, and the structure of the H{\small I} column density has a smoother distribution than that of the 1.1~mm emission.}
\item{The CO emission shows good spatial correlation with the 1.1~mm emission, and the 1.1~mm objects appear to trace the spatially resolved structure of the NANTEN CO objects.
CO molecules exist on the inner side of the self-shielding surface of $\mathrm{H_2}$ clouds \citep{1985ApJ...291..722T}, suggesting that a part of the 1.1~mm object consists of $\mathrm{H_2}$ gas.}
\item{The 24~$\micron$ emission is associated with most 1.1~mm peaks.
Since the 24~$\micron$ emission is a tracer of the star-formation activity, the 1.1~mm emission around the 24~$\micron$ sources corresponds to the envelope of high-density gasses that forms massive stars.}
\end{enumerate}

The 1.1mm objects can be classified into two types in terms of star-formation activity: the first classification encompasses the objects that are associated with H{\small II} regions, {\it Spitzer} and {\it Herschel} YSO candidates \citep{2007ApJ...655..212B,2014AJ....148..124S}, or bright 24~$\micron$ sources ($>10 \mathrm{MJy~sr^{-1}}$), and the second accounts for the objects which are not associated with these star-formation signatures.
Among the 44~objects, 34~objects are associated with signs of star formation, and the other 10~objects are not.
The 1.1~mm objects can also be classified into two types in terms of whether or not they are associated with the CO emission.
The existence of CO molecules in the 1.1~mm objects is judged by the detection of the CO ($J=1$--$0$) line using the data obtained by the Mopra 22-m \citep[beam size of $33\arcsec$ FWHM and typical virial mass of $\gtrsim6\times 10^3~M_\sun$,][]{2003MNRAS.338..609M, 2010ApJ...712.1248M, 2013IAUS..292..110M} and NANTEN 4-m \citep[$2\farcm6$ and $\gtrsim6\times 10^4~M_\sun$,][]{2001PASJ...53L..45M} telescopes.
The Mopra data is roughly comparable to the detection mass limit of the 1.1~mm objects ($\gtrsim 7\times10^3~M_\sun$).
Among the 44~objects, 23~objects are associated with the CO emission, and 21~objects are not.
Table \ref{table:sfco} shows the results of the classification by the existence of star-formation activity and CO detection in the 1.1~mm objects.

\subsubsection{Physical property comparisons with the previous GMC observations\label{sec622}}

\begin{deluxetable*}{lccccl}
\tabletypesize{\scriptsize}
\tablecaption{Numbers and median values of physical parameters of the 1.1 mm objects.\label{table:medians}}
\tablehead{&Numbers&$T_\mathrm{dust}$&$M_\mathrm{gas}$&Radius \\
&&(K)&($\times 10^3\mathrm{M_\sun}$)&(pc)}
\startdata
All&43&28.5&13.9&8.8 \\ \tableline
Starless&10&25.1&11.2&7.6 \\
Star forming&33&29.2&17.0&10.1
\enddata
\tablecomments{N88-1 is excluded from the statistics.}
\end{deluxetable*}
The physical properties estimated by the SED fits also revealed the nature of the 1.1~mm objects.
The masses and radii of the 1.1~mm objects are in the range of $4\times 10^3$--$3\times 10^5~M_\sun$ and up to 40~pc, respectively.
In addition, the hydrogen molecule densities of the 1.1~mm objects, assuming spherical symmetry, are a few $\times$ 10--100~$\mathrm{H_2\ cm^{-3}}$.
The statistical characteristics of the mass, size, and density of the 1.1~mm objects are roughly consistent with the GMC properties obtained by CO observations toward our galaxy \citep[e.g.,][]{1993prpl.conf..125B} and the LMC \citep{2008ApJS..178...56F}, supporting the idea that the detected 1.1~mm objects correspond to GMCs.

\begin{figure}
\plotone{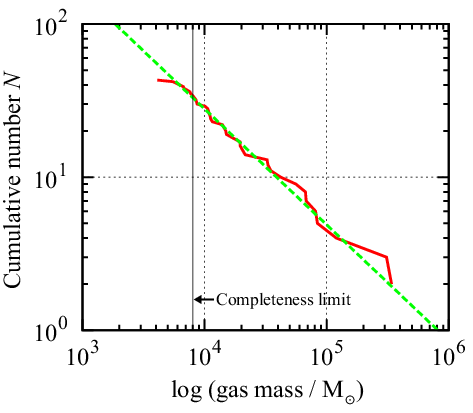}
\caption{The mass spectrum of the 1.1~mm objects is shown as a solid red line. The best-fitting power law above the completeness mass limit of $8\times 10^3~M_\sun$ is shown as a dashed green line, the function of which is $N=a (M_{\mathrm{gas}}/10^5~M_\sun)^{\alpha +1}$ with $a=4.888$ and $\alpha=-1.756$. \label{wide:massf}}
\end{figure}

The mass spectrum of the 1.1~mm objects is shown in Figure \ref{wide:massf}.
The cumulative number count of the 1.1~mm objects, $N$, was fitted by a power law with $dN/dM \propto M^\alpha$, where the power index $\alpha=-1.76\pm0.13$, with gas masses higher than $8\times10^3~M_{\sun}$.
Recently, van Kempen et al. (submitted to A\&A) revealed that the power index $\alpha=-2.0\pm0.2$, using the high-resolution CO ($J=2$--$1$) data in the SMC, and the result was also consistent with the power index of the 1.1~mm objects.
The power index is also consistent with that of the Large Magellanic Cloud \citep[$\alpha=-1.75\pm0.06$, ][]{2008ApJS..178...56F}.
On the other hand, the index is deeper than that of our galaxy \citep[$\alpha\approx-1.5$, e.g.,][]{1984A&A...133...99C, 1989ApJ...339..919S,1996ApJ...458..561D, 1996ApJ...466..282D}, meaning that it could be responsible for the differences in the environment of formation or dissipation of the GMCs in the SMC.

\subsubsection{The nature of the 1.1~mm objects}
\label{sec:nature}
We mentioned that the 1.1~mm objects correspond to GMC, but there is a possibility that some of the 1.1~mm objects consist of H{\small I} gas as an H{\small I} envelope or in the coexistence of H{\small I} and $\mathrm{H_2}$.
In Figure \ref{fig:aztec_on_HI}, the H{\small I} column density at the peak position of the 1.1~mm objects is of the order of $10^{21} \mathrm{atoms~cm^{-2}}$ with a resolution of $98\arcsec$, so the H{\small I} column density, with a resolution of $40\arcsec$, seems to be $\ga10^{21}~\mathrm{atoms~cm^{-2}}$.
Therefore, we should note that the H{\small I} column density of the 1.1~mm objects is of an order comparable to the column density estimated from the dust (see Table \ref{gmc:sedfit}), possibly implying that a large fraction of the mass of the 1.1~mm objects is dominated by an H{\small I} component.
This possibility is supported by recent simulation and theoretical studies \citep{2012MNRAS.421....9G,2012ApJ...759....9K} that point out the possibility of star formation in the H{\small I} dominated clouds in low-metallicity environments.

In conclusion, the 1.1~mm objects trace the dense gas regions that could form massive stars.
These ``dust-selected clouds'' also show similar physical properties to the GMCs in our galaxy and the LMC.

\subsubsection{Evolution of the 1.1~mm objects}
It seems reasonable to suppose that the detected 1.1~mm objects trace gas clouds massive and dense enough to form massive stars.
Here we examine the existence of the difference in the physical properties of classified objects.

\begin{figure*}
\plotone{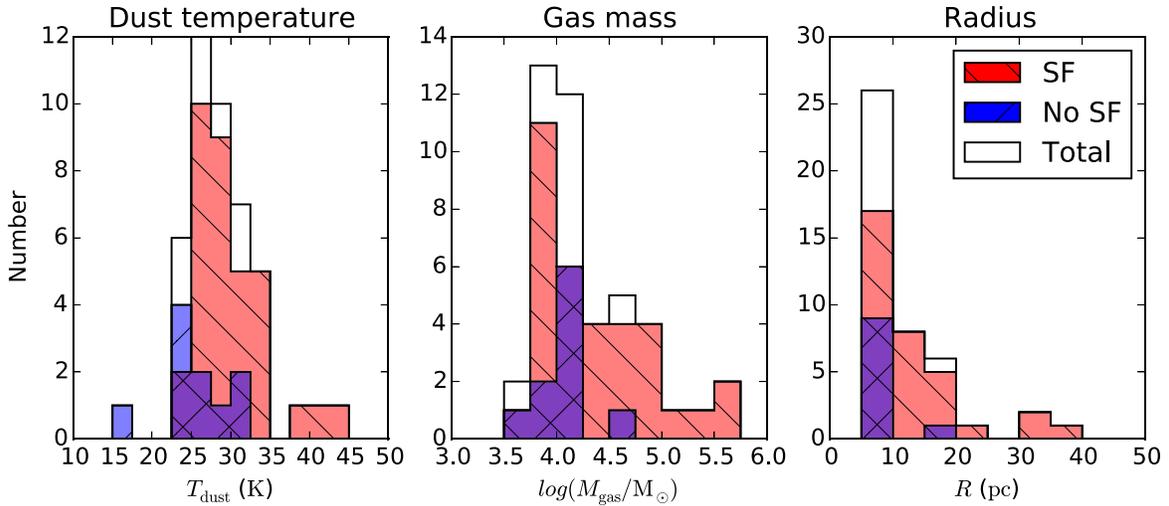}
\caption{The histograms of the physical properties of the 1.1~mm objects classified by the existence of star-formation activity. The overlap of the distributions is shown by a composite of colors and cross hatches. \label{histogram}}
\end{figure*}

\begin{figure}
\plotone{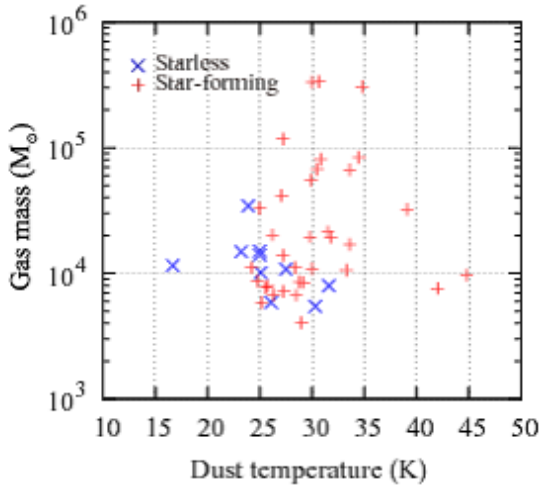}
\caption{Plots of the 1.1~mm objects on the plane where the ordinate and abscissa denote the dust temperature and gas mass, respectively. Blue cross and red plus show the starless and star-forming objects, respectively. \label{wide:TM}}
\end{figure}

Figure \ref{histogram} shows the histograms of the physical properties (dust temperature, gas mass, and radius) of the 1.1~mm objects, classified according to the existence of star-formation activity.
Table \ref{table:medians} shows the median values\footnote{N88 region is removed from our statistics.} of the physical parameters in some classification.
We note that the dust temperatures, gas masses, and radii of the star-forming objects are higher or larger than those of starless objects.
There are significant differences in the median values between the starless and star-forming clouds, with $p$-values of $0.01$, $0.20$, and $0.18$ for the dust temperature, dust mass, and radius, respectively, determined using the Mann-Whitney U test.
Figure \ref{wide:TM} also shows that the starless clouds concentrate in regions of low dust temperature and small gas mass.

As a result, the dust temperature, gas mass, and radius size differences between starless and star-forming clouds show the evolution stage of the GMCs. The same tendencies for the gas mass and size are shown by previous GMC studies based on CO line observations toward the LMC and M33 \citep{2009ApJS..184....1K, 2012ApJ...761...37M}.

\section{Summary}
\label{sec7}
The AzTEC/ASTE survey provided the first high-resolution (FWHM of $40\arcsec$) and high-sensitivity (1$\sigma$=5--12~mJy~beam$^{-1}$) image at a 1.1~mm wavelength toward the whole SMC.
A total of 44~extended 1.1~mm objects were identified in the FRUIT map above $\sim5\sigma$, and their physical properties were estimated by SED analysis using 1.1~mm and {\it Herschel} data.
The 1.1~mm objects displayed masses of $4\times 10^3$--$3\times 10^5~M_\sun$, and dust temperatures of 17--45~K.

Our conclusions on the nature of the 1.1~mm objects observed and discussed in this paper are summarized below:
\begin{enumerate}
\item{}The robust correlation between dust temperature and the ratio of the 24~$\micron$ luminosity (corrected for the PAH contribution) to the gas mass supports that the dust temperature of the 1.1~mm objects is mainly determined by the local star-formation activity.
\item{}The 1.1~mm objects showed good spatial correlation with the 24~$\micron$ and CO emission, and the estimated physical properties of the 1.1~mm objects revealed properties very similar to the well-known GMC properties, such as the gas mass, size, density, and mass function of our galaxy and the LMC.
\item{}The classification of the 1.1~mm objects in terms of star-formation activity revealed that the starless objects harbor lower dust temperatures, smaller gas masses, and smaller radii compared to those of the star-forming objects.
It is reasonable to conclude that the existence of star-formation activity reflects the evolutionary sequences.
\end{enumerate}

The newly identified dust-selected clouds, at 1.1~mm in this study, include significantly interesting samples for the investigation of the evolutionary phase of the GMCs in the low-metallicity environment.
In particular, high-sensitivity and high-resolution CO observations toward the CO-dark samples will reveal the formation timescale of CO molecules and the chemical and dynamical evolution process of GMCs in the low-metallicity environment, quantitatively.
Therefore, the high-sensitivity capability of ALMA will be essential for investigating the dust clouds discovered by our study.

To provide a better understanding of the evolutionary sequence of the dust-selected clouds, it is necessary to compare our result with the LMC and nearby galaxies as environments of differing metallicity.
This study demonstrated that the dust continuum observations from submillimeter telescopes and imaging arrays with direct detectors are very strong tools to search the sites of star formation in nearby galaxies. 
Although GMC identifications and physical property estimates using CO data are very well-established methods, dust continuum observations also provide a complementary approach.
In addition, the GMC study, using the dust continuum, will be an effective approach for investigating the metallicity effect in the ISM of nearby galaxies.

\acknowledgments
The ASTE project was driven by NRO/NAOJ, in collaboration with the University of Chile, and Japanese institutes including the University of Tokyo, Nagoya University, Osaka Prefecture University, Ibaraki University, Hokkaido University, and Joetsu University of Education.
Observations with ASTE were carried out remotely from Japan using NTT's GEMnet2 and its partner R\&E networks, which are based on AccessNova collaboration between the University of Chile, NTT Laboratories, and NAOJ.
This work is based on data products made with the {\it Spitzer} Space Telescope (JPL/Caltech under a contract with NASA), and has made use of SIMBAD, operated at CDS, Strasbourg, France.
This work was partly supported by MEXT KAKENHI Grant Number 20001003 and JSPS KAKENHI Grant Numbers 19403005, 22740127.
We appreciate the many fruitful discussions about the FRUIT procedure with M. Hiramatsu, N. Ikeda, Y. Otomo, Y. Shimajiri, and T. Tsukagoshi.
The authors also would like to thank the anonymous reviewer for their valuable comments and suggestions to improve our manuscript.

{\it Facilities:} \facility{ASTE/AzTEC}, \facility{ATCA}, \facility{{\it Herschel}/PACS}, \facility{{\it Herschel}/SPIRE}, \facility{Mopra}, \facility{NANTEN}, \facility{Parkes}, \facility{{\it Spitzer}/IRAC}, \facility{{\it Spitzer}/MIPS}

\bibliographystyle{aasjournal}
\bibliography{SMCI_takekoshi}

\appendix

\section{Result of the PCA cleaning}
\label{sec:PCA}
\begin{figure*}
\plotone{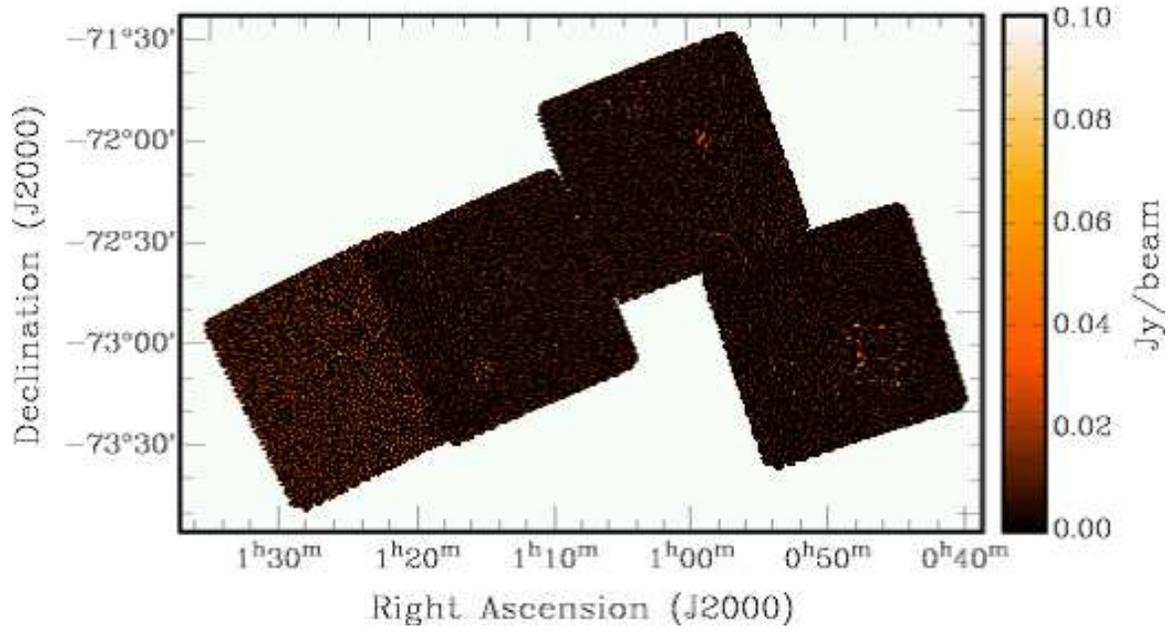}
\caption{The 1.1~mm continuum map of the SMC analyzed by PCA.\label{fig:aztec_pca}}
\end{figure*}

Figure \ref{fig:aztec_pca} shows the 1.1~mm continuum map analyzed using the PCA cleaning.
The map noise level is almost comparable to the FRUIT map ($1\sigma$ of  5--12~mJy~beam$^{-1}$).

The PCA objects are identified as local peaks of $\geq4.5\sigma$ in the signal-to-noise ratio maps.
We have identified 53~objects in total, including 28, 14, 10, and 1~objects in the SW, NE, Wing, and N88 fields, respectively.
The source catalog is shown in Table \ref{table:catalog_pca}.

We searched for the counterparts of the PCA objects using SIMBAD\footnote{http://simbad.u-strasbg.fr/simbad/} at the positions of the PCA objects within the beam size ($28\arcsec$), and corresponding emission nebulae were checked on the PCA images.
The identified objects are shown in Table \ref{table:catalog_pca}. 
Most of the PCA objects correspond to extended astronomical objects located in the SMC such as emission nebulae and molecular clouds.
The analysis for point sources is insufficient for these sources, thus, analysis for extended structures using FRUIT is important.
Some objects have counterparts other than the objects in the SMC.
SW-22 is identified as a Sc/Sb galaxy KHBG~27 \citep{2001PASP..113.1115K}, located in the nearby universe ($Z_{V-B}=0.0\pm0.2$).
Wing-4 is included in the 6dF Galaxy Survey catalog \citep{2009MNRAS.399..683J}, and the redshift of the galaxy is $z=0.06660$.
\citet{2011MNRAS.417.2651M} also classified this galaxy as a radio-loud active galactic nucleus, based on the optical identification of the Australia Telescope 20 GHz Survey sources.
Wing-7 can be explained by a strongly lensed submillimeter galaxy at a redshift of 1.4--3.9 \citep{2013ApJ...774L..30T}. 

\newpage
\section{Spectral energy distribution of the 1.1~mm objects}
\label{sec:sed}
\begin{figure}
\centering
\includegraphics[width=3.0in]{SMC_wide_SW_SED1.eps}
\includegraphics[width=3.0in]{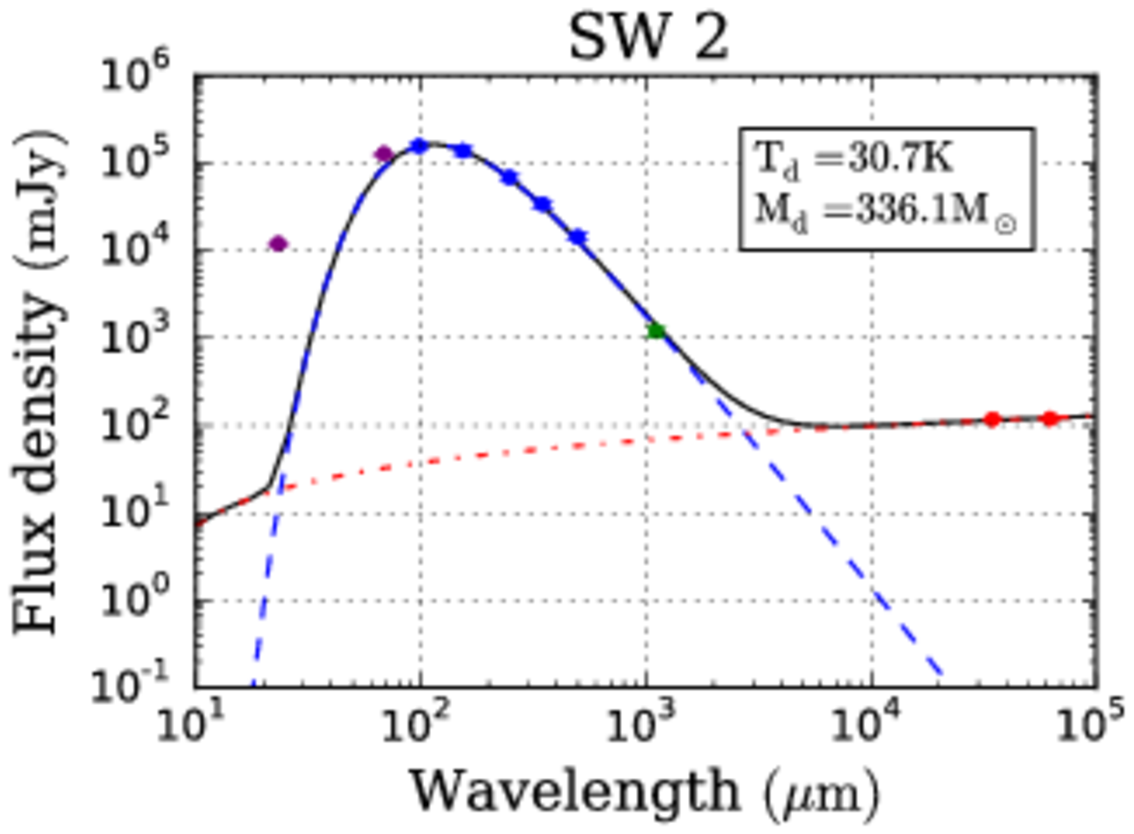}\\
\includegraphics[width=3.0in]{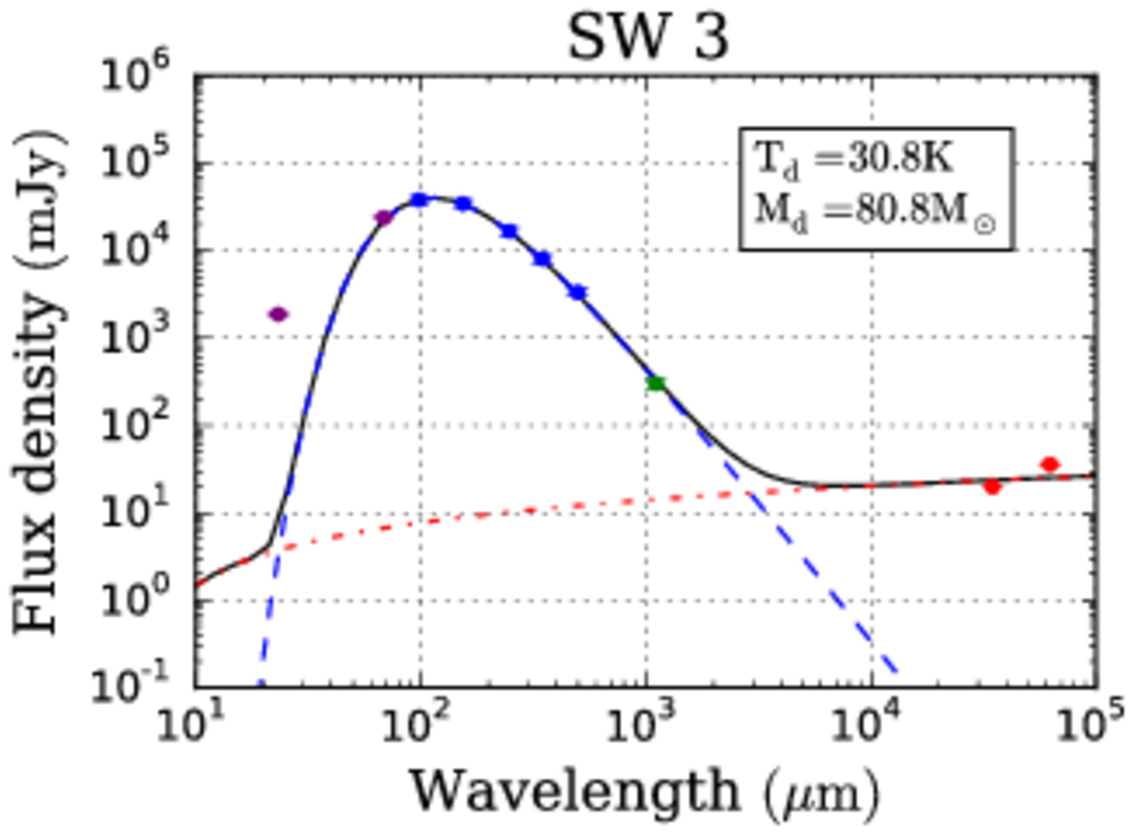}
\includegraphics[width=3.0in]{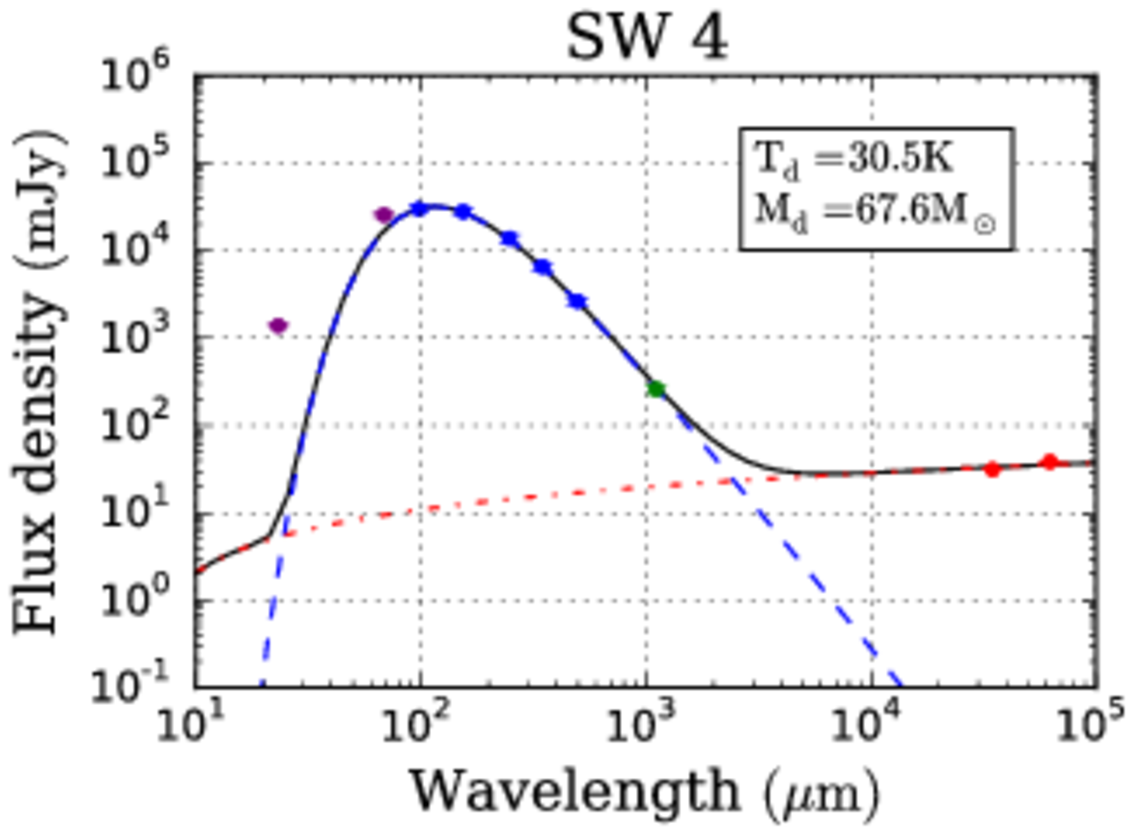}\\
\includegraphics[width=3.0in]{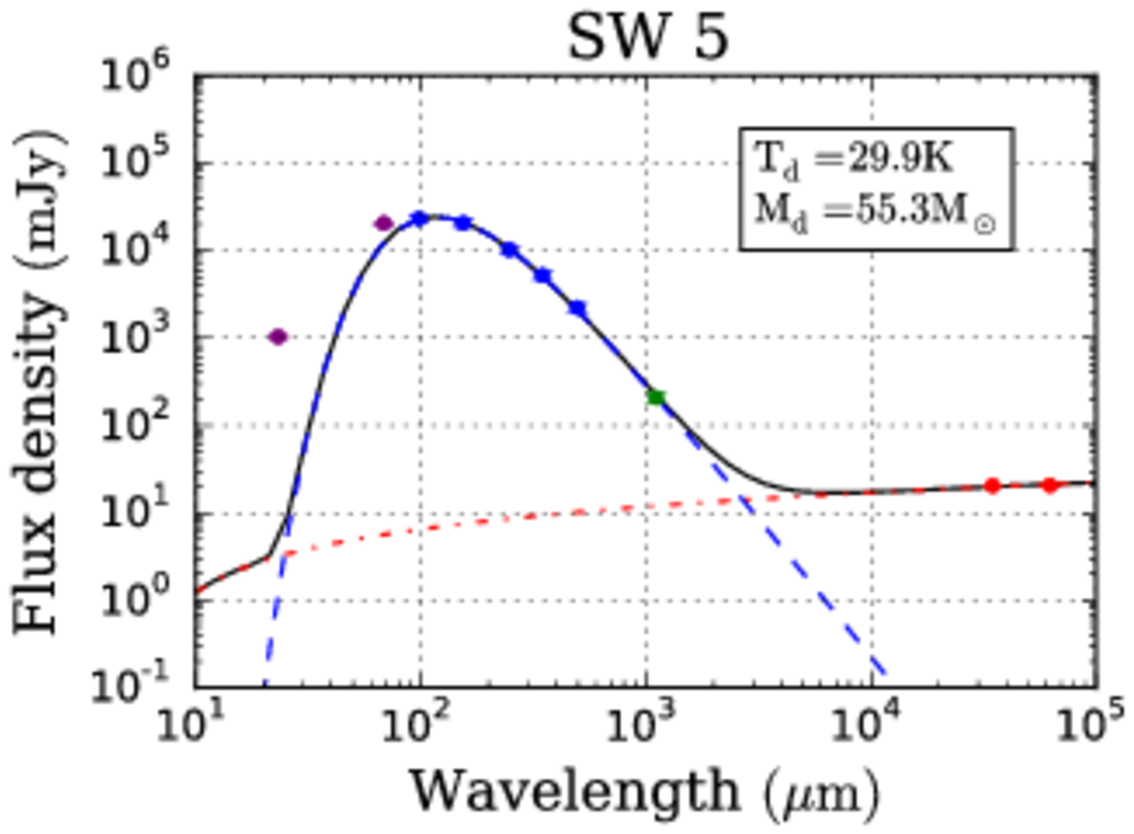}
\includegraphics[width=3.0in]{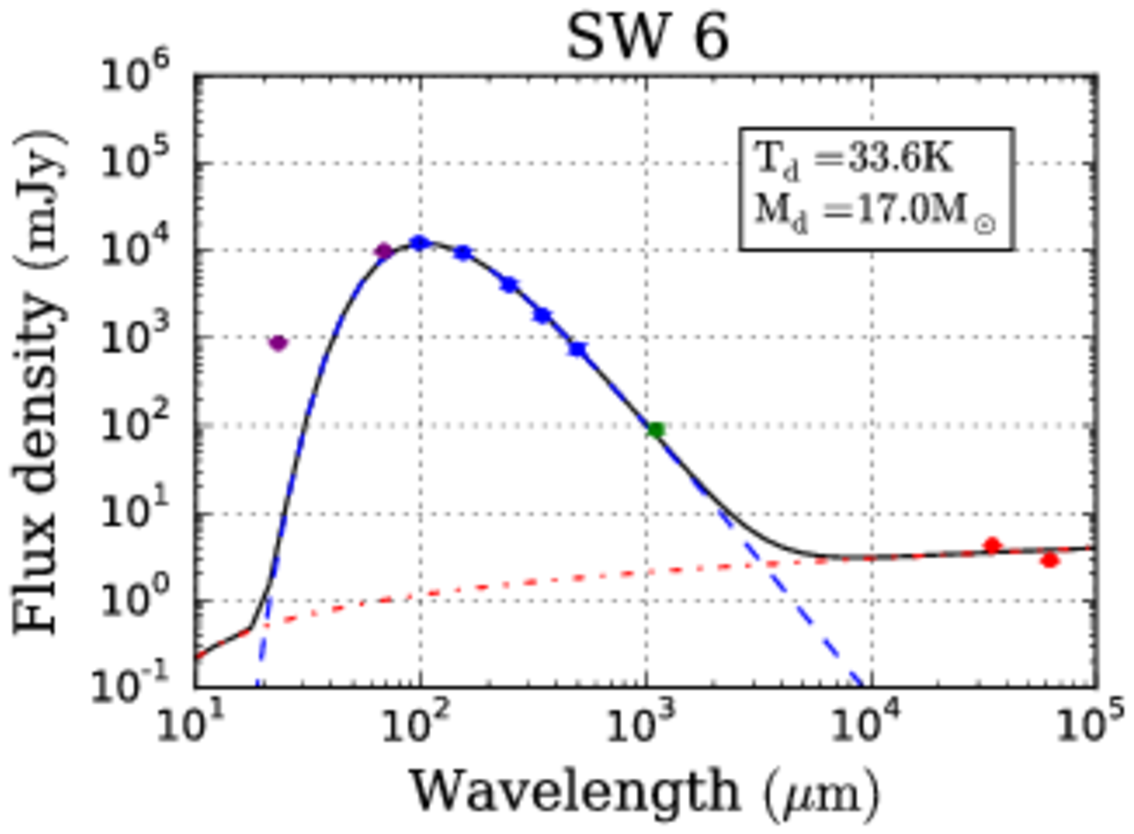}\\
\includegraphics[width=3.0in]{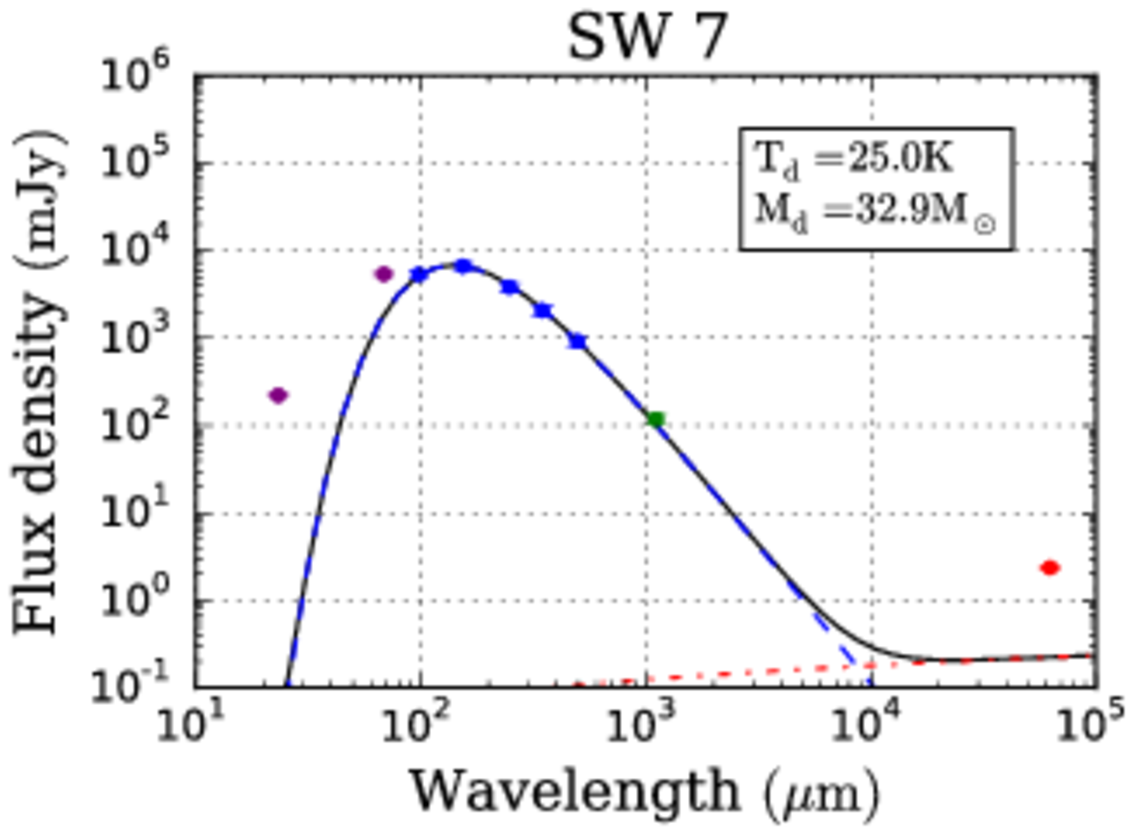}
\includegraphics[width=3.0in]{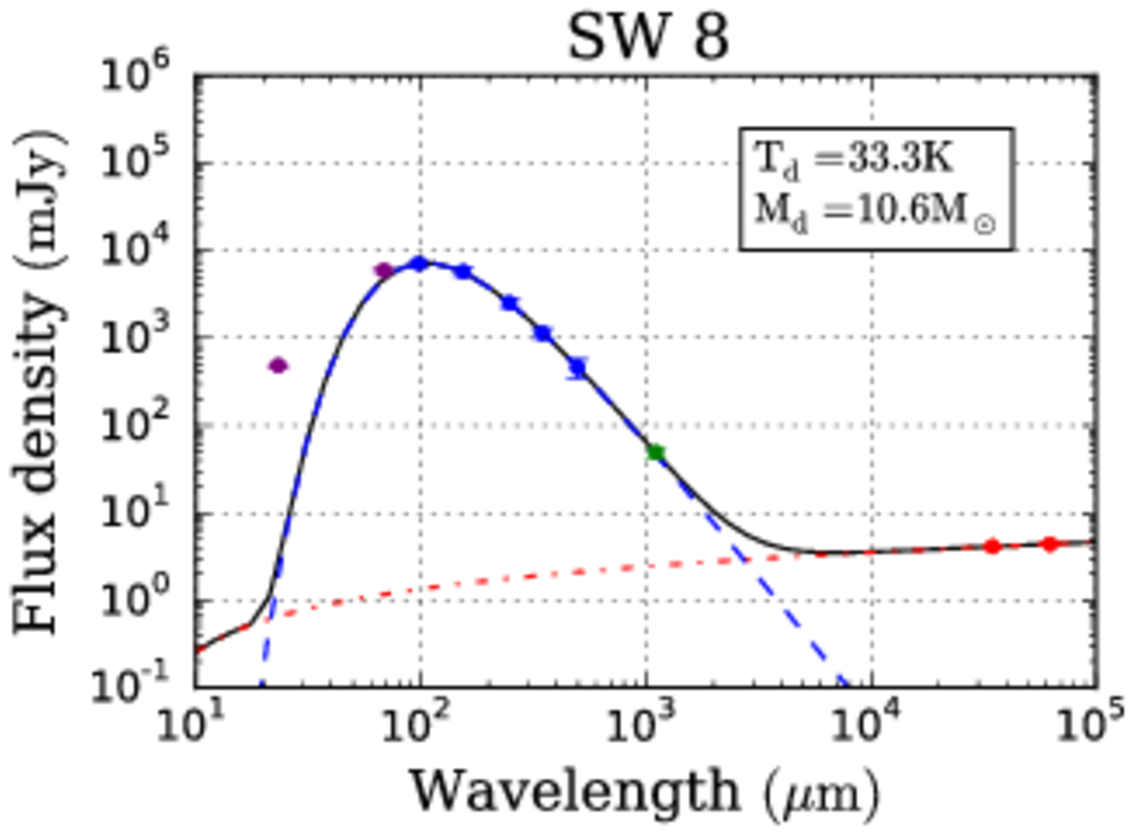}\\
\caption{The SEDs of AzTEC/SMC objects. The fitting points for cold dust SED are shown in green (1.1~mm) and blue (100, 160, 250, 350, and~500 $\micron$). The fitting points for free-free model SED are shown in red (4.8 and 8.64~GHz). The purple points (70 and 24~$\micron$) are not used for the fitting. The dashed blue and dashed-dotted red lines show the SED model for cold dust and free-free emission, respectively. The solid line shows the total SED for cold dust and free-free emission.}
\end{figure}

\begin{figure}
\figurenum{13}
\centering
\includegraphics[width=3.0in]{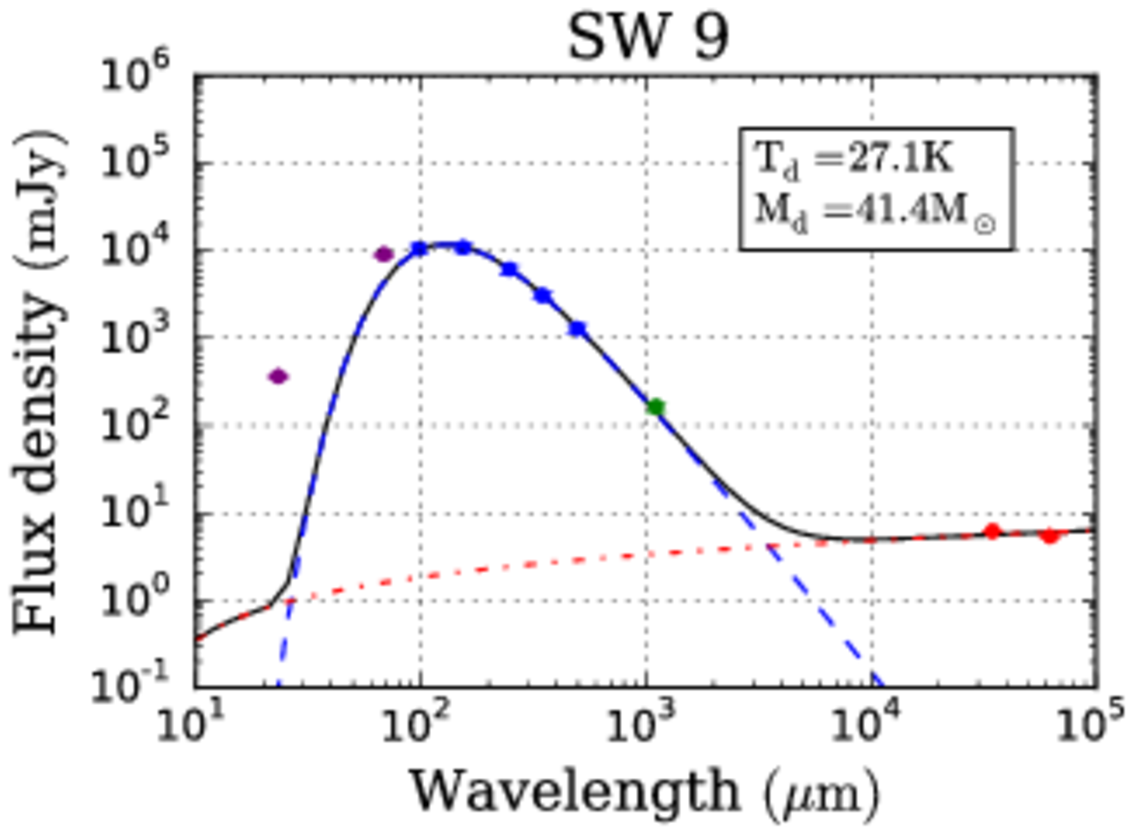}
\includegraphics[width=3.0in]{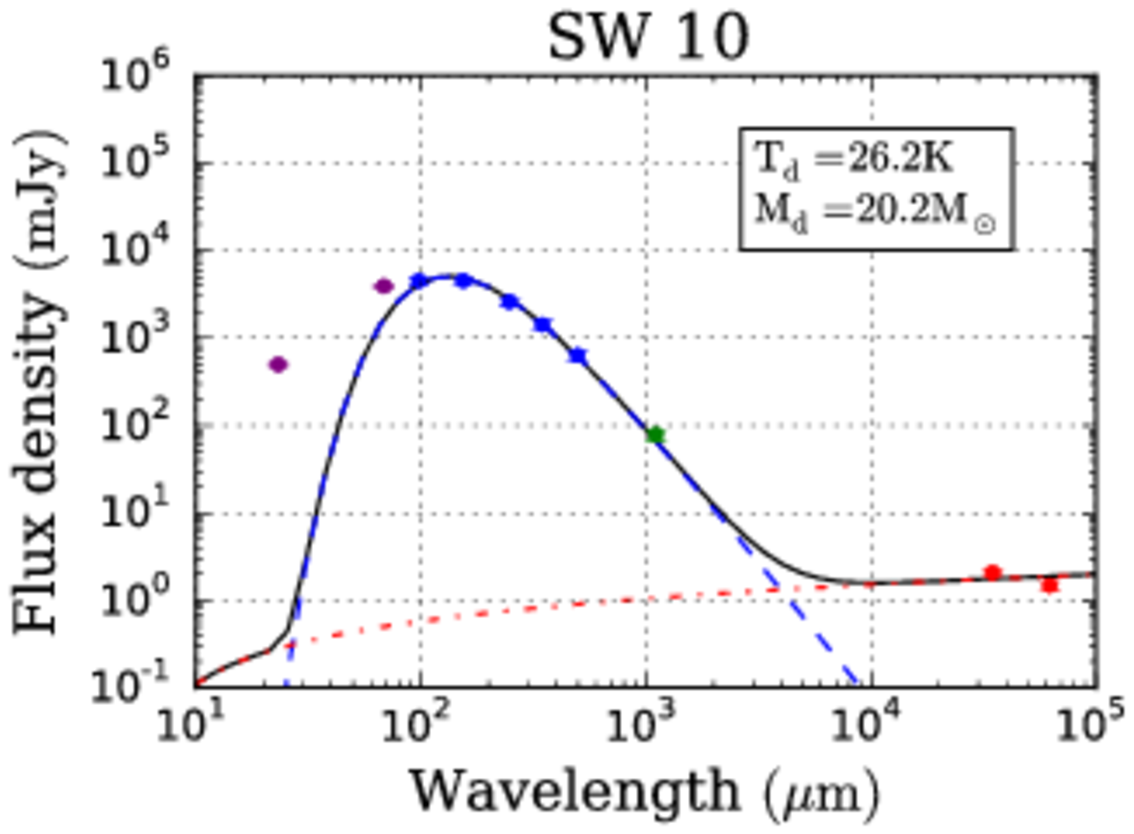}\\
\includegraphics[width=3.0in]{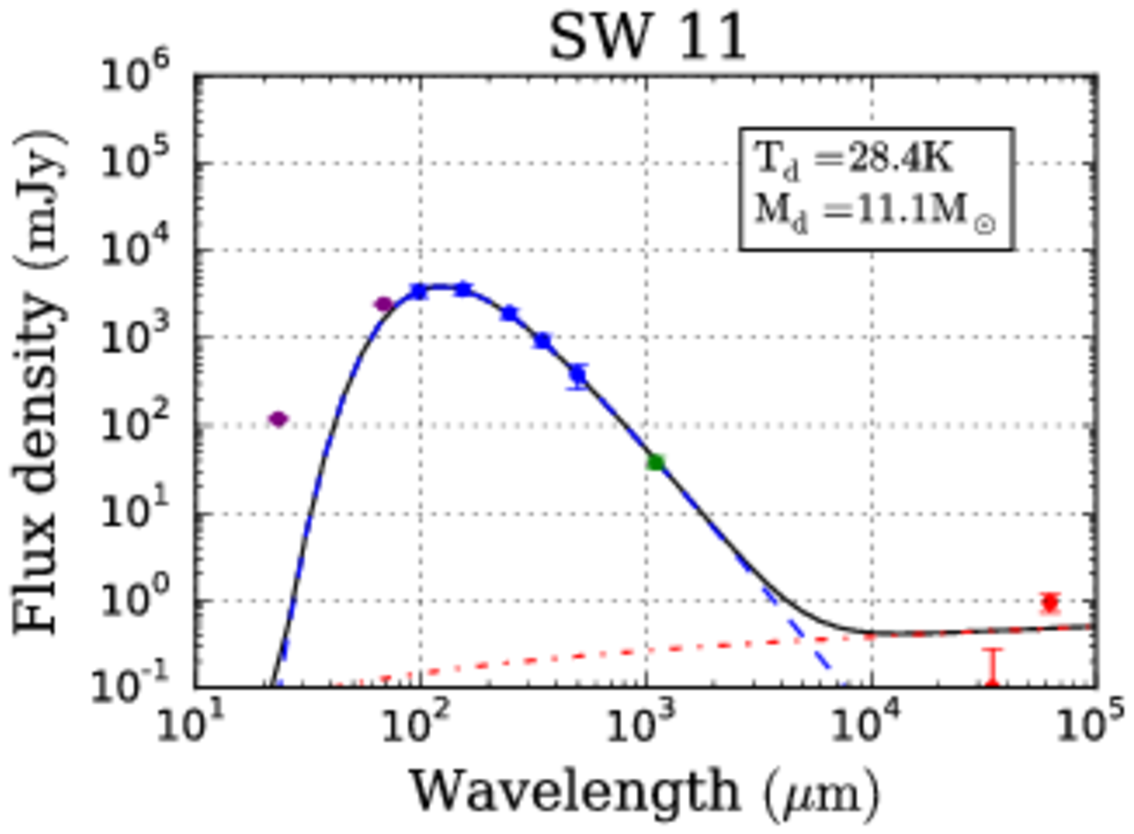}
\includegraphics[width=3.0in]{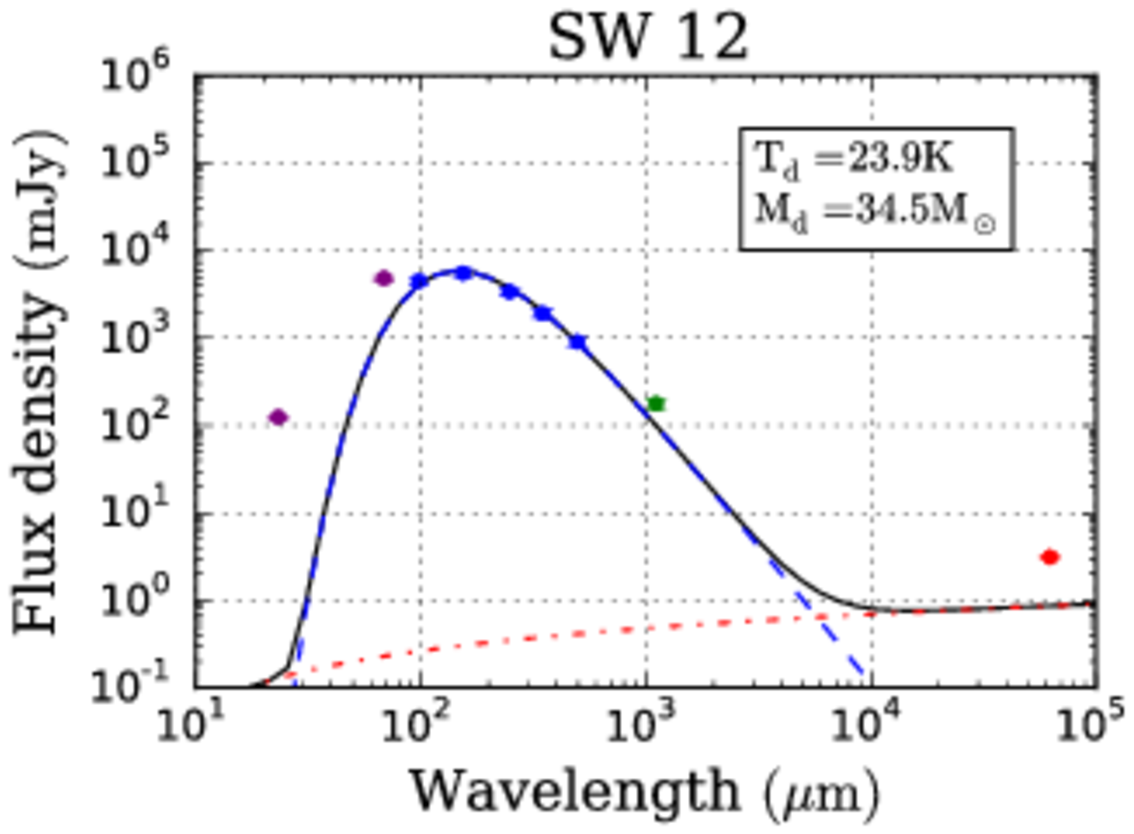}\\
\includegraphics[width=3.0in]{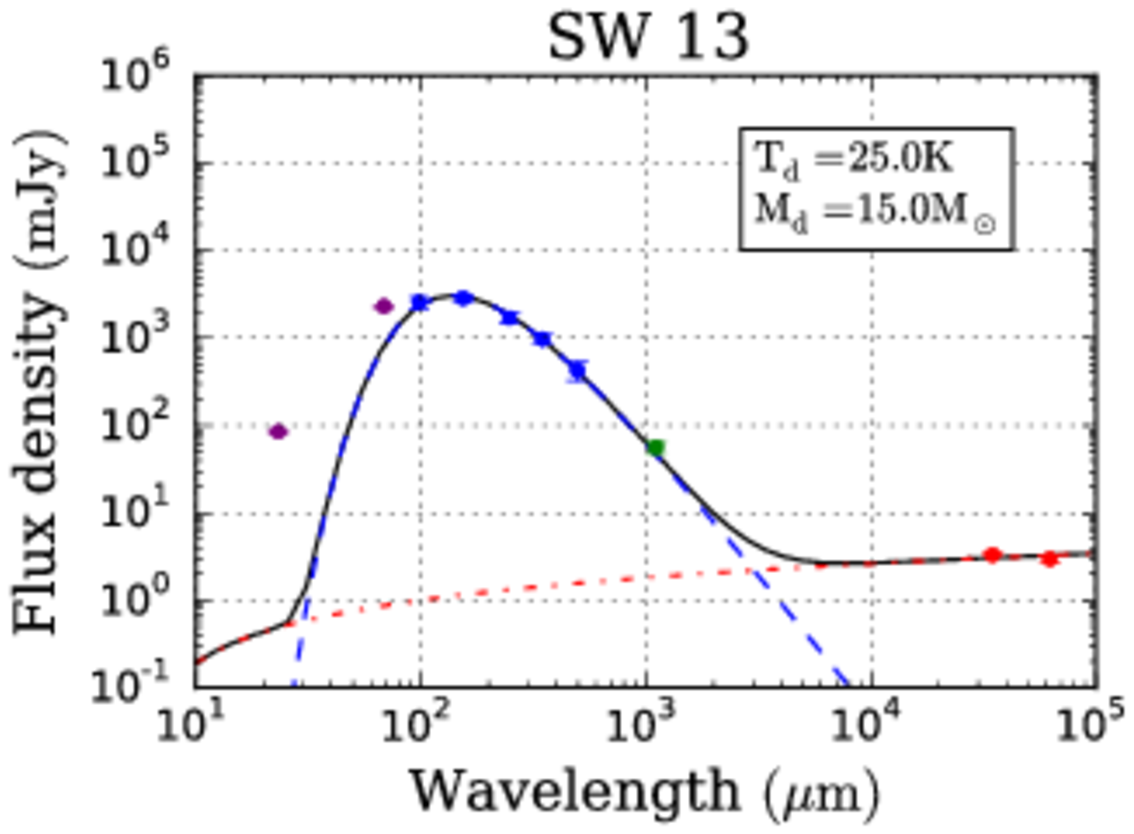}
\includegraphics[width=3.0in]{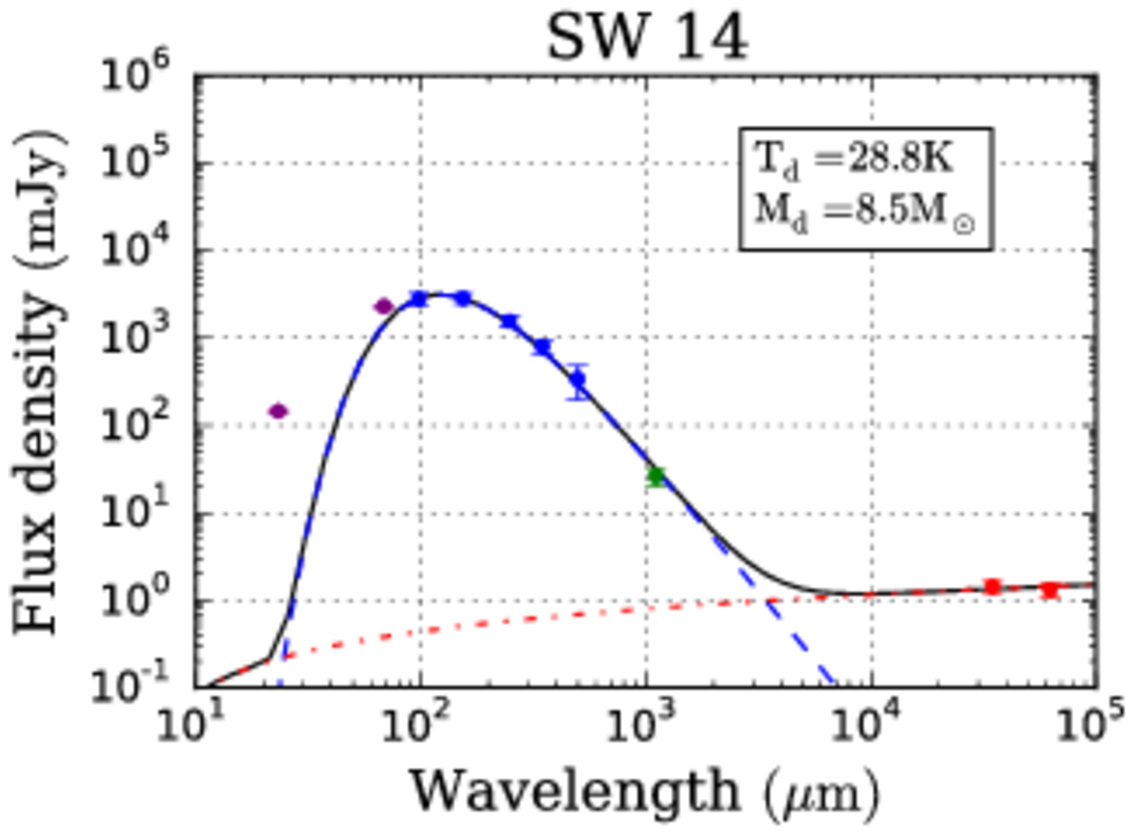}\\
\includegraphics[width=3.0in]{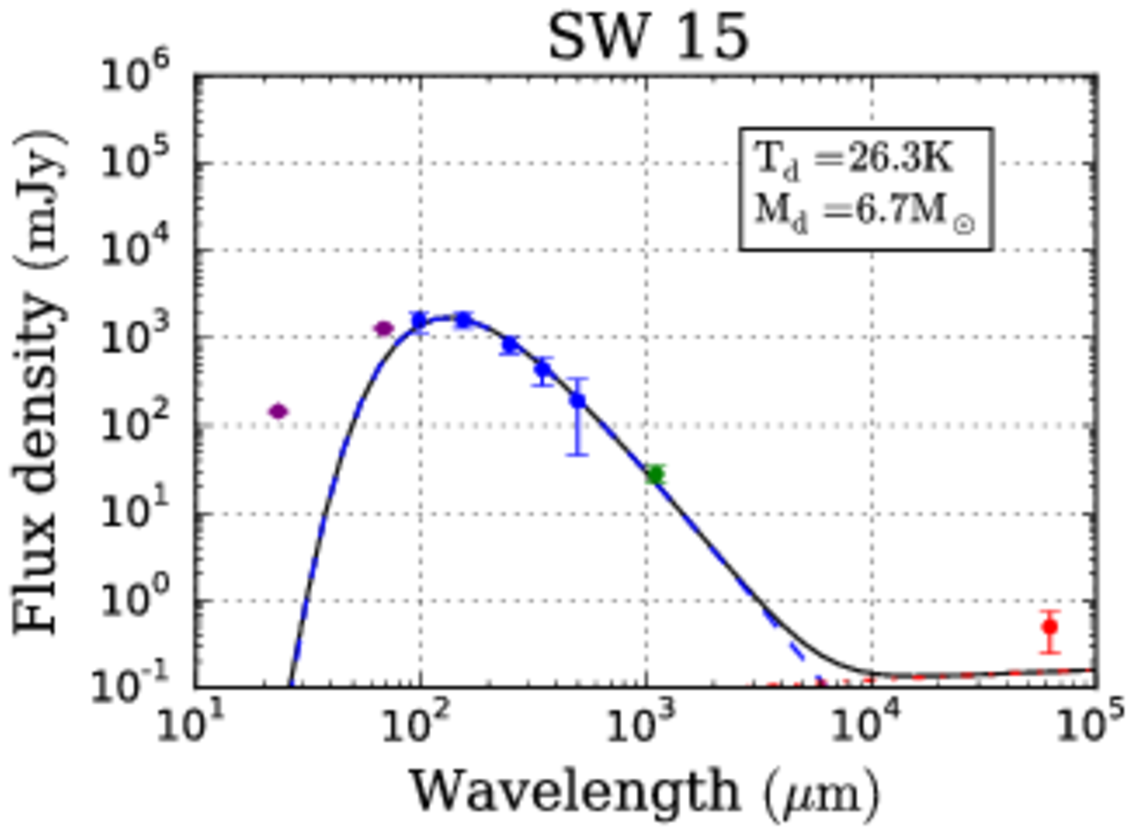}
\includegraphics[width=3.0in]{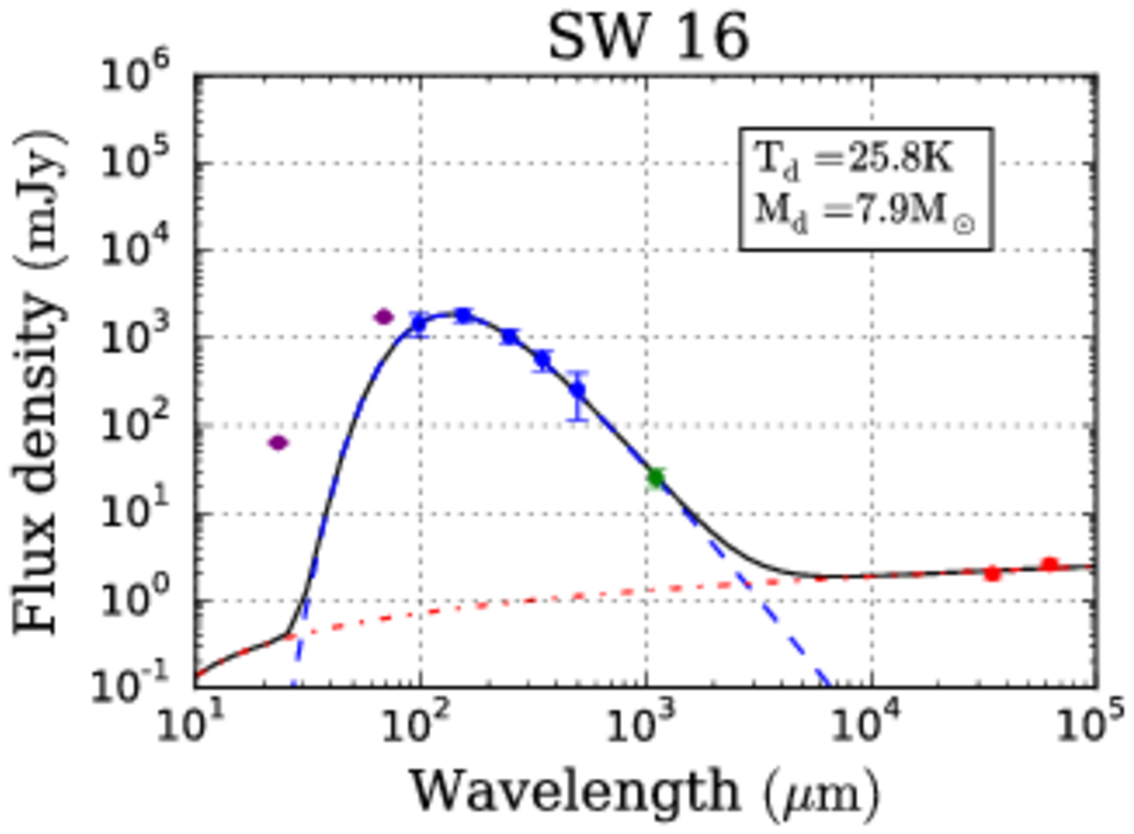}\\
\caption{{\it Continued.}}
\end{figure}

\begin{figure}
\figurenum{13}
\centering
\includegraphics[width=3.0in]{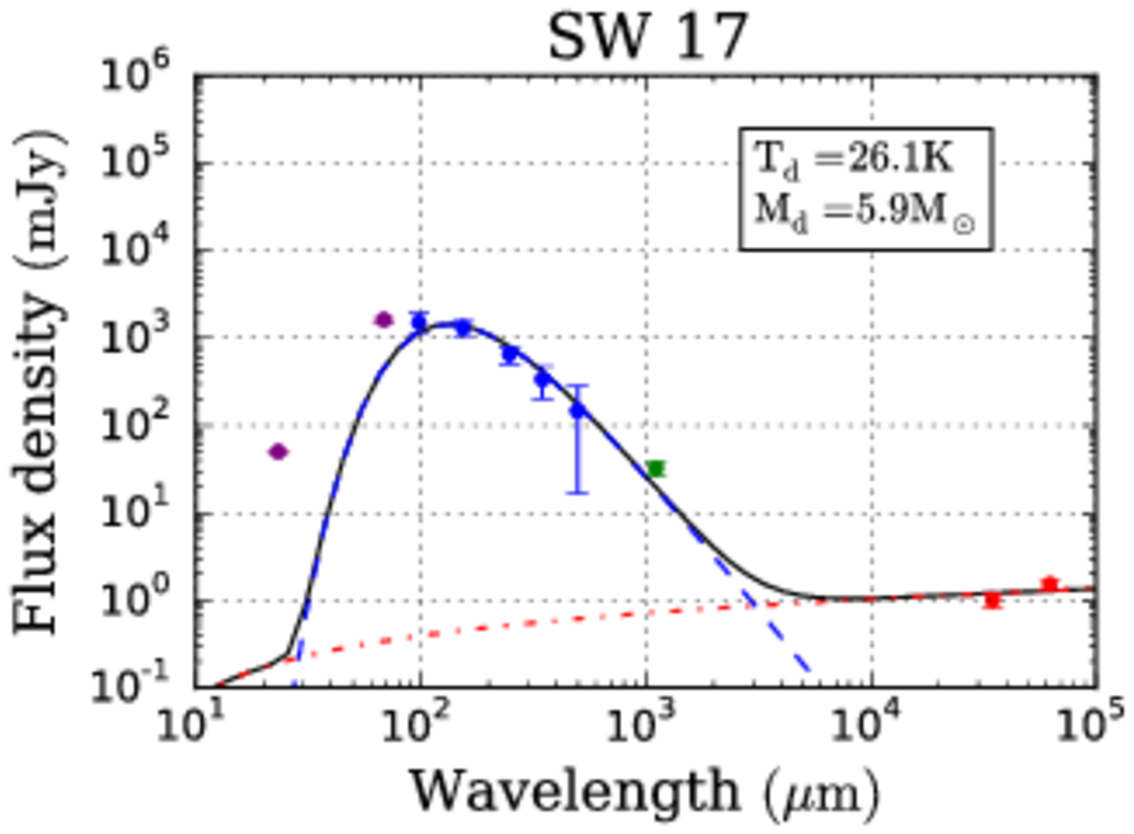}
\includegraphics[width=3.0in]{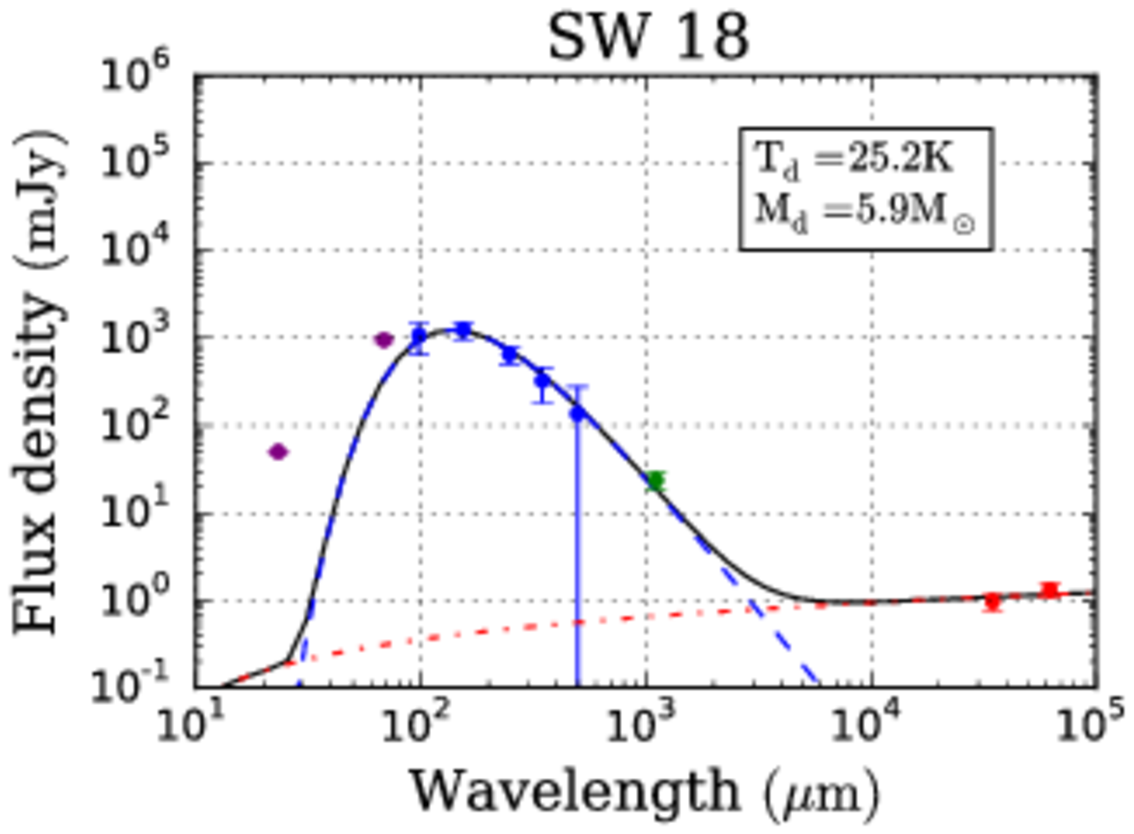}\\
\includegraphics[width=3.0in]{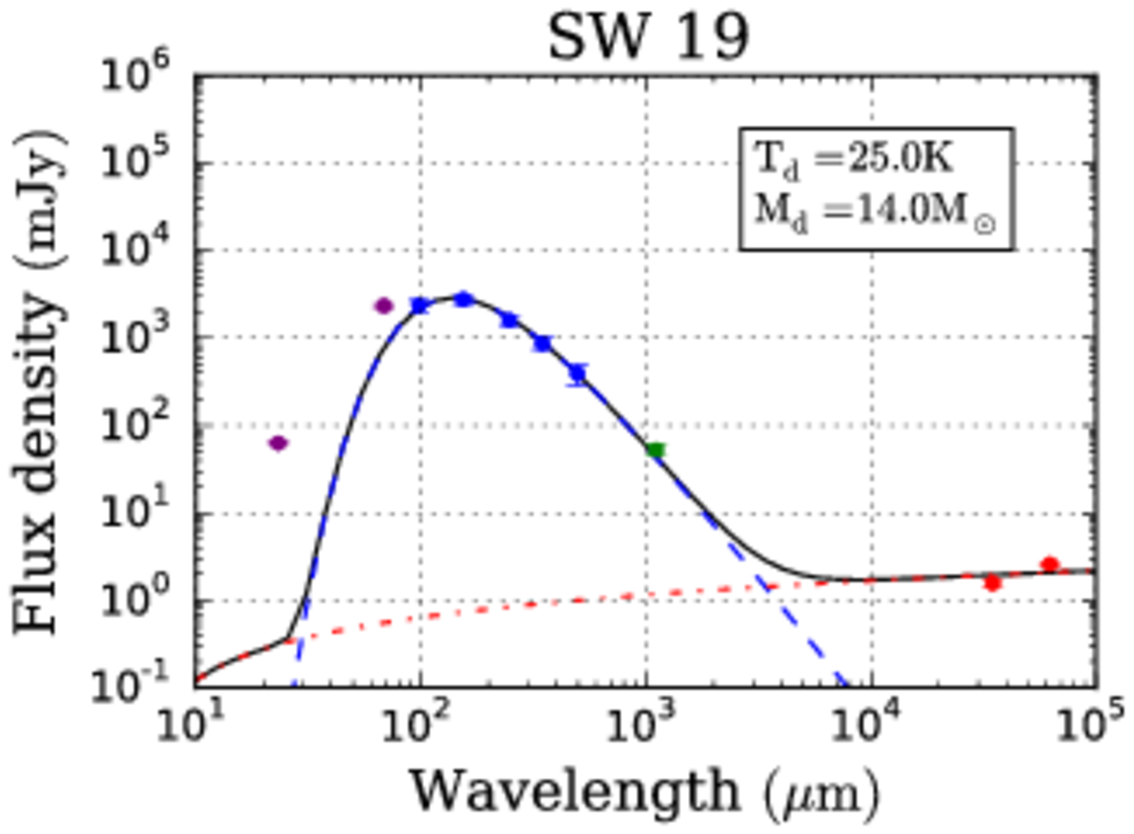}
\includegraphics[width=3.0in]{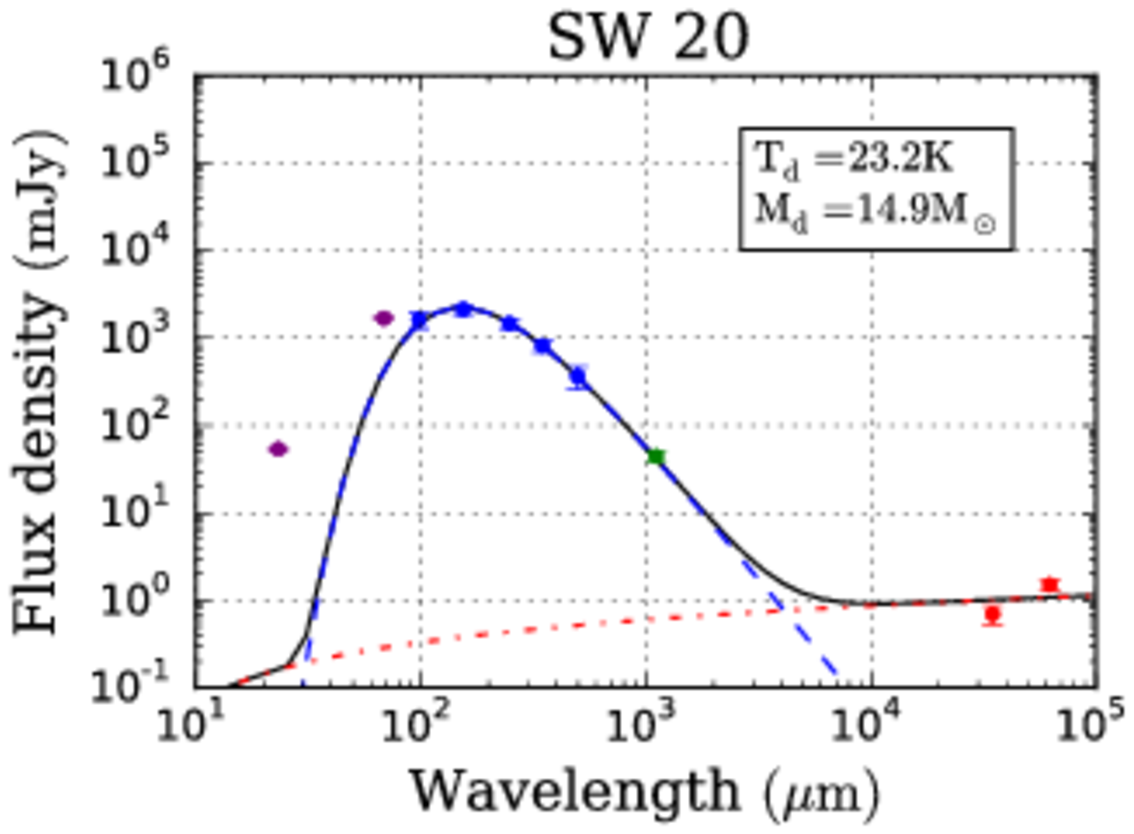}\\
\includegraphics[width=3.0in]{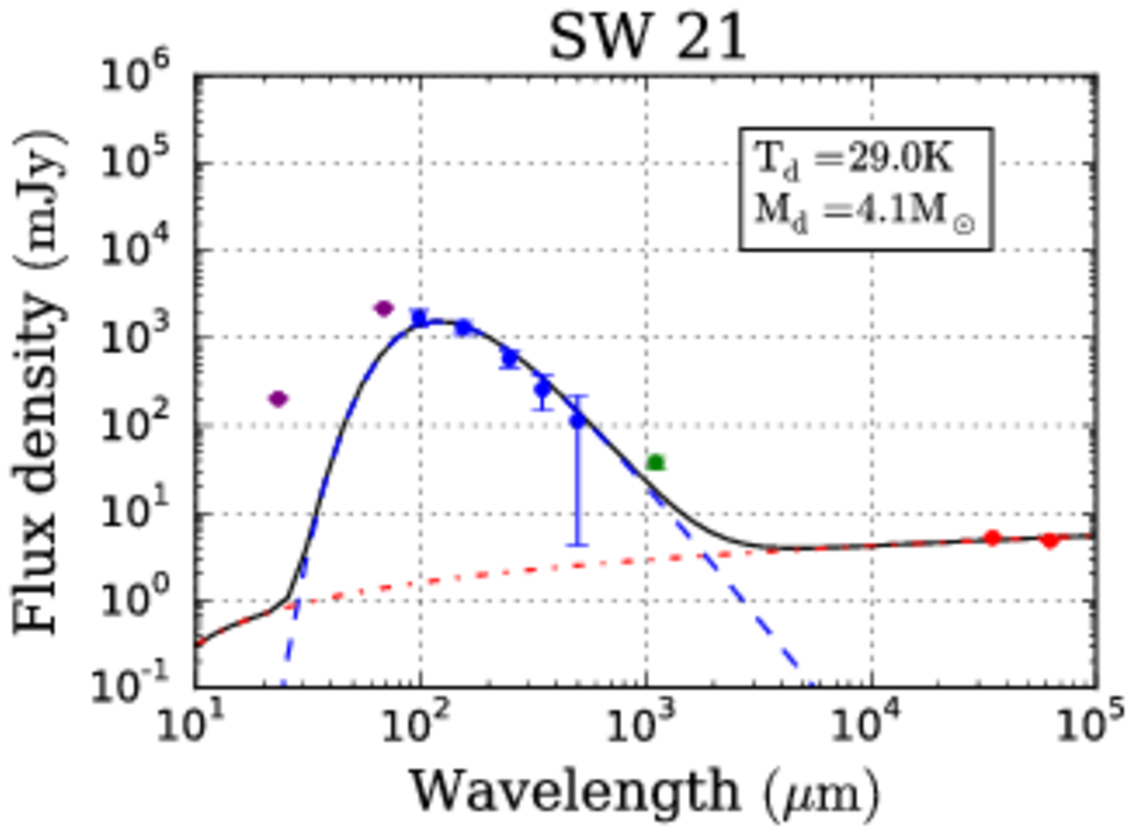}
\includegraphics[width=3.0in]{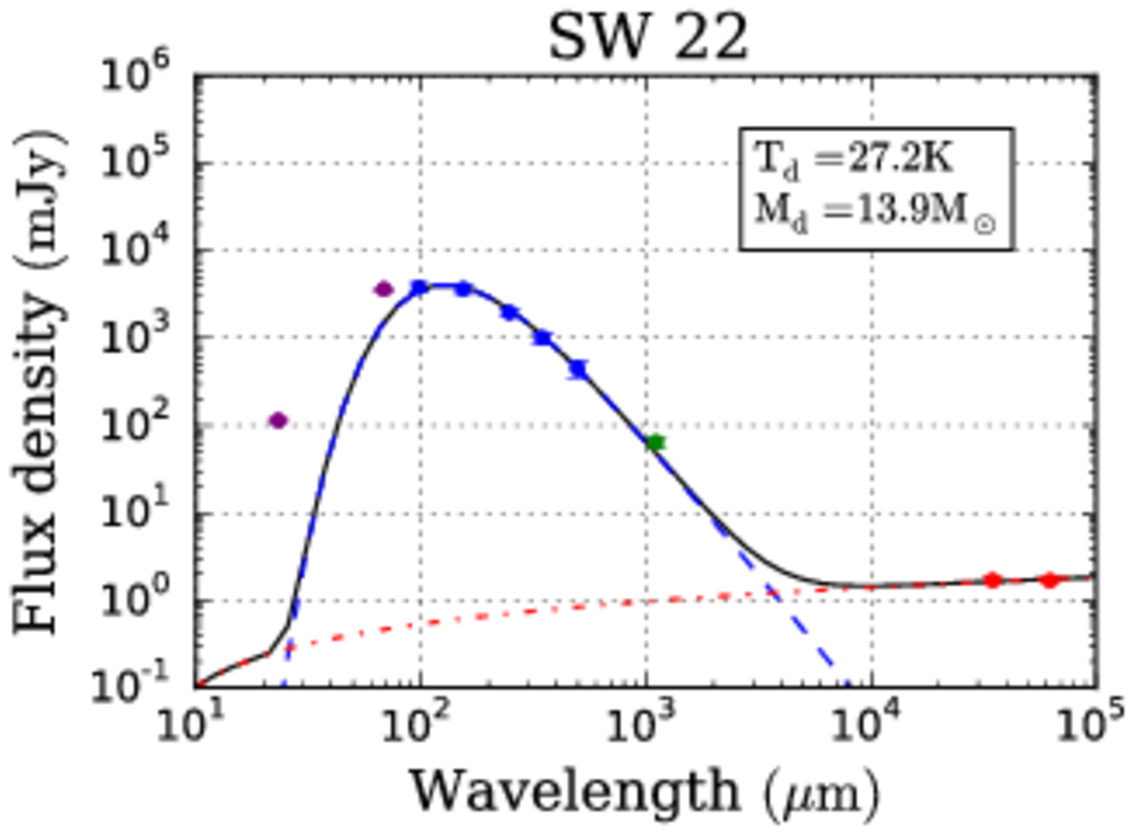}\\
\includegraphics[width=3.0in]{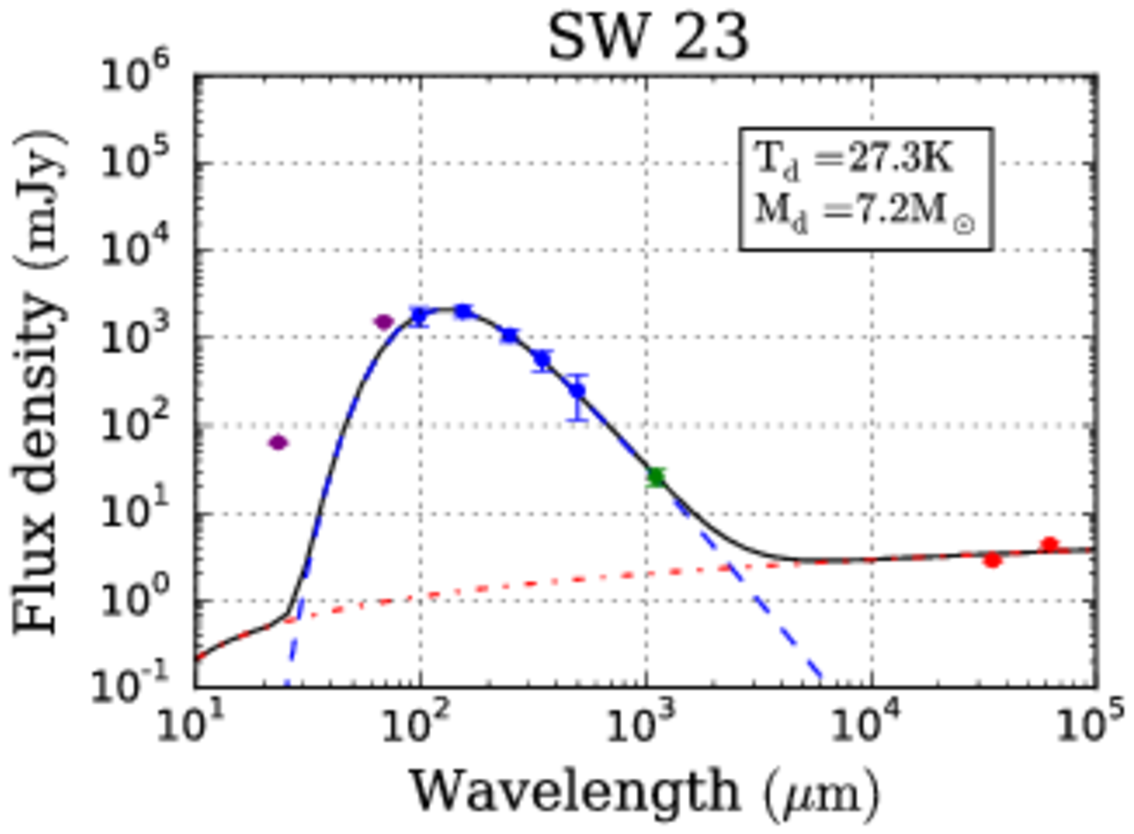}
\includegraphics[width=3.0in]{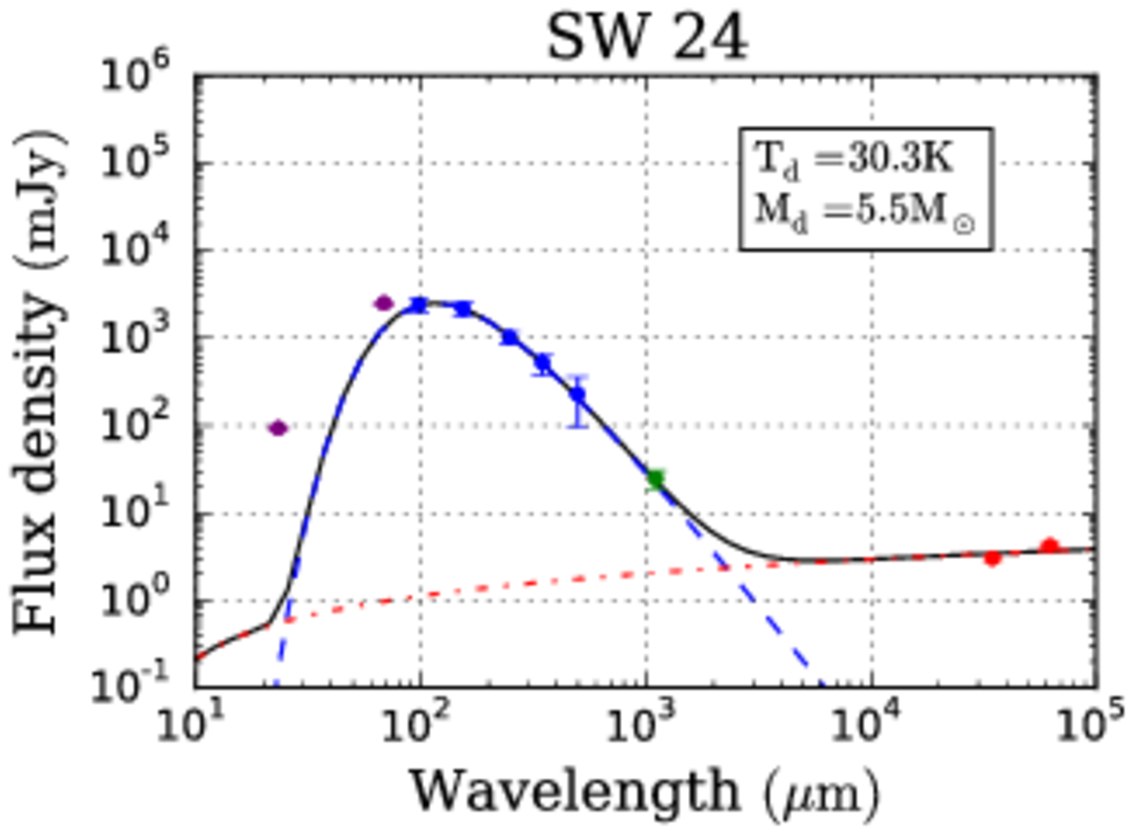}\\
\caption{{\it Continued.}}
\end{figure}

\begin{figure}
\figurenum{13}
\centering
\includegraphics[width=3.0in]{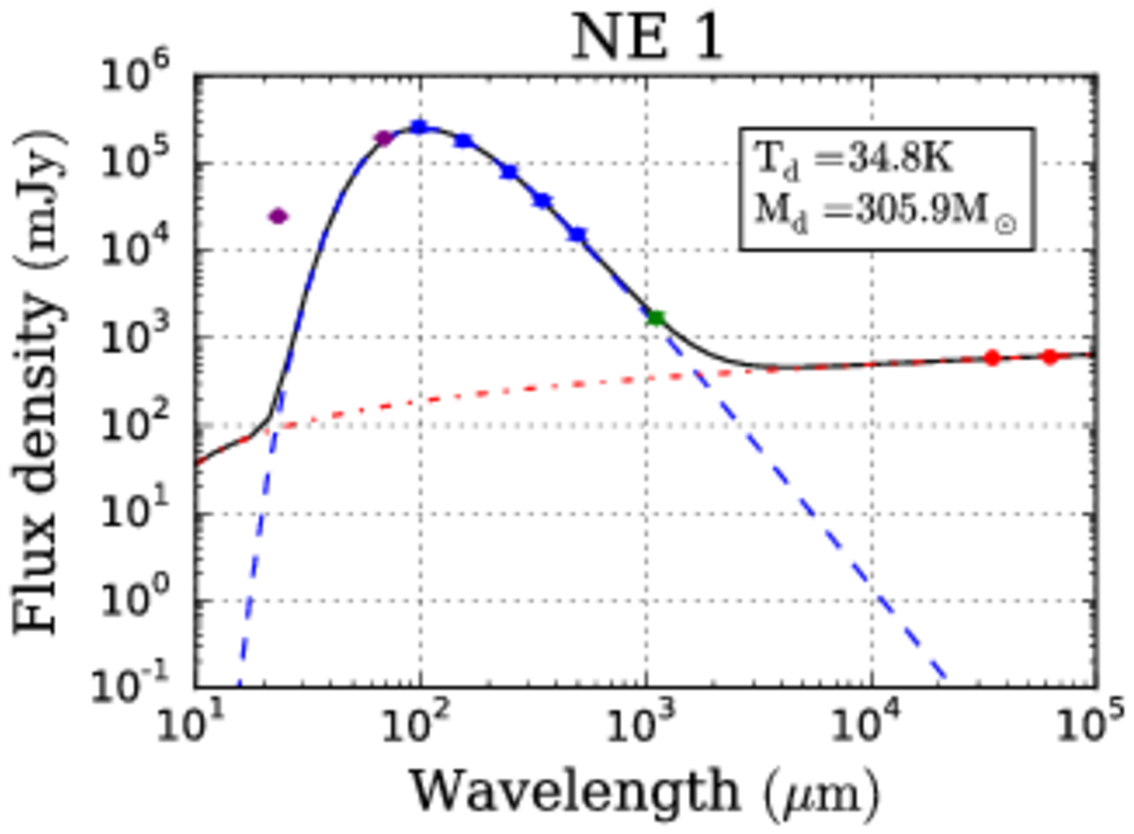}
\includegraphics[width=3.0in]{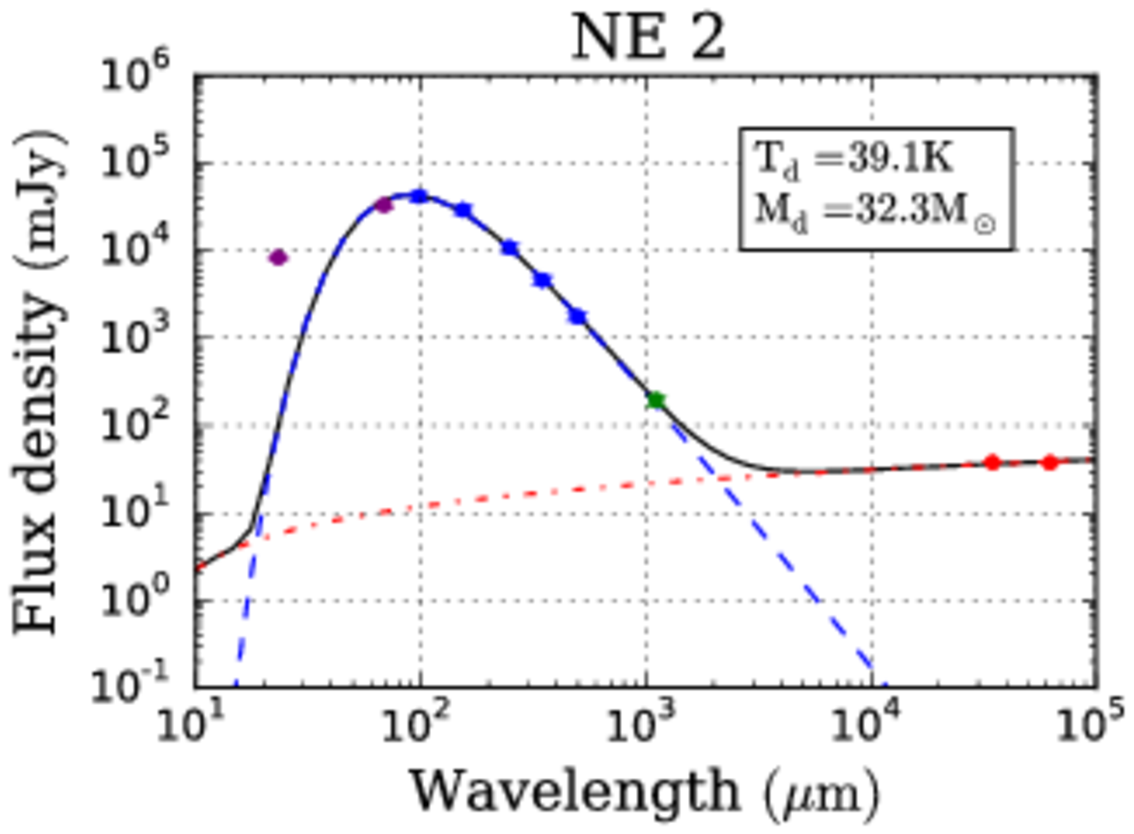}\\
\includegraphics[width=3.0in]{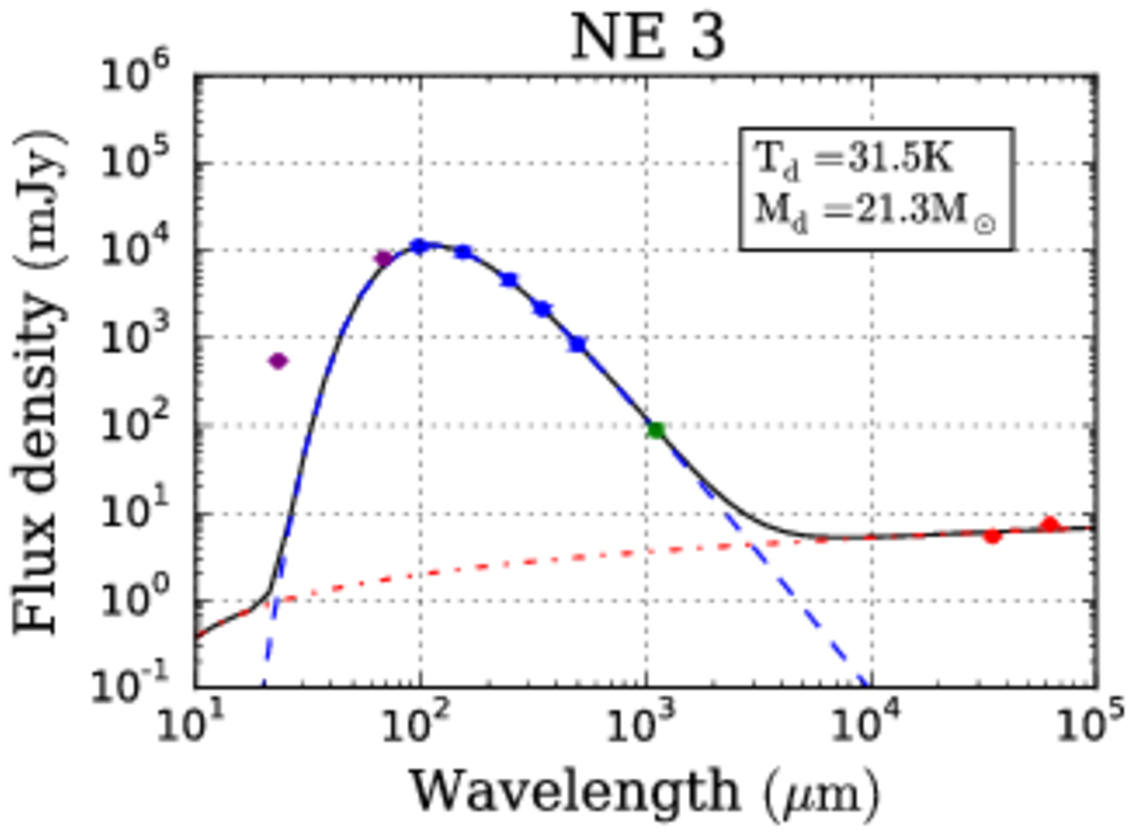}
\includegraphics[width=3.0in]{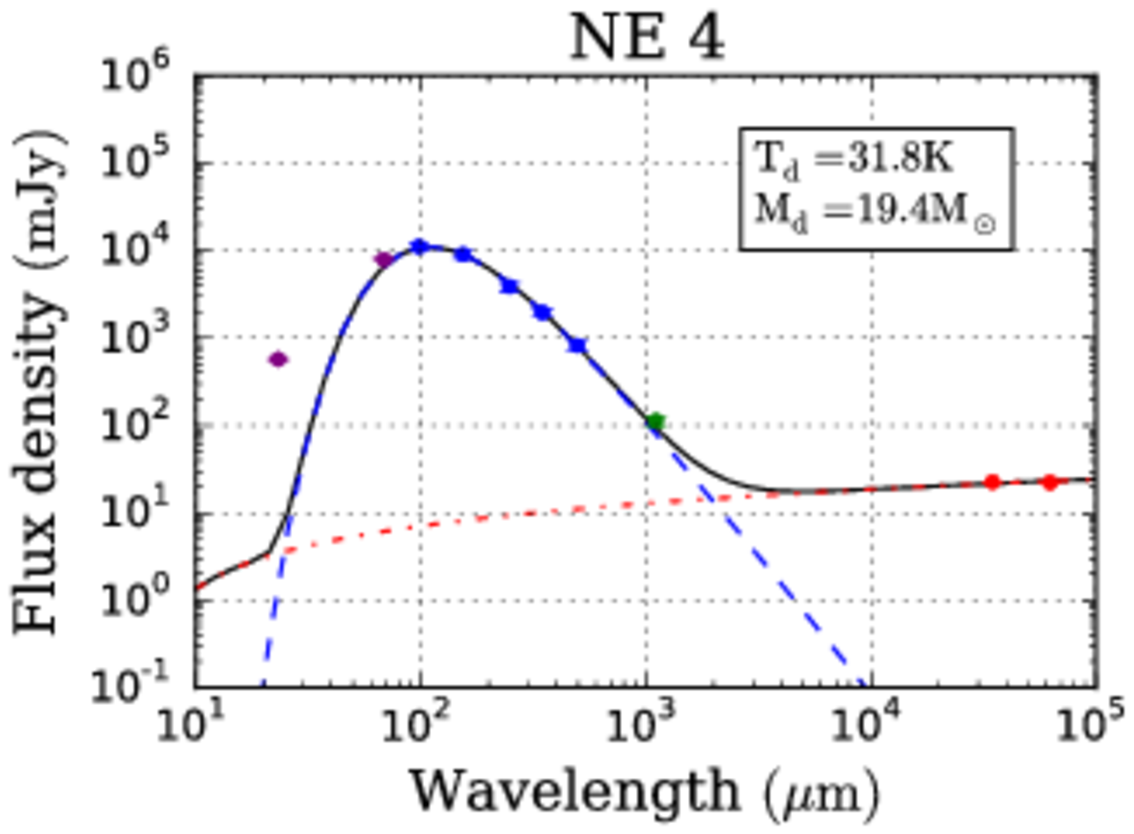}\\
\includegraphics[width=3.0in]{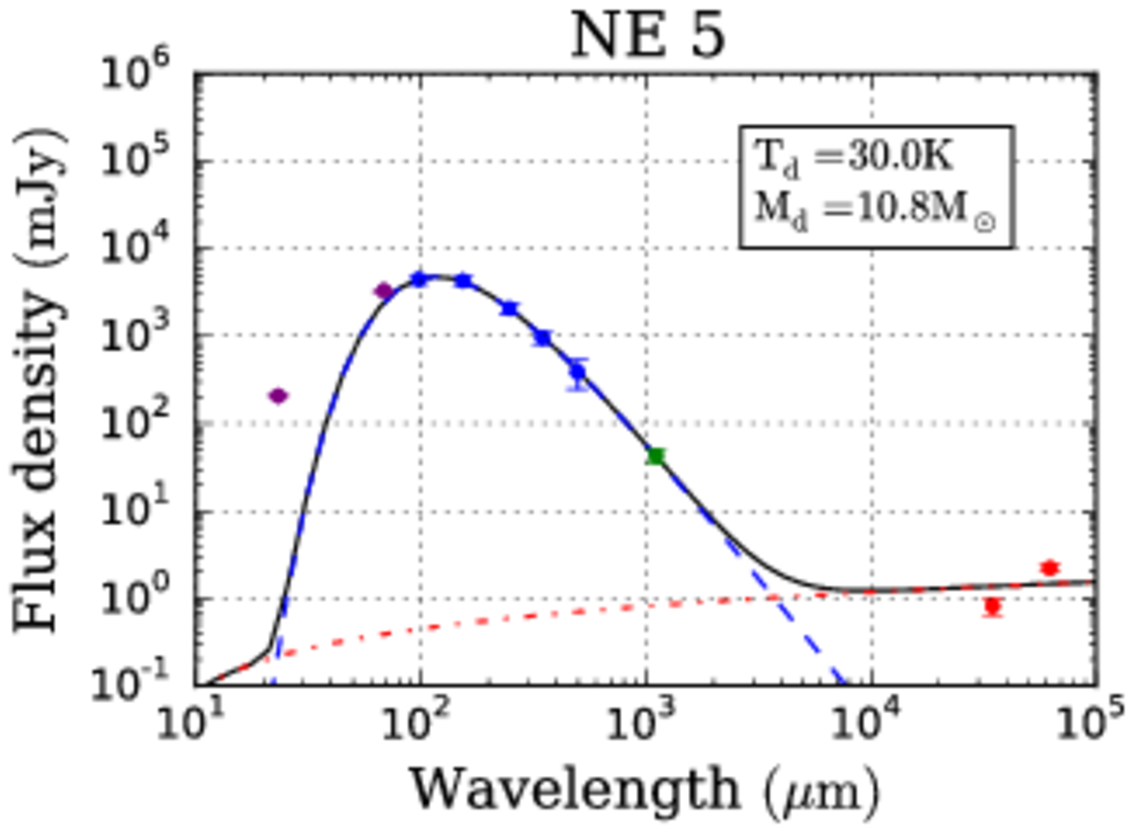}
\includegraphics[width=3.0in]{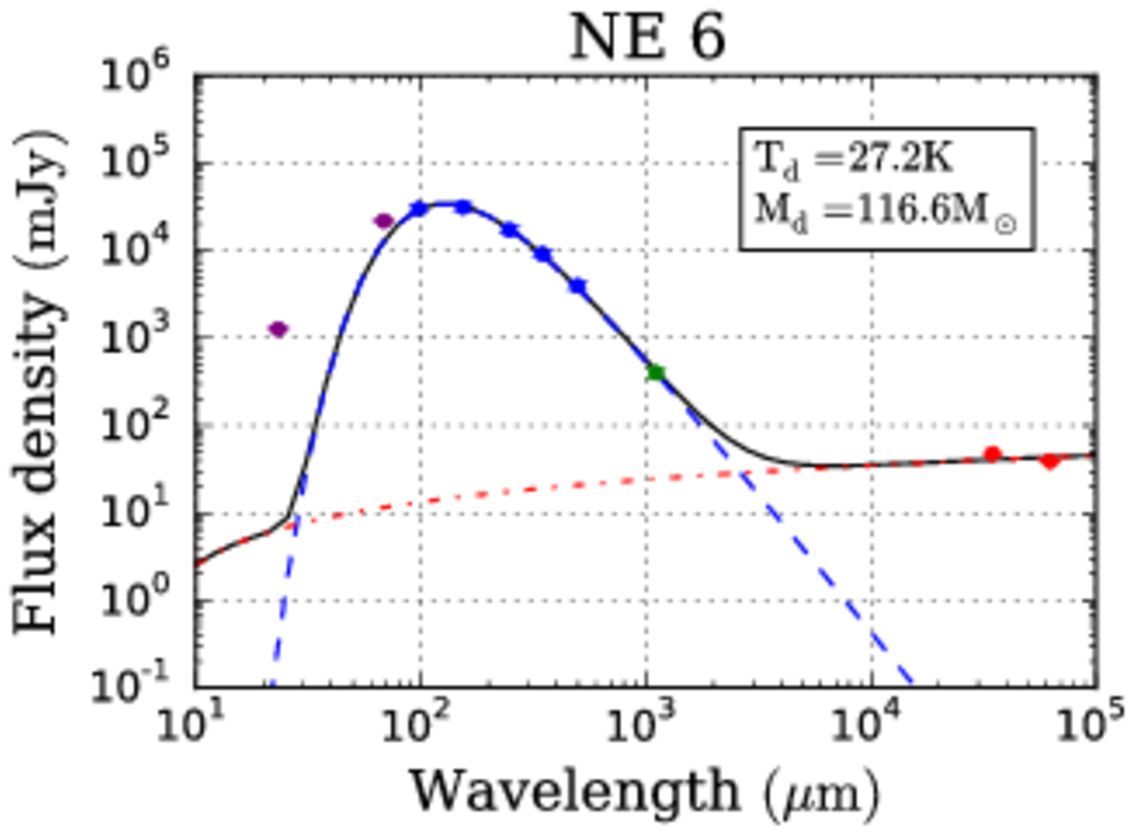}\\
\includegraphics[width=3.0in]{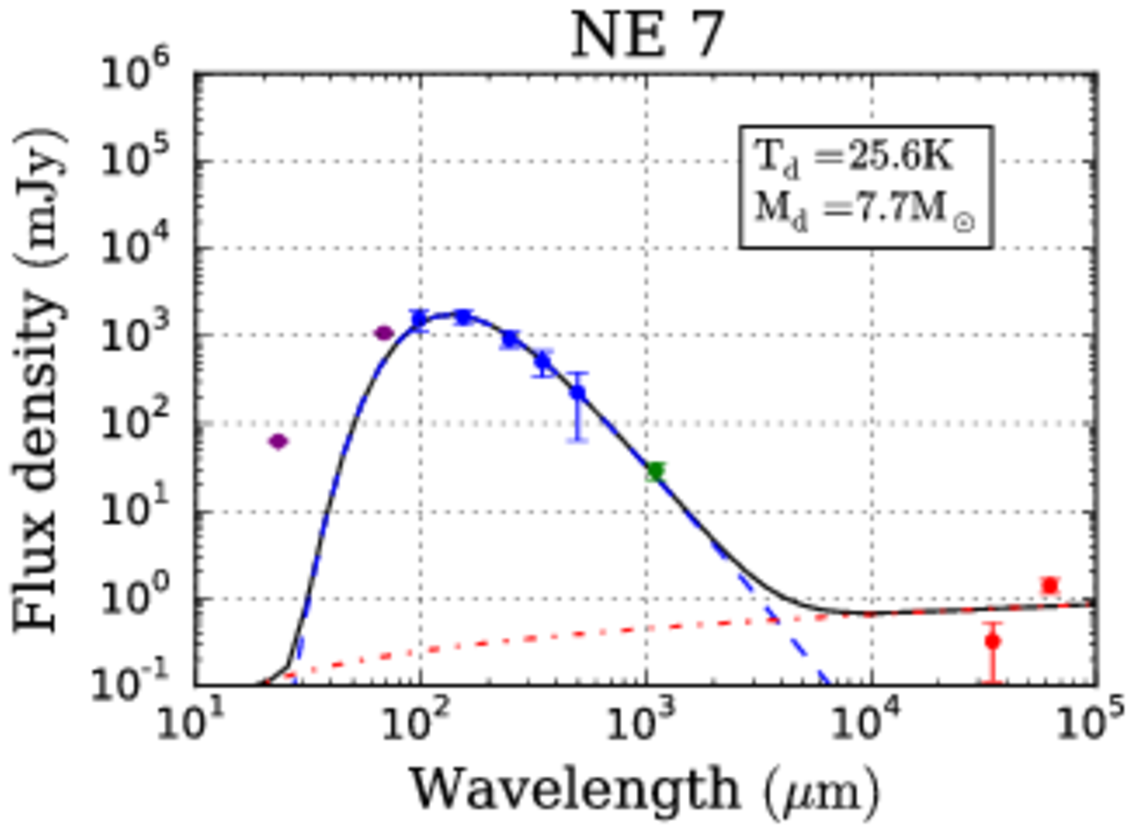}
\includegraphics[width=3.0in]{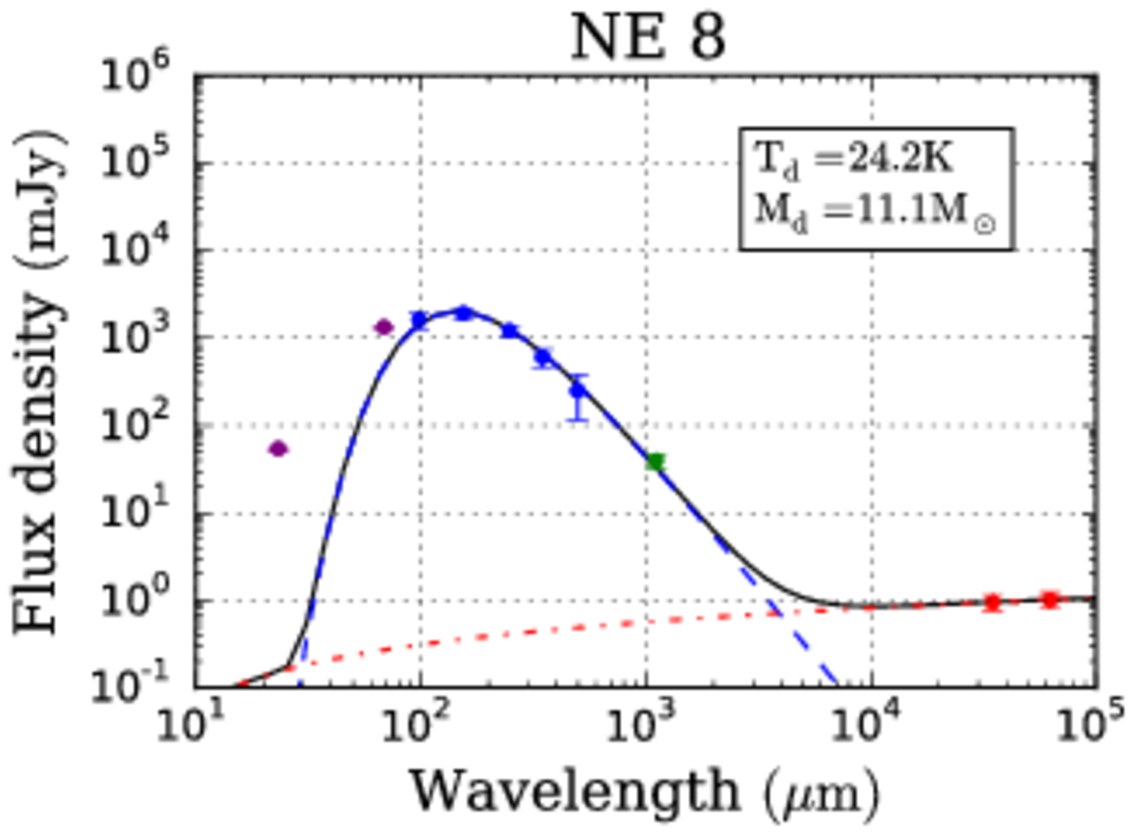}\\
\caption{{\it Continued.}}
\end{figure}

\begin{figure}
\figurenum{13}
\centering
\includegraphics[width=3.0in]{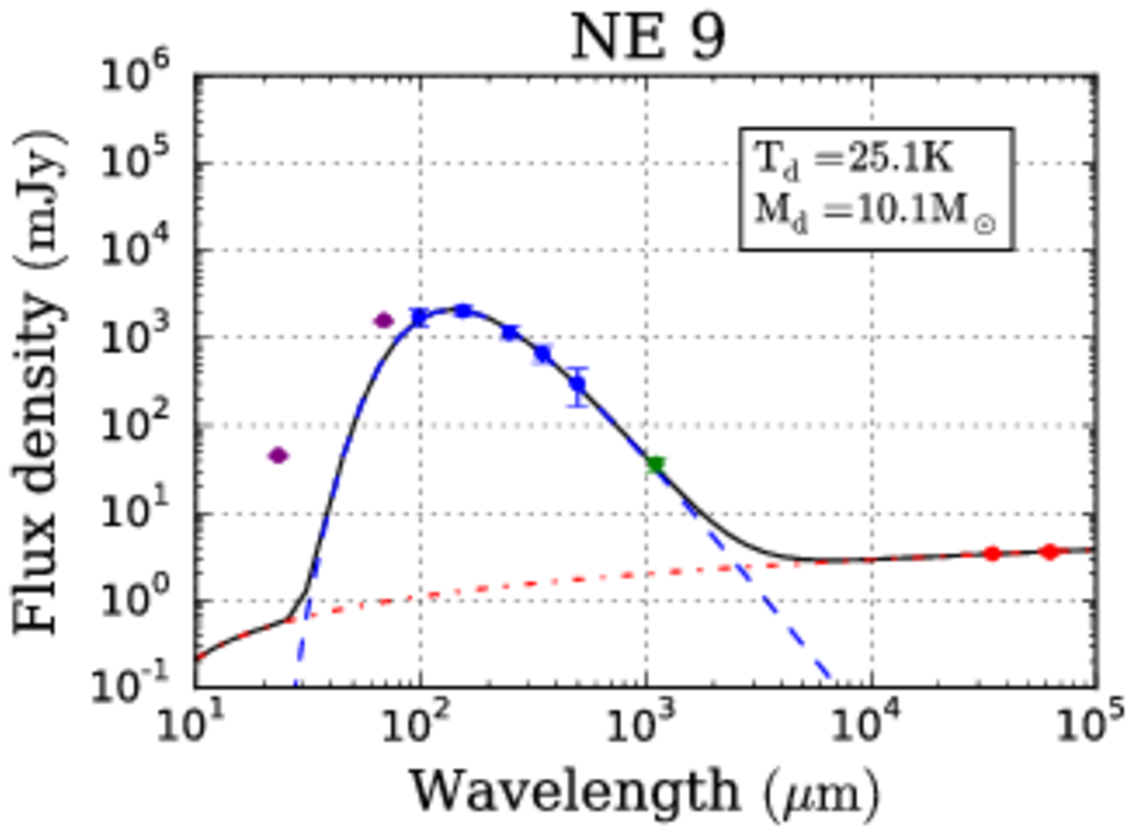}
\includegraphics[width=3.0in]{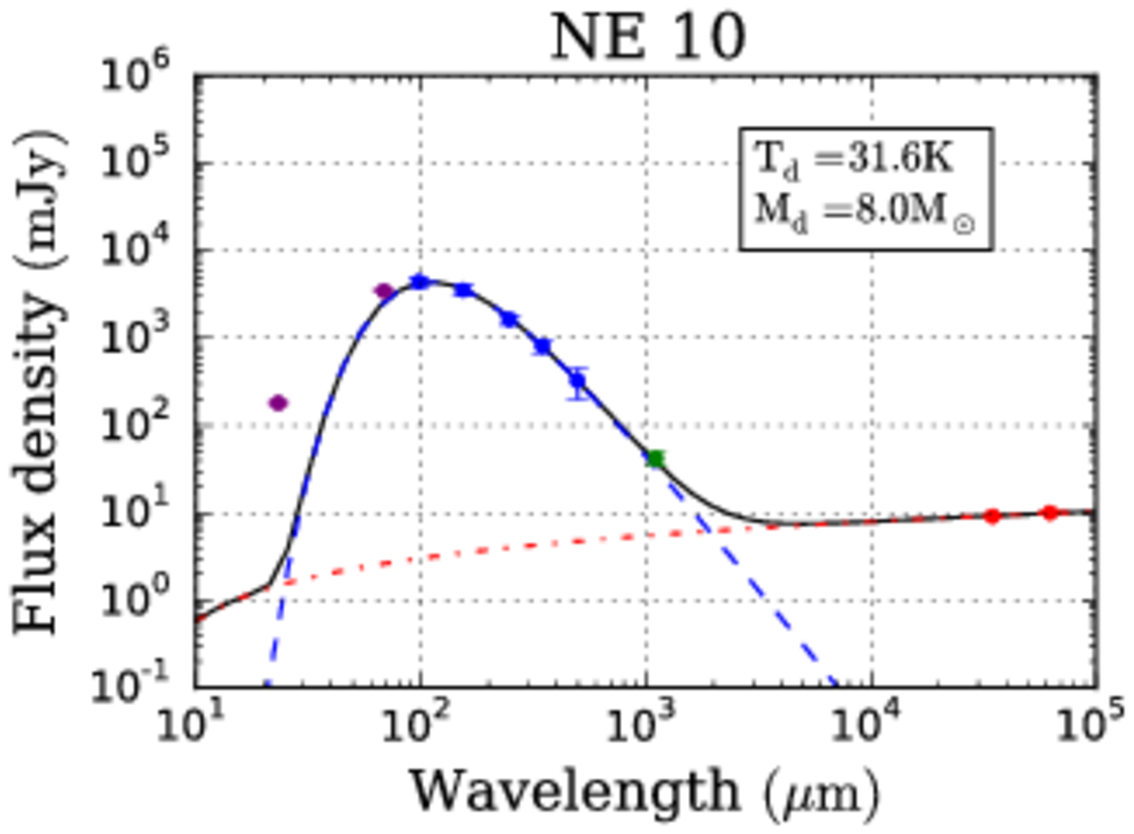}\\
\includegraphics[width=3.0in]{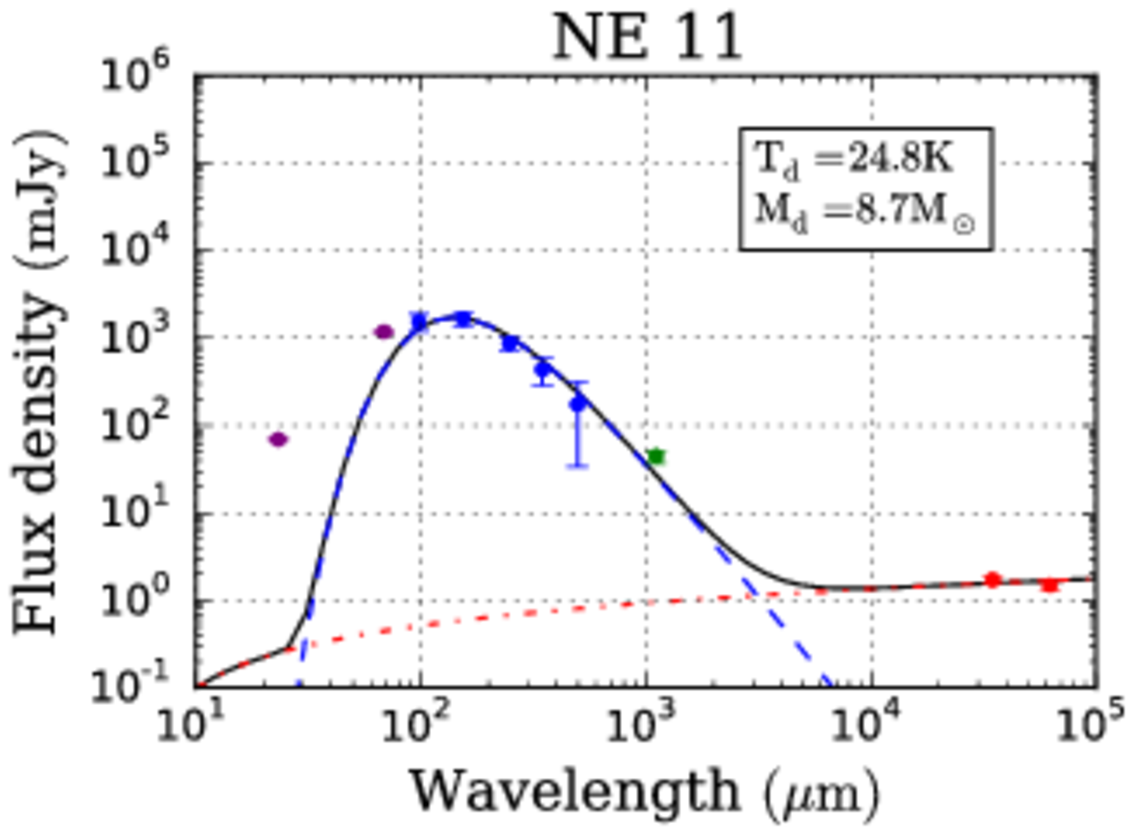}
\includegraphics[width=3.0in]{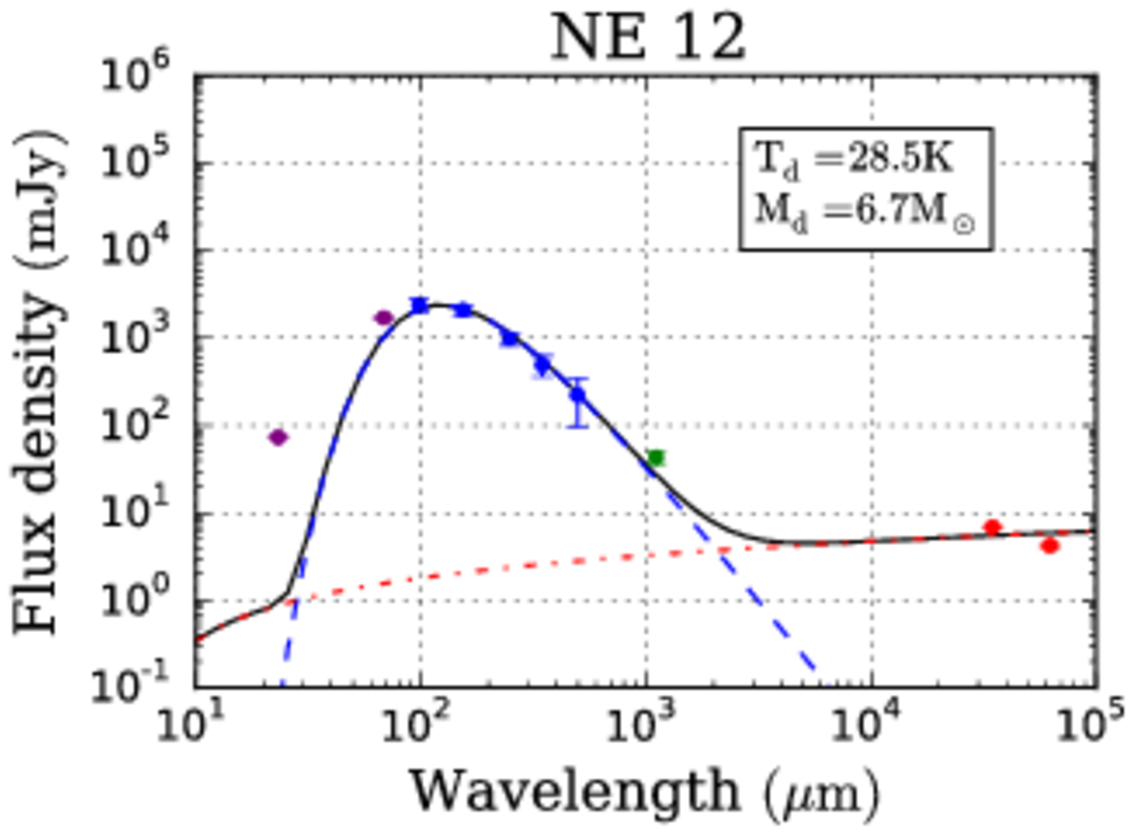}\\
\includegraphics[width=3.0in]{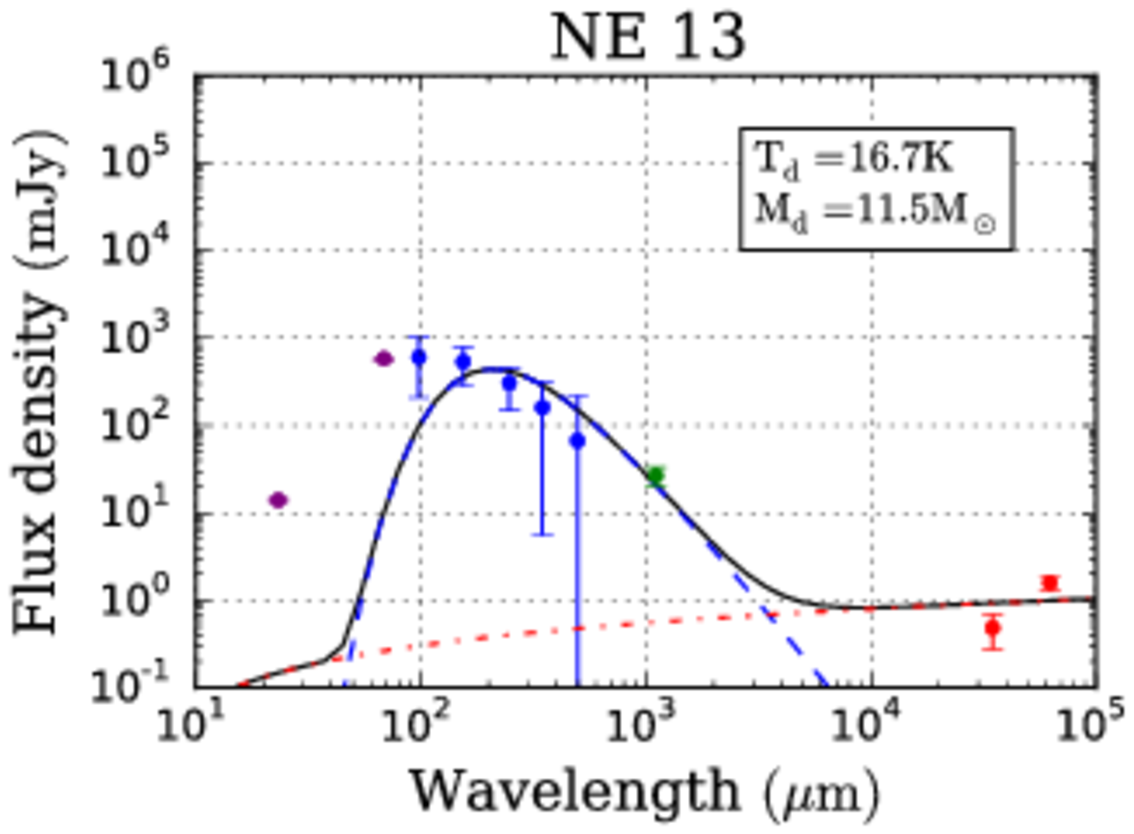}
\includegraphics[width=3.0in]{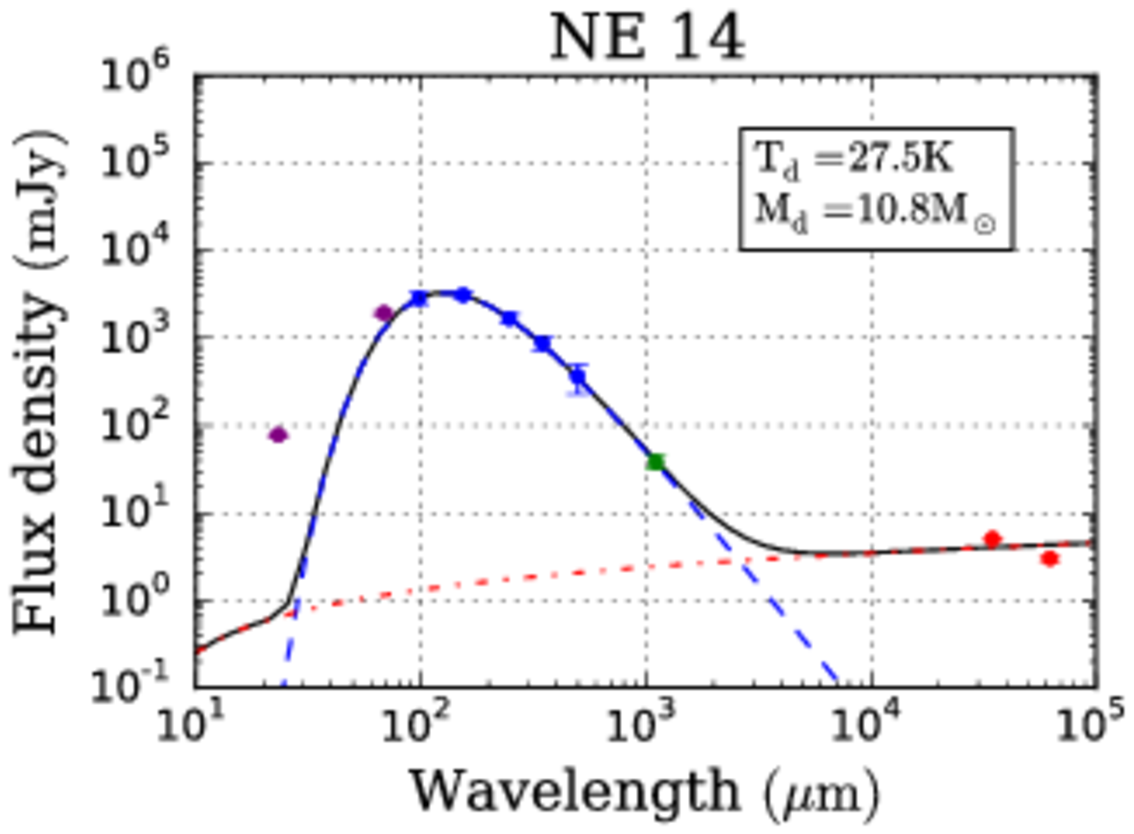}\\
\caption{{\it Continued.}}
\end{figure}

\begin{figure}
\figurenum{13}
\centering
\includegraphics[width=3.0in]{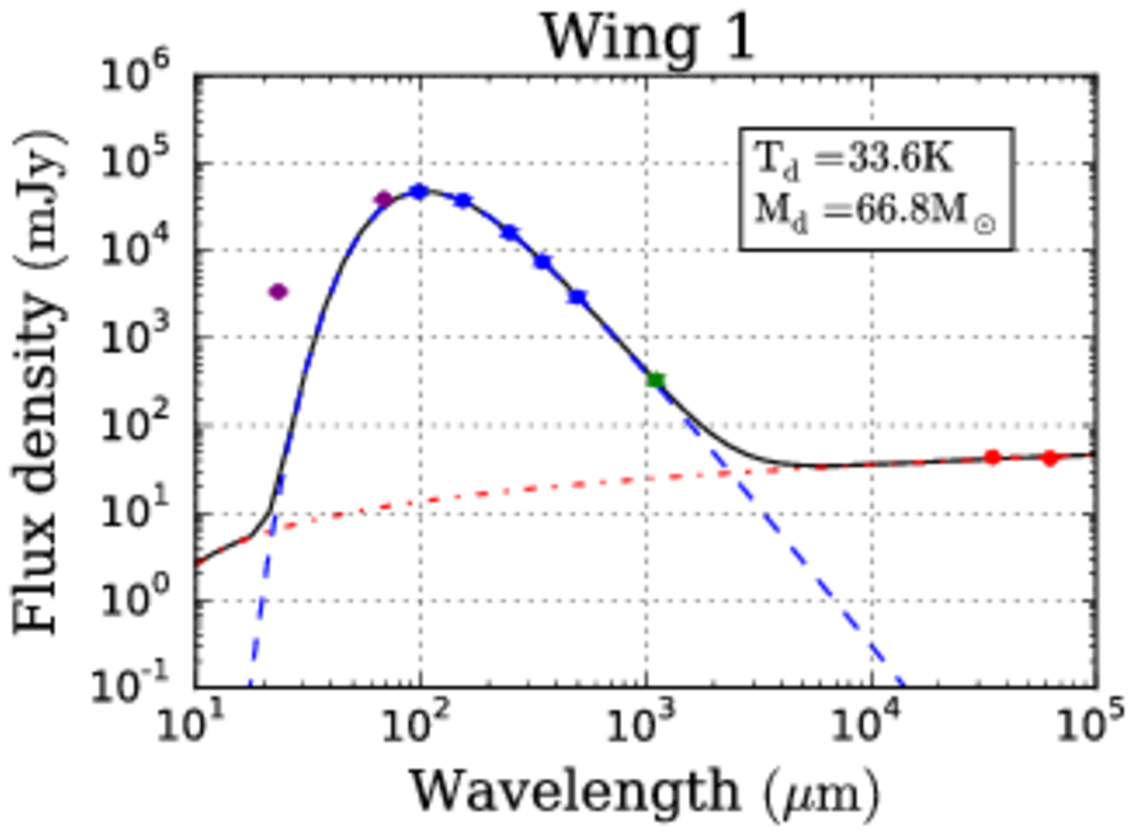}
\includegraphics[width=3.0in]{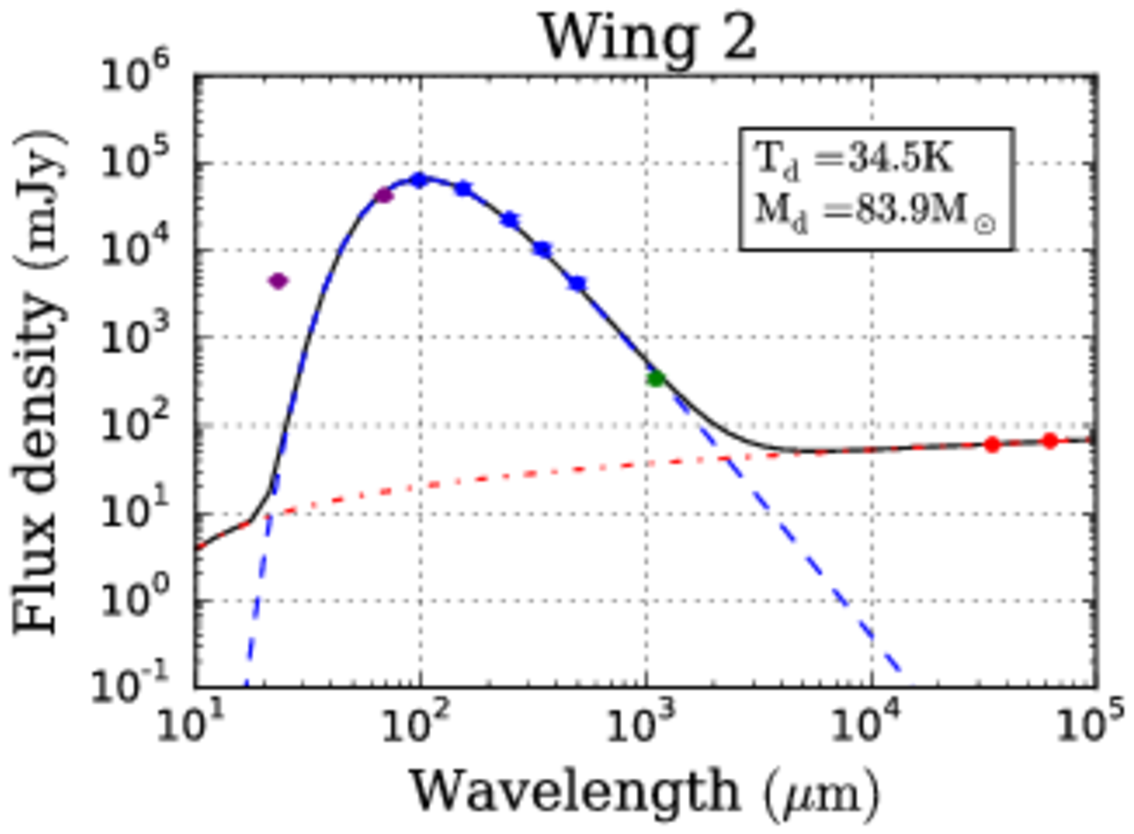}\\
\includegraphics[width=3.0in]{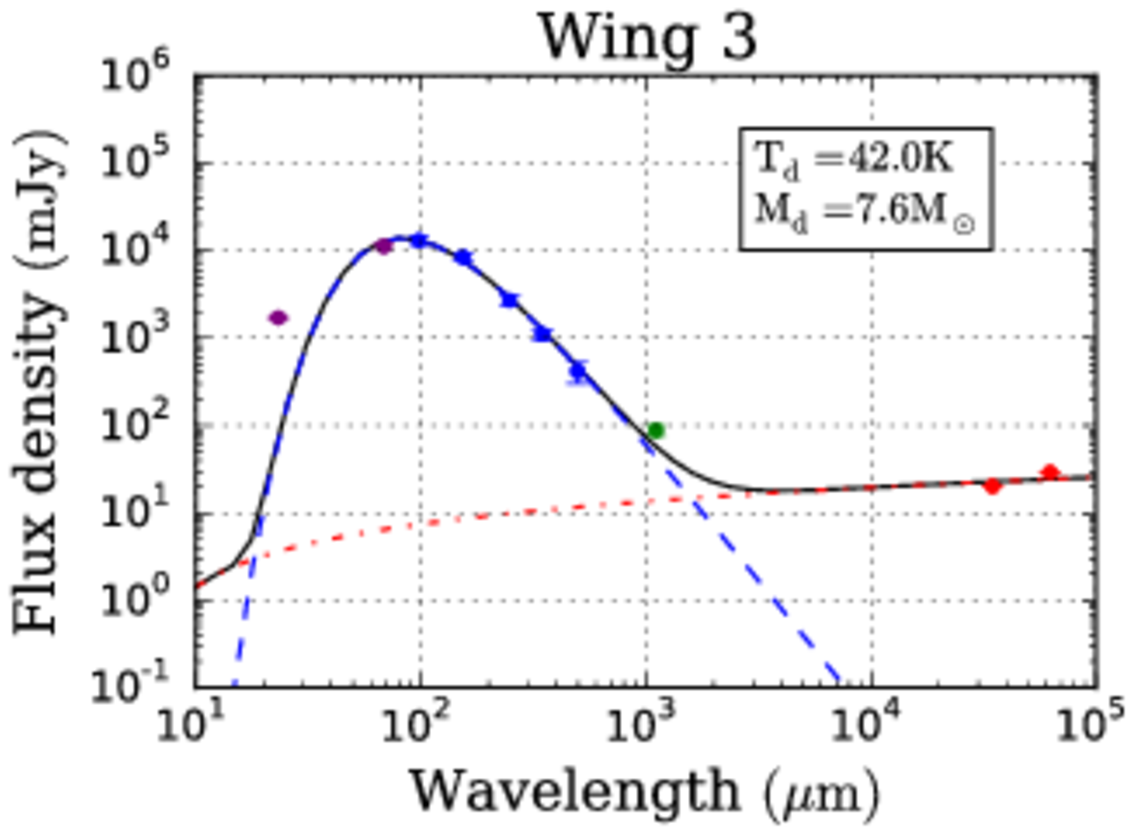}
\includegraphics[width=3.0in]{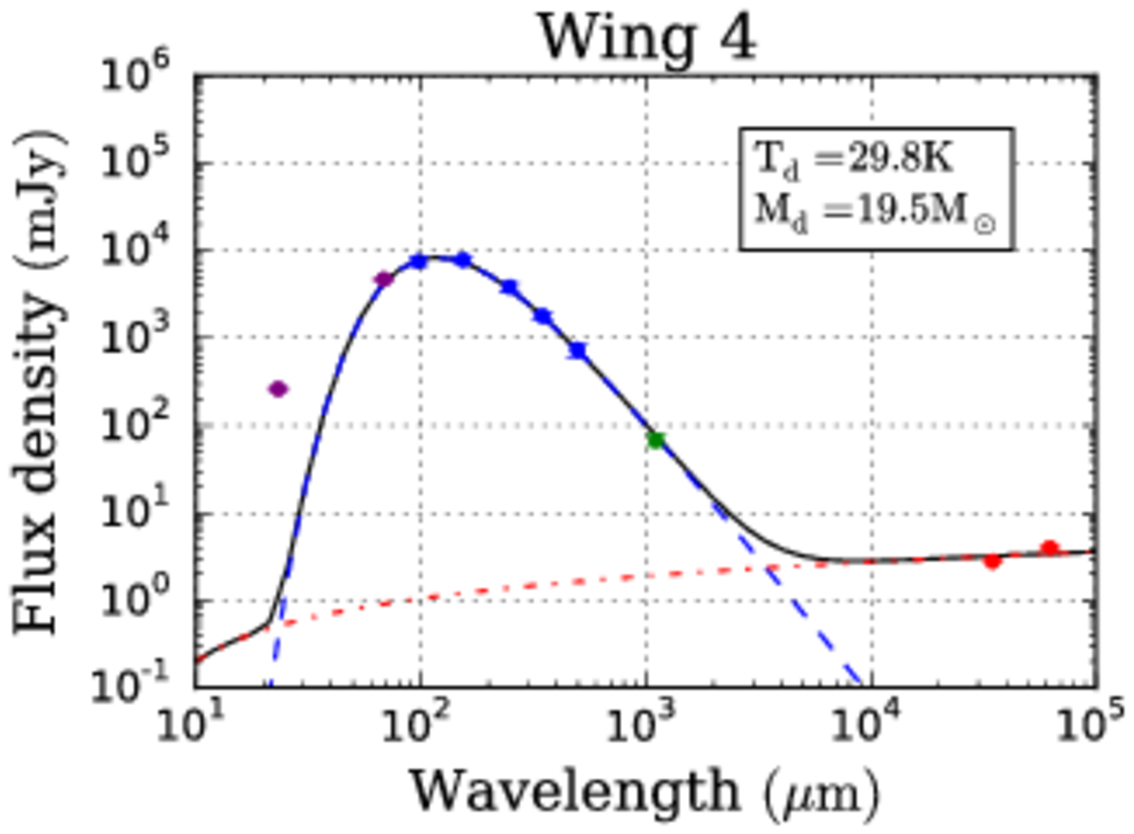}\\
\includegraphics[width=3.0in]{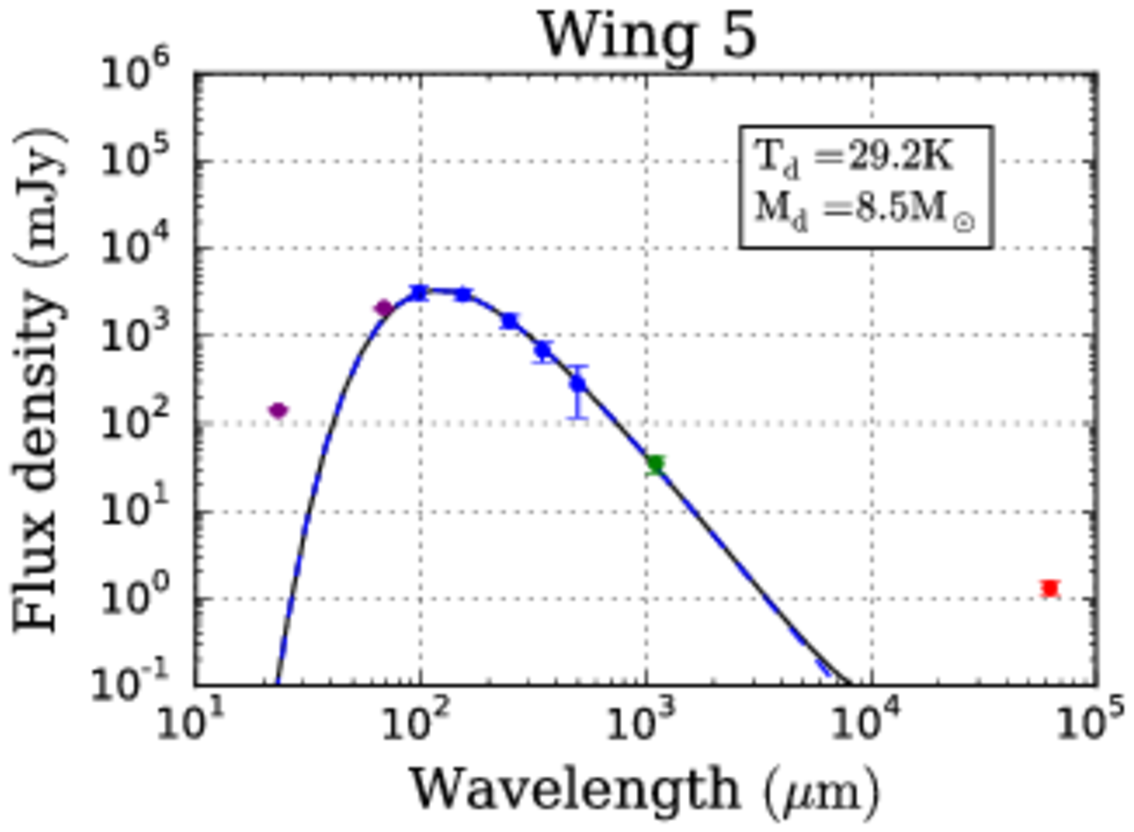}
\includegraphics[width=3.0in]{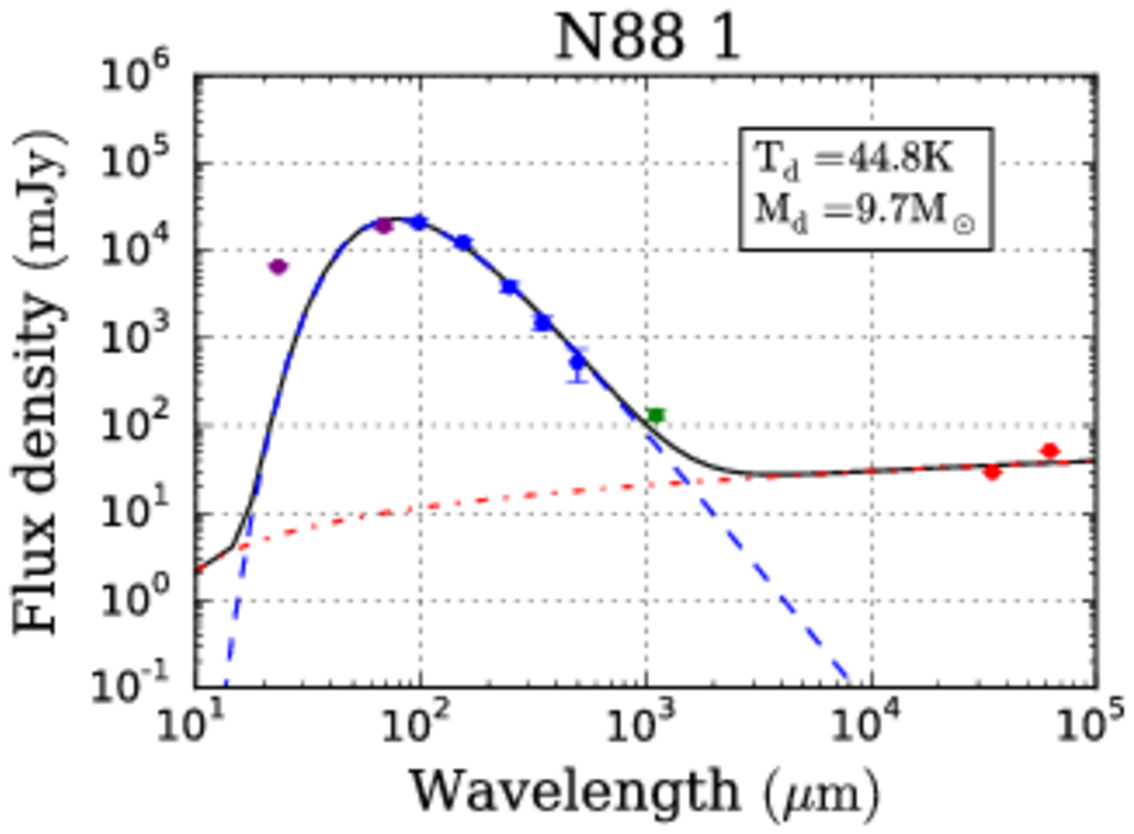}\\
\caption{{\it Continued.}}
\end{figure}

\newpage
\begin{deluxetable*}{llccccccccc}
\tabletypesize{\scriptsize}
\tablecaption{AzTEC/ASTE 1.1 mm extended source catalog of the SMC.\label{table:catalog_fruit}}
\tablehead{ID&  &$\alpha$&$\delta$&Peak flux&S/N&1.1 mm total flux&$R$&$R_{\mathrm{deconv.}}$\\
&&$(J2000)$&$(J2000)$&(mJy/beam)&&(mJy)&(pc)&(pc)}
\startdata
SW&1&00$^\circ$48$'$24.8$''$&-73$^d$05$^m$58$^s$&149.4 $\pm$5.2 &29.0 &1272.8 $\pm$162.9 &34.6 &34.1 \\
&2&00$^\circ$48$'$08.8$''$&-73$^d$14$^m$49$^s$&133.9 $\pm$4.9 &27.2 &1194.5 $\pm$152.9 &33.3 &32.8 \\
&3&00$^\circ$45$'$20.4$''$&-73$^d$22$^m$51$^s$&112.6 $\pm$5.0 &22.6 &300.9 $\pm$38.6 &17.1 &16.0 \\
&4&00$^\circ$46$'$40.0$''$&-73$^d$06$^m$08$^s$&97.0 $\pm$5.0 &19.3 &260.7 $\pm$33.4 &15.9 &14.8 \\
&5&00$^\circ$48$'$55.8$''$&-73$^d$09$^m$51$^s$&66.9 $\pm$5.1 &13.2 &207.8 $\pm$26.7 &16.1 &15.0 \\
&6&00$^\circ$45$'$04.1$''$&-73$^d$16$^m$41$^s$&61.9 $\pm$4.9 &12.5 &89.3 $\pm$11.8 &10.4 &8.6 \\
&7&00$^\circ$45$'$27.7$''$&-73$^d$18$^m$41$^s$&55.6 $\pm$5.2 &10.7 &117.3 $\pm$15.2 &12.1 &10.6 \\
&8&00$^\circ$49$'$29.1$''$&-73$^d$26$^m$27$^s$&54.1 $\pm$5.0 &10.9 &49.3 $\pm$7.3 &7.8 &5.2 \\
&9&00$^\circ$46$'$33.6$''$&-73$^d$15$^m$55$^s$&53.9 $\pm$5.2 &10.5 &164.0 $\pm$21.1 &14.7 &13.5 \\
&10&00$^\circ$52$'$38.6$''$&-73$^d$26$^m$32$^s$&53.2 $\pm$4.9 &10.9 &78.8 $\pm$10.5 &10.1 &8.2 \\
&11&00$^\circ$45$'$22.4$''$&-73$^d$12$^m$33$^s$&50.9 $\pm$4.9 &10.4 &37.4 $\pm$6.3 &6.9 &3.7 \\
&12&00$^\circ$52$'$46.9$''$&-73$^d$17$^m$51$^s$&48.8 $\pm$5.1 &9.7 &177.0 $\pm$22.7 &15.7 &14.5 \\
&13&00$^\circ$47$'$05.8$''$&-73$^d$22$^m$14$^s$&47.5 $\pm$5.0 &9.5 &56.0 $\pm$7.9 &8.8 &6.6 \\
&14&00$^\circ$46$'$25.3$''$&-73$^d$22$^m$11$^s$&47.2 $\pm$5.1 &9.2 &26.5 $\pm$6.1 &5.9 &1.0 \\
&15&00$^\circ$56$'$06.5$''$&-72$^d$47$^m$16$^s$&43.6 $\pm$5.4 &8.0 &27.7 $\pm$6.3 &6.1 &1.7 \\
&16&00$^\circ$48$'$15.4$''$&-73$^d$10$^m$32$^s$&42.9 $\pm$5.1 &8.5 &25.2 $\pm$5.9 &5.9 &1.0 \\
&17&00$^\circ$52$'$26.7$''$&-72$^d$41$^m$01$^s$&40.8 $\pm$5.4 &7.6 &32.4 $\pm$6.2 &6.8 &3.4 \\
&18&00$^\circ$46$'$48.2$''$&-73$^d$32$^m$04$^s$&39.8 $\pm$5.0 &8.0 &23.7 $\pm$5.8 &5.8 &0.3 \\
&19&00$^\circ$45$'$29.7$''$&-73$^d$08$^m$11$^s$&39.6 $\pm$5.1 &7.8 &52.6 $\pm$7.5 &8.8 &6.6 \\
&20&00$^\circ$48$'$04.0$''$&-73$^d$23$^m$06$^s$&39.1 $\pm$5.0 &7.8 &44.3 $\pm$6.8 &7.9 &5.4 \\
&21&00$^\circ$50$'$45.8$''$&-72$^d$47$^m$55$^s$&36.8 $\pm$5.1 &7.3 &37.6 $\pm$6.2 &7.6 &4.8 \\
&22&00$^\circ$45$'$41.7$''$&-73$^d$16$^m$38$^s$&36.7 $\pm$5.3 &6.9 &64.7 $\pm$8.9 &9.8 &7.9 \\
&23&00$^\circ$48$'$20.1$''$&-73$^d$19$^m$21$^s$&35.7 $\pm$5.0 &7.1 &25.8 $\pm$5.7 &6.2 &2.2 \\
&24&00$^\circ$48$'$51.6$''$&-73$^d$07$^m$26$^s$&35.2 $\pm$4.9 &7.2 &24.5 $\pm$5.7 &6.1 &1.7 \\\tableline
NE&1&00$^\circ$59$'$09.3$''$&-72$^d$10$^m$35$^s$&112.2 $\pm$6.1 &18.3 &1682.6 $\pm$215.4 &39.8 &39.4 \\
&2&01$^\circ$05$'$05.8$''$&-71$^d$59$^m$37$^s$&87.5 $\pm$6.2 &14.2 &195.9 $\pm$25.2 &14.4 &13.2 \\
&3&01$^\circ$02$'$52.2$''$&-71$^d$53$^m$46$^s$&58.9 $\pm$6.2 &9.5 &88.6 $\pm$11.9 &10.2 &8.4 \\
&4&01$^\circ$03$'$48.2$''$&-72$^d$03$^m$59$^s$&55.8 $\pm$6.1 &9.1 &112.9 $\pm$14.8 &11.9 &10.4 \\
&5&00$^\circ$57$'$57.3$''$&-72$^d$39$^m$24$^s$&55.0 $\pm$6.0 &9.2 &42.8 $\pm$7.3 &7.2 &4.2 \\
&6&01$^\circ$03$'$07.8$''$&-72$^d$03$^m$39$^s$&52.2 $\pm$5.9 &8.8 &400.2 $\pm$51.2 &23.0 &22.3 \\
&7&01$^\circ$06$'$03.1$''$&-72$^d$03$^m$35$^s$&49.7 $\pm$6.0 &8.3 &29.1 $\pm$6.9 &6.1 &1.7 \\
&8&01$^\circ$02$'$31.1$''$&-71$^d$56$^m$54$^s$&46.6 $\pm$5.8 &8.1 &39.8 $\pm$6.9 &7.2 &4.2 \\
&9&00$^\circ$58$'$32.8$''$&-72$^d$14$^m$54$^s$&42.6 $\pm$6.0 &7.1 &36.6 $\pm$6.7 &7.3 &4.4 \\
&10&01$^\circ$03$'$48.9$''$&-72$^d$02$^m$17$^s$&42.4 $\pm$5.9 &7.2 &42.1 $\pm$7.0 &7.8 &5.1 \\
&11&01$^\circ$08$'$41.7$''$&-71$^d$59$^m$12$^s$&41.7 $\pm$6.9 &6.0 &44.5 $\pm$7.6 &8.0 &5.5 \\
&12&01$^\circ$05$'$35.4$''$&-71$^d$59$^m$15$^s$&41.7 $\pm$5.9 &7.0 &43.4 $\pm$7.1 &7.8 &5.1 \\
&13&00$^\circ$58$'$26.3$''$&-72$^d$17$^m$60$^s$&39.6 $\pm$6.0 &6.6 &26.6 $\pm$6.5 &6.2 &2.2 \\
&14&01$^\circ$03$'$28.0$''$&-72$^d$01$^m$31$^s$&38.3 $\pm$6.0 &6.3 &38.2 $\pm$6.8 &7.4 &4.5 \\\tableline
Wing&1&01$^\circ$14$'$47.3$''$&-73$^d$19$^m$52$^s$&75.5 $\pm$6.4 &11.8 &330.7 $\pm$42.4 &19.1 &18.2 \\
&2&01$^\circ$14$'$05.6$''$&-73$^d$17$^m$04$^s$&73.0 $\pm$6.6 &11.1 &342.2 $\pm$43.8 &19.9 &19.0 \\
&3&01$^\circ$09$'$13.8$''$&-73$^d$11$^m$33$^s$&71.3 $\pm$6.7 &10.6 &87.9 $\pm$11.9 &10.1 &8.3 \\
&4&01$^\circ$14$'$21.0$''$&-73$^d$15$^m$41$^s$&58.1 $\pm$6.5 &8.9 &67.6 $\pm$9.6 &9.2 &7.1 \\
&5&01$^\circ$14$'$25.1$''$&-73$^d$14$^m$10$^s$&55.8 $\pm$6.5 &8.6 &34.8 $\pm$7.4 &6.5 &2.8 \\\tableline
N88&1&01$^\circ$24$'$08.6$''$&-73$^d$08$^m$60$^s$&148.1 $\pm$11.2 &13.2 &130.1 $\pm$18.3 &8.6 &6.4 
\enddata
\tablecomments{The columns give (1) source ID, (2) right ascension, (3) declination, (4) observed peak flux and noise level at 1.1 mm, (5) signal-to-noise ratio, (6) 1.1 mm total flux, and source radius (7) before and (8) after deconvolution.}
\end{deluxetable*}

\newpage
\begin{turnpage}
\begin{deluxetable*}{llccccccccccc}
\tabletypesize{\scriptsize}
\tablewidth{25cm}
\tablecaption{The total flux densities of the 1.1 mm extended objects. \label{table:catalog_flux}}
\tablehead{ID&&$S^\mathrm{Raw}_\mathrm{4.8 GHz}$&$S^\mathrm{Raw}_\mathrm{8.64 GHz}$&$S^\mathrm{FRUIT}_\mathrm{1.1 mm}$&$S^\mathrm{FRUIT}_{500 \micron}$&$S^\mathrm{FRUIT}_{350 \micron}$&$S^\mathrm{FRUIT}_{250 \micron}$&$S^\mathrm{FRUIT}_{160 \micron}$&$S^\mathrm{FRUIT}_{100 \micron}$&$S^\mathrm{Raw}_{70 \micron}$&$S^\mathrm{Raw}_{24 \micron}$&$S^\mathrm{Raw}_{8 \micron}$\\
&&(mJy)&(mJy)&(mJy)&(Jy)&(Jy)&(Jy)&(Jy)&(Jy)&(Jy)&(mJy)&(mJy)}
\startdata
SW&1&80.6 $\pm$0.8 &77.2 $\pm$0.8 &1272.8 $\pm$162.9 &12.88 $\pm$1.29 &30.59 $\pm$3.06 &61.27 $\pm$6.13 &127.09 $\pm$12.71 &137.64 $\pm$13.76 &117.96 $\pm$5.90 &4486.3 $\pm$179.5 &1819.7 $\pm$91.0 \\
&2&119.1 $\pm$1.2 &117.6 $\pm$1.2 &1194.5 $\pm$152.9 &14.29 $\pm$1.43 &33.85 $\pm$3.38 &68.29 $\pm$6.83 &137.38 $\pm$13.74 &155.52 $\pm$15.55 &127.36 $\pm$6.37 &11938.1 $\pm$477.5 &2690.2 $\pm$134.5 \\
&3&35.8 $\pm$0.4 &19.9 $\pm$0.2 &300.9 $\pm$38.6 &3.31 $\pm$0.33 &8.00 $\pm$0.80 &16.58 $\pm$1.66 &34.19 $\pm$3.42 &37.75 $\pm$3.78 &23.69 $\pm$1.18 &1864.1 $\pm$74.6 &744.0 $\pm$37.2 \\
&4&38.4 $\pm$0.4 &31.3 $\pm$0.3 &260.7 $\pm$33.4 &2.63 $\pm$0.27 &6.51 $\pm$0.65 &13.74 $\pm$1.38 &27.76 $\pm$2.78 &29.71 $\pm$2.97 &25.46 $\pm$1.27 &1378.3 $\pm$55.1 &867.9 $\pm$43.4 \\
&5&20.7 $\pm$0.2 &20.5 $\pm$0.2 &207.8 $\pm$26.7 &2.22 $\pm$0.23 &5.18 $\pm$0.52 &10.21 $\pm$1.02 &20.52 $\pm$2.05 &23.09 $\pm$2.31 &20.46 $\pm$1.02 &1031.0 $\pm$41.2 &391.5 $\pm$19.6 \\
&6&2.9 $\pm$0.1 &4.3 $\pm$0.1 &89.3 $\pm$11.8 &0.74 $\pm$0.11 &1.80 $\pm$0.20 &4.04 $\pm$0.41 &9.44 $\pm$0.95 &12.19 $\pm$1.24 &9.87 $\pm$0.49 &874.3 $\pm$35.0 &183.3 $\pm$9.2 \\
&7&2.4 $\pm$0.1 &-1.4 $\pm$0.1 &117.3 $\pm$15.2 &0.91 $\pm$0.11 &2.06 $\pm$0.22 &3.82 $\pm$0.39 &6.62 $\pm$0.67 &5.32 $\pm$0.57 &5.37 $\pm$0.27 &220.5 $\pm$8.8 &145.1 $\pm$7.3 \\
&8&4.4 $\pm$0.2 &4.1 $\pm$0.2 &49.3 $\pm$7.3 &0.46 $\pm$0.11 &1.13 $\pm$0.15 &2.51 $\pm$0.27 &5.73 $\pm$0.60 &7.07 $\pm$0.77 &5.90 $\pm$0.29 &484.7 $\pm$19.4 &143.6 $\pm$7.2 \\
&9&5.4 $\pm$0.1 &6.2 $\pm$0.1 &164.0 $\pm$21.1 &1.28 $\pm$0.14 &3.05 $\pm$0.31 &6.10 $\pm$0.61 &10.77 $\pm$1.08 &10.47 $\pm$1.06 &9.00 $\pm$0.45 &363.4 $\pm$14.5 &288.9 $\pm$14.4 \\
&10&1.5 $\pm$0.2 &2.1 $\pm$0.1 &78.8 $\pm$10.5 &0.63 $\pm$0.10 &1.42 $\pm$0.16 &2.60 $\pm$0.27 &4.52 $\pm$0.47 &4.49 $\pm$0.51 &3.91 $\pm$0.20 &496.6 $\pm$19.9 &79.1 $\pm$4.0 \\
&11&0.9 $\pm$0.2 &0.1 $\pm$0.2 &37.4 $\pm$6.3 &0.38 $\pm$0.12 &0.93 $\pm$0.15 &1.91 $\pm$0.22 &3.58 $\pm$0.42 &3.40 $\pm$0.48 &2.43 $\pm$0.12 &118.7 $\pm$4.7 &86.1 $\pm$4.3 \\
&12&3.1 $\pm$0.1 &-0.7 $\pm$0.1 &177.0 $\pm$22.7 &0.90 $\pm$0.10 &1.91 $\pm$0.20 &3.38 $\pm$0.34 &5.51 $\pm$0.56 &4.48 $\pm$0.47 &4.81 $\pm$0.24 &124.4 $\pm$5.0 &122.0 $\pm$6.1 \\
&13&3.0 $\pm$0.2 &3.3 $\pm$0.1 &56.0 $\pm$7.9 &0.43 $\pm$0.10 &0.97 $\pm$0.13 &1.69 $\pm$0.19 &2.85 $\pm$0.33 &2.57 $\pm$0.37 &2.30 $\pm$0.12 &84.9 $\pm$3.4 &76.4 $\pm$3.8 \\
&14&1.3 $\pm$0.3 &1.4 $\pm$0.2 &26.5 $\pm$6.1 &0.34 $\pm$0.14 &0.80 $\pm$0.16 &1.55 $\pm$0.21 &2.83 $\pm$0.38 &2.77 $\pm$0.49 &2.30 $\pm$0.11 &145.5 $\pm$5.8 &54.8 $\pm$2.7 \\
&15&0.5 $\pm$0.3 &-0.1 $\pm$0.2 &27.7 $\pm$6.3 &0.19 $\pm$0.15 &0.44 $\pm$0.15 &0.84 $\pm$0.17 &1.60 $\pm$0.30 &1.57 $\pm$0.42 &1.28 $\pm$0.06 &145.1 $\pm$5.8 &38.5 $\pm$1.9 \\
&16&2.6 $\pm$0.3 &2.0 $\pm$0.2 &25.2 $\pm$5.9 &0.26 $\pm$0.14 &0.57 $\pm$0.15 &1.05 $\pm$0.17 &1.83 $\pm$0.31 &1.45 $\pm$0.43 &1.74 $\pm$0.09 &63.4 $\pm$2.5 &39.0 $\pm$1.9 \\
&17&1.5 $\pm$0.2 &1.0 $\pm$0.2 &32.4 $\pm$6.2 &0.15 $\pm$0.13 &0.34 $\pm$0.13 &0.66 $\pm$0.15 &1.31 $\pm$0.26 &1.51 $\pm$0.38 &1.61 $\pm$0.08 &50.1 $\pm$2.0 &19.0 $\pm$0.9 \\
&18&1.3 $\pm$0.3 &1.0 $\pm$0.2 &23.7 $\pm$5.8 &0.14 $\pm$0.14 &0.32 $\pm$0.14 &0.65 $\pm$0.15 &1.24 $\pm$0.28 &1.07 $\pm$0.42 &0.95 $\pm$0.05 &50.0 $\pm$2.0 &22.4 $\pm$1.1 \\
&19&2.6 $\pm$0.2 &1.6 $\pm$0.1 &52.6 $\pm$7.5 &0.39 $\pm$0.10 &0.87 $\pm$0.13 &1.59 $\pm$0.18 &2.76 $\pm$0.32 &2.35 $\pm$0.36 &2.33 $\pm$0.12 &62.8 $\pm$2.5 &58.7 $\pm$2.9 \\
&20&1.5 $\pm$0.2 &0.7 $\pm$0.2 &44.3 $\pm$6.8 &0.37 $\pm$0.11 &0.81 $\pm$0.13 &1.46 $\pm$0.18 &2.12 $\pm$0.28 &1.62 $\pm$0.34 &1.68 $\pm$0.08 &53.7 $\pm$2.1 &58.9 $\pm$2.9 \\
&21&4.8 $\pm$0.2 &5.2 $\pm$0.2 &37.6 $\pm$6.2 &0.11 $\pm$0.11 &0.26 $\pm$0.11 &0.58 $\pm$0.12 &1.32 $\pm$0.24 &1.70 $\pm$0.36 &2.17 $\pm$0.11 &201.4 $\pm$8.1 &27.0 $\pm$1.4 \\
&22&1.7 $\pm$0.2 &1.7 $\pm$0.1 &64.7 $\pm$8.9 &0.45 $\pm$0.10 &1.00 $\pm$0.13 &1.95 $\pm$0.21 &3.63 $\pm$0.39 &3.80 $\pm$0.45 &3.61 $\pm$0.18 &114.2 $\pm$4.6 &79.9 $\pm$4.0 \\
&23&4.3 $\pm$0.2 &2.9 $\pm$0.2 &25.8 $\pm$5.7 &0.25 $\pm$0.13 &0.57 $\pm$0.14 &1.08 $\pm$0.17 &2.02 $\pm$0.31 &1.80 $\pm$0.42 &1.52 $\pm$0.08 &63.9 $\pm$2.6 &39.8 $\pm$2.0 \\
&24&4.2 $\pm$0.3 &3.0 $\pm$0.2 &24.5 $\pm$5.7 &0.23 $\pm$0.13 &0.52 $\pm$0.14 &1.01 $\pm$0.17 &2.16 $\pm$0.32 &2.42 $\pm$0.46 &2.49 $\pm$0.12 &92.8 $\pm$3.7 &39.1 $\pm$2.0 \\\tableline
NE&1&607.7 $\pm$6.1 &589.9 $\pm$5.9 &1682.6 $\pm$215.4 &15.24 $\pm$1.52 &37.25 $\pm$3.73 &78.87 $\pm$7.89 &179.40 $\pm$17.94 &256.52 $\pm$25.65 &191.39 $\pm$9.57 &24577.0 $\pm$983.1 &2733.6 $\pm$136.7 \\
&2&37.4 $\pm$0.4 &37.7 $\pm$0.4 &195.9 $\pm$25.2 &1.75 $\pm$0.19 &4.57 $\pm$0.46 &10.74 $\pm$1.08 &29.07 $\pm$2.91 &41.44 $\pm$4.15 &32.89 $\pm$1.64 &8303.7 $\pm$332.1 &1088.0 $\pm$54.4 \\
&3&7.4 $\pm$0.2 &5.4 $\pm$0.1 &88.6 $\pm$11.9 &0.84 $\pm$0.13 &2.14 $\pm$0.24 &4.61 $\pm$0.47 &9.61 $\pm$0.97 &11.15 $\pm$1.14 &8.10 $\pm$0.41 &547.5 $\pm$21.9 &169.6 $\pm$8.5 \\
&4&22.3 $\pm$0.3 &22.5 $\pm$0.2 &112.9 $\pm$14.8 &0.82 $\pm$0.12 &1.95 $\pm$0.21 &3.87 $\pm$0.40 &8.98 $\pm$0.91 &11.06 $\pm$1.12 &7.87 $\pm$0.39 &572.0 $\pm$22.9 &114.5 $\pm$5.7 \\
&5&2.2 $\pm$0.2 &0.8 $\pm$0.2 &42.8 $\pm$7.3 &0.39 $\pm$0.14 &0.96 $\pm$0.17 &2.05 $\pm$0.25 &4.27 $\pm$0.48 &4.43 $\pm$0.56 &3.29 $\pm$0.16 &206.2 $\pm$8.2 &65.0 $\pm$3.3 \\
&6&39.0 $\pm$0.4 &47.7 $\pm$0.5 &400.2 $\pm$51.2 &3.94 $\pm$0.40 &9.11 $\pm$0.91 &17.15 $\pm$1.72 &31.26 $\pm$3.13 &29.82 $\pm$2.98 &21.73 $\pm$1.09 &1264.0 $\pm$50.6 &469.5 $\pm$23.5 \\
&7&1.4 $\pm$0.3 &0.3 $\pm$0.2 &29.1 $\pm$6.9 &0.23 $\pm$0.16 &0.51 $\pm$0.17 &0.93 $\pm$0.19 &1.65 $\pm$0.31 &1.55 $\pm$0.43 &1.08 $\pm$0.05 &62.5 $\pm$2.5 &30.9 $\pm$1.5 \\
&8&1.0 $\pm$0.2 &1.0 $\pm$0.2 &39.8 $\pm$6.9 &0.25 $\pm$0.13 &0.60 $\pm$0.14 &1.21 $\pm$0.18 &1.91 $\pm$0.29 &1.61 $\pm$0.37 &1.33 $\pm$0.07 &54.3 $\pm$2.2 &38.5 $\pm$1.9 \\
&9&3.6 $\pm$0.2 &3.4 $\pm$0.2 &36.6 $\pm$6.7 &0.30 $\pm$0.14 &0.67 $\pm$0.15 &1.16 $\pm$0.18 &2.04 $\pm$0.30 &1.74 $\pm$0.38 &1.58 $\pm$0.08 &45.5 $\pm$1.8 &23.3 $\pm$1.2 \\
&10&10.0 $\pm$0.2 &9.2 $\pm$0.2 &42.1 $\pm$7.0 &0.32 $\pm$0.13 &0.81 $\pm$0.15 &1.63 $\pm$0.20 &3.51 $\pm$0.40 &4.38 $\pm$0.54 &3.43 $\pm$0.17 &180.5 $\pm$7.2 &50.6 $\pm$2.5 \\
&11&1.5 $\pm$0.2 &1.7 $\pm$0.2 &44.5 $\pm$7.6 &0.17 $\pm$0.14 &0.44 $\pm$0.15 &0.87 $\pm$0.17 &1.66 $\pm$0.27 &1.53 $\pm$0.35 &1.18 $\pm$0.06 &69.2 $\pm$2.8 &30.7 $\pm$1.5 \\
&12&4.2 $\pm$0.2 &6.9 $\pm$0.2 &43.4 $\pm$7.1 &0.22 $\pm$0.13 &0.49 $\pm$0.13 &0.98 $\pm$0.16 &2.08 $\pm$0.29 &2.40 $\pm$0.39 &1.70 $\pm$0.08 &73.2 $\pm$2.9 &30.7 $\pm$1.5 \\
&13&1.6 $\pm$0.2 &0.5 $\pm$0.2 &26.6 $\pm$6.5 &0.07 $\pm$0.15 &0.16 $\pm$0.16 &0.31 $\pm$0.16 &0.54 $\pm$0.26 &0.60 $\pm$0.39 &0.58 $\pm$0.03 &14.0 $\pm$0.6 &9.5 $\pm$0.5 \\
&14&3.0 $\pm$0.2 &5.1 $\pm$0.2 &38.2 $\pm$6.8 &0.36 $\pm$0.14 &0.88 $\pm$0.16 &1.67 $\pm$0.21 &3.09 $\pm$0.38 &2.87 $\pm$0.44 &1.92 $\pm$0.10 &77.9 $\pm$3.1 &42.5 $\pm$2.1 \\\tableline
Wing&1&42.2 $\pm$0.4 &43.5 $\pm$0.4 &330.7 $\pm$42.4 &2.95 $\pm$0.30 &7.44 $\pm$0.75 &16.06 $\pm$1.61 &37.21 $\pm$3.72 &47.04 $\pm$4.71 &38.72 $\pm$1.94 &3383.6 $\pm$135.3 &531.9 $\pm$26.6 \\
&2&66.9 $\pm$0.7 &60.2 $\pm$0.6 &342.2 $\pm$43.8 &4.20 $\pm$0.42 &10.43 $\pm$1.04 &22.64 $\pm$2.27 &50.71 $\pm$5.07 &63.72 $\pm$6.37 &42.44 $\pm$2.12 &4484.7 $\pm$179.4 &864.5 $\pm$43.2 \\
&3&29.0 $\pm$0.3 &20.2 $\pm$0.2 &87.9 $\pm$11.9 &0.42 $\pm$0.11 &1.10 $\pm$0.15 &2.68 $\pm$0.29 &8.40 $\pm$0.86 &12.98 $\pm$1.32 &11.16 $\pm$0.56 &1707.6 $\pm$68.3 &177.8 $\pm$8.9 \\
&4&4.0 $\pm$0.2 &2.8 $\pm$0.1 &67.6 $\pm$9.6 &0.74 $\pm$0.14 &1.79 $\pm$0.21 &3.83 $\pm$0.40 &7.78 $\pm$0.80 &7.48 $\pm$0.79 &4.69 $\pm$0.23 &261.2 $\pm$10.4 &131.1 $\pm$6.6 \\
&5&1.3 $\pm$0.2 &-0.9 $\pm$0.2 &34.8 $\pm$7.4 &0.28 $\pm$0.16 &0.68 $\pm$0.18 &1.49 $\pm$0.22 &2.96 $\pm$0.39 &3.14 $\pm$0.49 &2.10 $\pm$0.10 &140.6 $\pm$5.6 &41.1 $\pm$2.1 \\\tableline
N88&1&51.5 $\pm$0.5 &29.4 $\pm$0.3 &130.1 $\pm$18.3 &0.53 $\pm$0.22 &1.48 $\pm$0.26 &3.83 $\pm$0.44 &12.40 $\pm$1.27 &20.77 $\pm$2.10 &19.01 $\pm$0.95 &6525.8 $\pm$261.0 &365.7 $\pm$18.3 
\enddata
\end{deluxetable*}

\end{turnpage}
\newpage
\begin{deluxetable*}{llcccccc}
\tabletypesize{\scriptsize}
\tablecaption{Physical properties of the 1.1 mm extended objects.\label{gmc:sedfit}}
\tablehead{ID&&$T_\mathrm{dust}$&$M_\mathrm{dust}$&$M_\mathrm{gas}$&$n_\mathrm{H_2}$&$N_\mathrm{H_2}$&1.1 mm free-free\\	
&&(K)&($\mathrm{M_\odot}$)&($\times 10^3 \mathrm{M_\odot}$)&($\mathrm{H_2/cm^3}$)&($\mathrm{\times 10^{21} H_2/cm^2}$)&}
\startdata
SW&1&$30.0 \pm1.0 $&$332.1 \pm32.7 $&$332.1 \pm32.7 $&$28.8 \pm2.8 $&$4.1 \pm0.4 $&4.0 ~\%\\
&2&$30.7 \pm1.1 $&$336.1 \pm33.2 $&$336.1 \pm33.2 $&$32.4 \pm3.2 $&$4.4 \pm0.4 $&6.4 ~\%\\
&3&$30.8 \pm1.1 $&$80.8 \pm8.0 $&$80.8 \pm8.0 $&$58.3 \pm5.8 $&$4.1 \pm0.4 $&5.3 ~\%\\
&4&$30.5 \pm1.0 $&$67.6 \pm6.7 $&$67.6 \pm6.7 $&$60.4 \pm6.0 $&$3.9 \pm0.4 $&8.5 ~\%\\
&5&$29.9 \pm1.0 $&$55.3 \pm5.5 $&$55.3 \pm5.5 $&$47.2 \pm4.7 $&$3.1 \pm0.3 $&6.4 ~\%\\
&6&$33.6 \pm1.4 $&$17.0 \pm2.0 $&$17.0 \pm2.0 $&$53.8 \pm6.2 $&$2.3 \pm0.3 $&2.6 ~\%\\
&7&$25.0 \pm0.8 $&$32.9 \pm3.4 $&$32.9 \pm3.4 $&$66.5 \pm6.8 $&$3.3 \pm0.3 $&0.1 ~\%\\
&8&$33.3 \pm1.5 $&$10.6 \pm1.4 $&$10.6 \pm1.4 $&$79.1 \pm10.6 $&$2.5 \pm0.3 $&5.6 ~\%\\
&9&$27.1 \pm0.9 $&$41.4 \pm4.1 $&$41.4 \pm4.1 $&$46.9 \pm4.7 $&$2.8 \pm0.3 $&2.3 ~\%\\
&10&$26.2 \pm0.9 $&$20.2 \pm2.4 $&$20.2 \pm2.4 $&$70.5 \pm8.2 $&$2.9 \pm0.3 $&1.5 ~\%\\
&11&$28.4 \pm1.3 $&$11.1 \pm1.6 $&$11.1 \pm1.6 $&$121.2 \pm17.7 $&$3.4 \pm0.5 $&0.8 ~\%\\
&12&$23.9 \pm0.7 $&$34.5 \pm3.6 $&$34.5 \pm3.6 $&$32.2 \pm3.4 $&$2.1 \pm0.2 $&0.3 ~\%\\
&13&$25.0 \pm1.0 $&$15.0 \pm2.1 $&$15.0 \pm2.1 $&$78.7 \pm10.9 $&$2.9 \pm0.4 $&3.7 ~\%\\
&14&$28.8 \pm1.7 $&$8.5 \pm1.6 $&$8.5 \pm1.6 $&$147.3 \pm27.2 $&$3.6 \pm0.7 $&3.4 ~\%\\
&15&$26.3 \pm2.2 $&$6.7 \pm1.8 $&$6.7 \pm1.8 $&$107.2 \pm28.8 $&$2.7 \pm0.7 $&0.3 ~\%\\
&16&$25.8 \pm1.9 $&$7.9 \pm1.9 $&$7.9 \pm1.9 $&$137.1 \pm32.1 $&$3.3 \pm0.8 $&5.8 ~\%\\
&17&$26.1 \pm2.3 $&$5.9 \pm1.7 $&$5.9 \pm1.7 $&$68.3 \pm19.8 $&$1.9 \pm0.5 $&2.5 ~\%\\
&18&$25.2 \pm2.6 $&$5.9 \pm1.9 $&$5.9 \pm1.9 $&$106.2 \pm34.4 $&$2.5 \pm0.8 $&3.1 ~\%\\
&19&$25.0 \pm1.1 $&$14.0 \pm2.0 $&$14.0 \pm2.0 $&$73.2 \pm10.5 $&$2.7 \pm0.4 $&2.5 ~\%\\
&20&$23.2 \pm1.1 $&$14.9 \pm2.3 $&$14.9 \pm2.3 $&$106.3 \pm16.8 $&$3.5 \pm0.5 $&1.5 ~\%\\
&21&$29.0 \pm2.6 $&$4.1 \pm1.2 $&$4.1 \pm1.2 $&$33.6 \pm9.7 $&$1.0 \pm0.3 $&8.7 ~\%\\
&22&$27.2 \pm1.1 $&$13.9 \pm1.8 $&$13.9 \pm1.8 $&$52.2 \pm6.9 $&$2.1 \pm0.3 $&1.7 ~\%\\
&23&$27.3 \pm1.9 $&$7.2 \pm1.6 $&$7.2 \pm1.6 $&$106.1 \pm23.3 $&$2.7 \pm0.6 $&8.8 ~\%\\
&24&$30.3 \pm2.2 $&$5.5 \pm1.3 $&$5.5 \pm1.3 $&$87.2 \pm20.0 $&$2.2 \pm0.5 $&9.4 ~\%\\\tableline
NE&1&$34.8 \pm1.5 $&$305.9 \pm33.7 $&$305.9 \pm33.7 $&$17.3 \pm1.9 $&$2.8 \pm0.3 $&23.0 ~\%\\
&2&$39.1 \pm1.8 $&$32.3 \pm3.7 $&$32.3 \pm3.7 $&$38.7 \pm4.4 $&$2.3 \pm0.3 $&12.4 ~\%\\
&3&$31.5 \pm1.2 $&$21.3 \pm2.4 $&$21.3 \pm2.4 $&$71.9 \pm8.1 $&$3.0 \pm0.3 $&4.6 ~\%\\
&4&$31.8 \pm1.3 $&$19.4 \pm2.3 $&$19.4 \pm2.3 $&$40.8 \pm4.8 $&$2.0 \pm0.2 $&12.8 ~\%\\
&5&$30.0 \pm1.4 $&$10.8 \pm1.6 $&$10.8 \pm1.6 $&$105.3 \pm15.9 $&$3.1 \pm0.5 $&2.1 ~\%\\
&6&$27.2 \pm0.8 $&$116.6 \pm11.2 $&$116.6 \pm11.2 $&$34.2 \pm3.3 $&$3.2 \pm0.3 $&6.8 ~\%\\
&7&$25.6 \pm2.1 $&$7.7 \pm2.1 $&$7.7 \pm2.1 $&$123.0 \pm33.0 $&$3.1 \pm0.8 $&1.8 ~\%\\
&8&$24.2 \pm1.5 $&$11.1 \pm2.2 $&$11.1 \pm2.2 $&$108.2 \pm21.6 $&$3.2 \pm0.6 $&1.6 ~\%\\
&9&$25.1 \pm1.5 $&$10.1 \pm2.0 $&$10.1 \pm2.0 $&$92.6 \pm18.4 $&$2.8 \pm0.6 $&6.2 ~\%\\
&10&$31.6 \pm1.7 $&$8.0 \pm1.3 $&$8.0 \pm1.3 $&$61.1 \pm10.2 $&$1.9 \pm0.3 $&14.7 ~\%\\
&11&$24.8 \pm1.7 $&$8.7 \pm2.1 $&$8.7 \pm2.1 $&$60.8 \pm14.6 $&$2.0 \pm0.5 $&2.4 ~\%\\
&12&$28.5 \pm1.9 $&$6.7 \pm1.4 $&$6.7 \pm1.4 $&$51.3 \pm11.0 $&$1.6 \pm0.4 $&8.5 ~\%\\
&13&$16.7 \pm2.6 $&$11.5 \pm5.4 $&$11.5 \pm5.4 $&$171.0 \pm80.1 $&$4.4 \pm2.1 $&2.3 ~\%\\
&14&$27.5 \pm1.3 $&$10.8 \pm1.8 $&$10.8 \pm1.8 $&$96.5 \pm15.7 $&$2.9 \pm0.5 $&7.1 ~\%\\\tableline
Wing&1&$33.6 \pm1.3 $&$66.8 \pm7.0 $&$66.8 \pm7.0 $&$34.3 \pm3.6 $&$2.7 \pm0.3 $&8.4 ~\%\\
&2&$34.5 \pm1.3 $&$83.9 \pm8.7 $&$83.9 \pm8.7 $&$38.0 \pm3.9 $&$3.1 \pm0.3 $&12.0 ~\%\\
&3&$42.0 \pm2.6 $&$7.6 \pm1.2 $&$7.6 \pm1.2 $&$26.1 \pm4.1 $&$1.1 \pm0.2 $&17.3 ~\%\\
&4&$29.8 \pm1.1 $&$19.5 \pm2.2 $&$19.5 \pm2.2 $&$90.1 \pm10.3 $&$3.4 \pm0.4 $&3.2 ~\%\\
&5&$29.2 \pm1.7 $&$8.5 \pm1.7 $&$8.5 \pm1.7 $&$112.4 \pm22.2 $&$3.0 \pm0.6 $&0.1 ~\%\\\tableline
N88&1&$44.8 \pm3.2 $&$9.7 \pm1.7 $&$9.7 \pm1.7 $&$53.5 \pm9.5 $&$1.9 \pm0.3 $&18.0 ~\%
\enddata
\tablecomments{The columns give (1) Source ID, (2) dust temperature, (3) total dust mass, (4) total gas mass, (5) $\mathrm{H_2}$ gas density, (6) $\mathrm{H_2}$ column density, and (7) free-free contribution ratio to 1.1 mm total flux. Total gas mass includes the atomic and molecular phases of hydrogen, and the other gas components such as He. $\mathrm{H_2}$ gas density and column density are estimated assuming that all hydrogen component consists of $\mathrm{H_2}$ molecules. }
\end{deluxetable*}
\newpage
\begin{deluxetable*}{llcc}
\tabletypesize{\scriptsize}
\tablecaption{Association with the signs of star formation activity and CO emission.\label{table:sfco}}
\tablehead{ID&&Star formation&CO emission}
\startdata
SW&1&N27\tablenotemark{a}/DEM~S40\tablenotemark{b}/S3MC~72--76,78,79\tablenotemark{c}/24 $\micron$&NANTEN/MOPRA\\
&2&N20-23,25,26/DEM~S35-38/S3MC~57,58,61--64,67/24 $\micron$&NANTEN/MOPRA\\
&3&N13AB/DEM~S16/S3MC~16,19/24 $\micron$&NANTEN/MOPRA\\
&4&N12A/DEM~S23/S3MC~16,19/24 $\micron$&NANTEN/MOPRA\\
&5&N30A/DEM~S45/S3MC~37,38/24 $\micron$&NANTEN/MOPRA\\
&6&N11/24 $\micron$&\\
&7&S3MC~23/24 $\micron$&NANTEN/MOPRA\\
&8&N33/S3MC~95,97/24 $\micron$&\\
&9&S3MC~43/24 $\micron$&MOPRA\\
&10&N51/DEM~S72/S3MC~126,127/24 $\micron$&\\
&11&S3MC~17,18,20/24 $\micron$&NANTEN/MOPRA\\
&12&&\\
&13&&\\
&14&24 $\micron$&NANTEN/MOPRA\\
&15&S3MC~146,147/24 $\micron$&\\
&16&DEM~S39&\\
&17&&\\
&18&24 $\micron$&\\
&19&&\\
&20&&NANTEN/MOPRA\\
&21&S3MC~111/24 $\micron$&\\
&22&S3MC~22&\\
&23&S3MC~70/24 $\micron$&\\
&24&&\\\tableline
NE&1&N66ABC/DEM~S103/S3MC~188--192,194--198,200,203,204/24 $\micron$&NANTEN/MOPRA\\
&2&N78ABD/DEM~S27/S3MC~243,245,246,248/24 $\micron$&MOPRA\\
&3&N77AB/DEM~S117ab/S3MC~222,223/24 $\micron$&NANTEN/MOPRA\\
&4&N76A/S3MC~237/24 $\micron$&\\
&5&N62/DEM~S93/S3MC~171,173,175/24 $\micron$&\\
&6&S3MC~225--227,229,231/24 $\micron$&NANTEN/MOPRA\\
&7&S3MC~256/24 $\micron$&\\
&8&DEM~S116/S3MC~218&MOPRA\\
&9&&\\
&10&&\\
&11&24 $\micron$&MOPRA\\
&12&S3MC~251/24 $\micron$&MOPRA\\
&13&&\\
&14&&NANTEN/MOPRA\\\tableline
Wing&1&N84ABD/DEM~S151,152/S3MC~278,280/24 $\micron$&NANTEN/MOPRA\\
&2&N83ABC/DEM~S147,148/S3MC~269,271,272/24 $\micron$&NANTEN/MOPRA\\
&3&N81/DEM~S138/S3MC~264/24 $\micron$&\\
&4&N84C/DEM~S149/24 $\micron$&NANTEN/MOPRA\\
&5&DEM~S150/S3MC~276/24 $\micron$&NANTEN/MOPRA\\\tableline
N88&1&N88/DEM~S161/24 $\micron$&MOPRA
\enddata
\tablenotetext{a}{The Henize Catalogue of SMC Emission Nebulae \citep{1956ApJS....2..315H}.}
\tablenotetext{b}{Magellanic Cloud Emission-line Survey DEM Catalog: Small Magellanic Cloud \citep{1976MmRAS..81...89D}.}
\tablenotetext{c}{S$^3$MC Young Stellar Object Catalog \citep[Table 4]{2007ApJ...655..212B}.}
\end{deluxetable*}

\newpage
\begin{deluxetable*}{ccccccl}
\tabletypesize{\scriptsize}
\tablecaption{AzTEC/ASTE 1.1 mm PCA source catalog of the SMC. \label{table:catalog_pca}}
\tablehead{ID&  &$\alpha$&$\delta$&Peak flux&S/N&Counterpart objects\\
&&$(J2000)$&$(J2000)$&(mJy/beam)&&}
\startdata
SW&1&0$^\circ$48$'$08.8$''$&-73$^d$14$^m$49$^s$&97.5$\pm$5.5&17.6&SW-2\tablenotemark{a},~N25/26\tablenotemark{b},~S38\tablenotemark{c},~SMCB-2~3\tablenotemark{d}\\
&2&0$^\circ$48$'$24.1$''$&-73$^d$05$^m$55$^s$&97.2$\pm$5.8&16.7&SW-3,~N27,~S40\\
&3&0$^\circ$45$'$21.1$''$&-73$^d$22$^m$51$^s$&95.6$\pm$5.6&17.0&SW-1,~N13,~S16\\
&4&0$^\circ$46$'$41.4$''$&-73$^d$06$^m$05$^s$&76.4$\pm$5.6&13.6&SW-4,~N12A,~LIRS~36~1\tablenotemark{e},~S23\\
&5&0$^\circ$48$'$06.3$''$&-73$^d$17$^m$52$^s$&65.0$\pm$5.7&11.5&SW-2,~N23,~SMCB-2~6\\
&6&0$^\circ$47$'$54.9$''$&-73$^d$17$^m$17$^s$&51.2$\pm$5.6&9.2&SW-2,~N21/22/23,~S34,~SMCB-2~2\\
&7&0$^\circ$48$'$55.8$''$&-73$^d$09$^m$54$^s$&49.6$\pm$5.7&8.7&SW-6,~N30A,~S45\\
&8&0$^\circ$45$'$03.4$''$&-73$^d$16$^m$41$^s$&48.2$\pm$5.5&8.7&SW-5,~N11\\
&9&0$^\circ$45$'$23.1$''$&-73$^d$12$^m$33$^s$&43.7$\pm$5.5&7.9&SW-11,~S17\\
&10&0$^\circ$49$'$29.1$''$&-73$^d$26$^m$33$^s$&41.0$\pm$5.6&7.3&SW-8,~N33\\
&11&0$^\circ$45$'$29.8$''$&-73$^d$18$^m$41$^s$&38.5$\pm$5.8&6.6&SW-7,~SMCB-1~1\\
&12&0$^\circ$48$'$18.2$''$&-73$^d$10$^m$27$^s$&35.7$\pm$5.7&6.3&SW-16,~S39\\
&13&0$^\circ$47$'$49.8$''$&-73$^d$15$^m$16$^s$&35.5$\pm$5.8&6.2&SW-2,~N20\\
&14&0$^\circ$46$'$26.0$''$&-73$^d$22$^m$09$^s$&32.9$\pm$5.8&5.7&SW-14,~SMCB-1~2,~3\\
&15&0$^\circ$48$'$05.4$''$&-73$^d$23$^m$07$^s$&32.2$\pm$5.7&5.7&SW-20\\
&16&0$^\circ$49$'$45.9$''$&-73$^d$10$^m$32$^s$&31.9$\pm$5.7&5.6&N34,~S50\\
&17&0$^\circ$44$'$18.4$''$&-73$^d$33$^m$01$^s$&31.6$\pm$6.1&5.2&SW-13\\
&18&0$^\circ$47$'$06.5$''$&-73$^d$22$^m$14$^s$&31.1$\pm$5.6&5.5&SW-10,~OGLE-CL~SMC~187\tablenotemark{f}~(OB~cluster\tablenotemark{g})\\
&19&0$^\circ$56$'$07.1$''$&-72$^d$47$^m$22$^s$&29.8$\pm$6.1&4.9&S3MC~J005606.83-724743.15~(YSO?)\\
&20&0$^\circ$52$'$37.9$''$&-73$^d$26$^m$38$^s$&29.7$\pm$5.5&5.4&N51,~S72\\
&21&0$^\circ$54$'$03.7$''$&-73$^d$19$^m$39$^s$&29.6$\pm$5.5&5.4&S3MC~J005403.36-731938.30\tablenotemark{h}~(YSO)\\
&22&0$^\circ$46$'$34.4$''$&-73$^d$15$^m$46$^s$&28.9$\pm$5.8&5.0&SW-9,~KHBG~27\tablenotemark{i}~(Sc/Sd~Galaxy)\\
&23&0$^\circ$44$'$57.2$''$&-73$^d$10$^m$09$^s$&28.6$\pm$5.7&5.0&SW-15,~N10,~S11\\
&24&0$^\circ$52$'$47.6$''$&-73$^d$17$^m$54$^s$&26.6$\pm$5.7&4.7&-\\
&25&0$^\circ$49$'$46.2$''$&-73$^d$24$^m$32$^s$&26.6$\pm$5.6&4.7&SW-12,~S3MC~J004946.18-732426.49~(YSO?)\\
&26&0$^\circ$45$'$41.6$''$&-73$^d$13$^m$53$^s$&26.2$\pm$5.7&4.6&S3MC~J004542.11-731344.86~(YSO?)\\
&27&0$^\circ$46$'$51.4$''$&-73$^d$15$^m$22$^s$&25.7$\pm$5.7&4.5&SW-9,~2MASS~J00465185-7315248\tablenotemark{j}~(YSO?)\\
&28&0$^\circ$48$'$08.7$''$&-73$^d$08$^m$49$^s$&25.5$\pm$5.6&4.5&S3MC~J004809.65-730832.42~(YSO?)\\\tableline
NE&1&1$^\circ$05$'$05.7$''$&-71$^d$59$^m$34$^s$&77.0$\pm$6.8&11.2&NE-2\tablenotemark{a},~N78A/B,~S126\\
&2&0$^\circ$58$'$54.2$''$&-72$^d$09$^m$57$^s$&63.7$\pm$6.6&9.6&NE-1,~N66,~S103,~NGC~346\tablenotemark{k}\\
&3&0$^\circ$59$'$08.0$''$&-72$^d$11$^m$05$^s$&55.1$\pm$7.1&7.8&NE-1,~N66A,~S103,~NGC~346\\
&4&1$^\circ$02$'$50.3$''$&-71$^d$53$^m$40$^s$&51.2$\pm$6.9&7.4&NE-3,~N77A/B,~S117\\
&5&0$^\circ$57$'$58.7$''$&-72$^d$39$^m$21$^s$&49.2$\pm$6.7&7.3&NE-5,~N62,~S93\\
&6&0$^\circ$59$'$15.2$''$&-72$^d$11$^m$20$^s$&47.9$\pm$6.7&7.1&NE-1,~N66A,~S103,~NGC~346\\
&7&1$^\circ$06$'$03.1$''$&-72$^d$03$^m$35$^s$&45.6$\pm$6.7&6.8&NE-7,~N78C,~S130\\
&8&1$^\circ$02$'$31.7$''$&-71$^d$56$^m$54$^s$&39.3$\pm$6.5&6.0&NE-8,~N75,~S116\\
&9&1$^\circ$03$'$48.8$''$&-72$^d$03$^m$59$^s$&37.6$\pm$6.9&5.5&NE-1,~N76A,~S123\\
&10&0$^\circ$59$'$24.1$''$&-72$^d$08$^m$20$^s$&37.5$\pm$6.5&5.8&NE-1,~N66,~S103,~NGC~346\\
&11&0$^\circ$59$'$19.0$''$&-72$^d$09$^m$14$^s$&37.4$\pm$6.8&5.5&NE-4,~N66,~S103,~NGC~346\\
&12&0$^\circ$58$'$47.8$''$&-72$^d$13$^m$06$^s$&33.4$\pm$6.7&5.0&NE-1,~N66,~S103,~NGC~346\\
&13&1$^\circ$05$'$43.5$''$&-72$^d$03$^m$11$^s$&32.8$\pm$7.0&4.7&N78C,~S130\\
&14&0$^\circ$58$'$45.2$''$&-72$^d$12$^m$45$^s$&32.5$\pm$6.9&4.7&NE-1,~N66,~S103,~NGC~346\\\tableline
Wing&1&1$^\circ$09$'$13.1$''$&-73$^d$11$^m$36$^s$&58.6$\pm$7.6&7.8&Wing-3\tablenotemark{a},~N81,~S138\\
&2&1$^\circ$14$'$04.9$''$&-73$^d$17$^m$01$^s$&54.8$\pm$7.4&7.4&Wing-2,~N83C,~S147\\
&3&1$^\circ$14$'$48.1$''$&-73$^d$19$^m$54$^s$&52.7$\pm$7.2&7.3&Wing-1,~N84B,~S152\\
&4&1$^\circ$11$'$32.6$''$&-73$^d$02$^m$13$^s$&48.9$\pm$7.2&6.8&6dFGS~gJ011132.5-730210~(galaxy,~AGN)\tablenotemark{l},~MM~J01115-7302\tablenotemark{m}\\
&5&1$^\circ$14$'$37.6$''$&-73$^d$18$^m$30$^s$&49.0$\pm$7.2&6.8&Wing-1,~N84A,~S151\\
&6&1$^\circ$13$'$52.0$''$&-73$^d$15$^m$55$^s$&47.4$\pm$7.2&6.6&Wing-2,~N83B,~S148\\
&7&1$^\circ$07$'$03.6$''$&-73$^d$02$^m$00$^s$&43.1$\pm$7.6&5.6&MM~J01071-7302\tablenotemark{m}~(strongly-lensed~submillimeter~galaxy)\\
&8&1$^\circ$14$'$25.1$''$&-73$^d$14$^m$10$^s$&39.2$\pm$7.3&5.4&Wing-5,~S3MC~J011421.63-731353.6~(YSO?)\\
&9&1$^\circ$14$'$21.7$''$&-73$^d$15$^m$38$^s$&38.7$\pm$7.3&5.3&Wing-4,~N84C,~S149\\
&10&1$^\circ$13$'$49.1$''$&-73$^d$18$^m$05$^s$&34.1$\pm$7.3&4.7&Wing-2,~N83A,~S147\\\tableline
N88&1&1$^\circ$24$'$08.6$''$&-73$^d$09$^m$03$^s$&112.5$\pm$12.7&8.9&N88-1\tablenotemark{a},~N88,~S161\\
\enddata
\tablecomments{The columns give (1) source ID, (2) right ascension, (3) declination, (4) observed peak flux and noise level, (5) signal-to-noise ratio, and (6) counterpart objects.}
\tablenotetext{a}{AzTEC/ASTE 1.1 mm extended source catalog (Table \ref{table:catalog_fruit}).}
\tablenotetext{b}{The Henize Catalogue of SMC Emission Nebulae \citep{1956ApJS....2..315H}.}
\tablenotetext{c}{Magellanic Cloud Emission-line Survey DEM Catalog: Small Magellanic Cloud \citep{1976MmRAS..81...89D}.}
\tablenotetext{d}{Molecular Clouds in the SMC \citep{1993A&A...271....9R}.}
\tablenotetext{e}{Molecular Clouds in the SMC \citep{2008ApJ...686..948B}.}
\tablenotetext{f}{Star clusters in the SMC \citep{2000AJ....119.1214B}.}
\tablenotetext{g}{\citet{2004AJ....127.1632O}.}
\tablenotetext{h}{The Spitzer Survey of the SMC Catalog \citep{2007ApJ...655..212B}.}
\tablenotetext{i}{HST photometry of galaxy clusters behind the SMC \citep{2001PASP..113.1115K}.}
\tablenotetext{j}{2MASS All Sky Catalog of point sources \citep{2003tmc..book.....C}.}
\tablenotetext{k}{the New General Catalogue of Nebulae and Clusters of Stars (NGC).}
\tablenotetext{l}{The 6dF Galaxy Survey Catalog\citep{2009MNRAS.399..683J}.}
\tablenotetext{m}{\citet{2013ApJ...774L..30T}.}

\end{deluxetable*}

\end{document}